\documentclass[iop,numberedappendix,appendixfloats,twocolappendix]{emulateapj}
\usepackage{color}
\usepackage{subfigure}
\usepackage{natbib}
\usepackage{amsmath}
\usepackage{multirow}
\shorttitle{Dwarf\ Galaxy\ Dark\ Matter\ Density\ Profiles}
\shortauthors{Adams\ et\ al.}
\slugcomment{Accepted by \apj}

\begin{document}
\title{Dwarf Galaxy Dark Matter Density Profiles Inferred from Stellar and Gas Kinematics\altaffilmark{*}
}
\author{Joshua J.~Adams\altaffilmark{1,2}, Joshua D.~Simon\altaffilmark{1}, 
Maximilian H.~Fabricius\altaffilmark{3}, Remco C. E.~van den Bosch\altaffilmark{4}, 
John C.~ Barentine\altaffilmark{5}, 
Ralf Bender\altaffilmark{3,6}, Karl Gebhardt\altaffilmark{5,7}, Gary J.~Hill \altaffilmark{5,7,8}, Jeremy D.~Murphy\altaffilmark{9}, 
R.~A.~Swaters\altaffilmark{10}, 
Jens Thomas\altaffilmark{3}, Glenn van de Ven\altaffilmark{4}, 
} 
\altaffiltext{*}{This paper includes data obtained at The McDonald Observatory of The University of Texas at Austin.}
\altaffiltext{1}{Observatories of the Carnegie Institution of Science, 813 Santa Barbara Street, Pasadena, CA 91101}
\altaffiltext{2}{Current address: ASML US, 77 Danbury Rd, Wilton, CT 06897; jja439@gmail.com}
\altaffiltext{3}{Max-Planck Institut f\"{u}r extraterrestrische Physik, Giessenbachstra\ss e, D-85741 Garching bei M\"{u}nchen, Germany}
\altaffiltext{4}{Max-Planck Institut f\"{u}r Astronomie, K\"{o}nigstuhl 17, 69117 Heidelberg, Germany}
\altaffiltext{5}{Department of Astronomy, University of Texas at Austin, 2515 Speedway, Stop C1400, Austin, Texas 78712-1205, USA}
\altaffiltext{6}{Universit\"{a}ts-Sternwarte, Ludwig-Maximilians-Universit\"{a}t, Scheinerstrasse 1, D-81679 M\"{u}nchen, Germany}
\altaffiltext{7}{Texas Cosmology Center, University of Texas at Austin, 1 University Station C1400, Austin, TX 78712, USA}
\altaffiltext{8}{McDonald Observatory, 2515 Speedway, Stop C1402, Austin, Texas 78712-1205, USA}
\altaffiltext{9}{Department of Astrophysical Sciences, Princeton University, 4 Ivy Lane, Peyton Hall, Princeton, NJ 08544, USA}
\altaffiltext{10}{National Optical Astronomy Observatory, 950 North Cherry Avenue, Tucson, AZ 85719, USA}
\begin{abstract}
We present new constraints on the density profiles of dark matter (DM) halos in seven nearby 
dwarf galaxies from measurements of their integrated stellar light 
and gas kinematics. 
The gas kinematics of low mass galaxies frequently suggest that they contain 
constant density DM cores, while N-body simulations instead predict a cuspy 
profile. We present a data set of high resolution integral field 
spectroscopy on seven galaxies and measure the stellar and 
gas kinematics simultaneously. Using Jeans modeling on our full sample, we 
examine whether gas kinematics in general produce shallower density profiles than are 
derived from the stars. Although 2/7 galaxies show some localized differences 
in their rotation curves between the two tracers, 
estimates of the central logarithmic slope of the DM density profile, 
$\gamma$, are generally robust. 
The mean and standard deviation of the logarithmic slope for the population 
are $\gamma=0.67\pm0.10$ when measured in the stars and $\gamma=0.58\pm0.24$ when 
measured in the gas. We also find that the halos are not under-concentrated 
at the radii of half their maximum velocities. 
Finally, we search for correlations of the DM density profile with stellar 
velocity anisotropy 
and other baryonic properties. Two popular mechanisms to explain cored DM halos are an exotic DM component or feedback models 
that strongly couple the energy of supernovae into repeatedly driving out gas and dynamically heating the DM halos. While 
such models do not yet have falsifiable predictions that we can measure, we investigate correlations that may eventually 
be used to test models. We do not find a secondary parameter that strongly correlates with the central DM density slope, but 
we do find some weak correlations. The central DM density slope weakly 
correlates with the abundance of $\alpha$ elements in the 
stellar population, anti-correlates 
with HI fraction, and anti-correlates with vertical orbital anisotropy. 
We expect, if anything, the opposite of these three trends for feedback models. Determining the importance of these correlations 
will require further model developments and larger observational samples.
\end{abstract}


\section{Introduction}
\label{sec:intro}
\par Rotationally supported galaxies have historically been important objects for revealing and characterizing 
dark matter (DM), starting with the asymptotically flat rotation curves seen by \cite{Freem70} and 
\cite{Rubin70}. Several following works 
\citep{Roberts75,Bosma78,Rubin78a,Rubin78b,Rubin80,Bosma81a,Bosma81b} strengthened the case for DM in disky galaxies to the point of scientific consensus. 
DM characterization brought further surprises. For twenty years since \cite{Flore94} and \cite{Moore94}, there 
has been tension between theoretically expected and 
observed distributions of 
DM in the central regions of late-type dwarf galaxies. This ``core-cusp'' problem is that 
N-body simulations predict that cold dark matter (CDM) settles into a cuspy distribution with density rising to the 
smallest observable or simulatable radii, while kinematic observations 
often favor approximately constant density cores at a common scale of $\sim$ 1 kpc at the center of galaxies. 
Large investments in computational models of galaxies have led to several plausible physical mechanisms to 
create DM cores. Meanwhile, more and better observations have been pursued to retire systematic risks particular to 
some analysis methods and tracers, characterize enough systems to make statistical statements, and search for additional observables that 
could constrain the theoretical models. Even the simple statement that the ``core-cusp'' 
problem is unsolved may be disputed by 
some workers in this field, but our present study will adopt this agnostic stance.
\par This paper makes an empirical study of mass distributions in late-type dwarf galaxies with a kinematic 
tracer rarely employed in this subject, stellar kinematics via spectroscopy of 
integrated light, in addition to the more 
traditionally used emission-line kinematics of nebular gas. We present data and mass models for seven 
such galaxies observed with a wide-field integral field spectrograph at optical wavelengths. This study builds on our results  
from \cite{Adams12}, where one such galaxy was studied in both tracers at a lower spectral resolution. That work used the 
gaseous kinematics in NGC~2976 to constrain the DM to a strongly cored profile, as also found by 
previous groups with HI and H$\alpha$ data sets on the same galaxy \citep{Simo03,deBl08}. 
However, the best-fit solutions to the axisymmetric 
Jeans equations and the stellar kinematics instead indicated, with sizable uncertainty and covariance with the stellar 
mass-to-light ratio, that a cuspy DM distribution exists in NGC~2976. With $\gamma$ being the logarithmic density slope 
and $\gamma=1$ being 
the canonical value for a cuspy distribution from the Navarro, Frenk, and White distribution \citep[NFW;][]{Nava96a}, we found the best fit 
from the stellar tracers at $\gamma=0.9$, while the fully cored ($\gamma=0$) fit was disfavored at 2.5$\sigma$ significance. We will 
expound on our interpretation for this discord, in light of the newer data and a reexamination 
of the luminosity profile, in \S \ref{sec_N2976comp}. While the new data 
presented in this paper have higher spectral resolution, essential for resolving the stellar velocity dispersion, and signal-to-noise, 
the DM parameter constraints only modestly tighten compared to the \cite{Adams12} results. The reason for the persistent 
uncertainty is that the stellar 
mass-to-light ratio is uncertain and even in low-mass, late-type galaxies 
the ``disk-halo degeneracy'' \citep{vanA86} prevents a unique decomposition 
of the rotation curve. 
\par One important observational constraint for the theoretical models is the presence or 
absence of cores over a large 
range of halo masses. Dwarf spheroidal galaxies (dSph) have a long literature of DM profile constraints, although conclusions 
are controversial and subject to debates on modeling systematics. dSph studies quite often infer DM cores 
\citep{Walke11,Amori12,Jarde12,Amori13,Amori13b}, although \cite{Jarde13} find a cusp in Draco with a non-parametric 
density profile in a Schwarzschild model, \cite{Bredd13} use Schwarzschild 
models of Sculptor to conclude 
that the density profile is unconstrained with current data, and 
\cite{Richa14} argue that Sculptor may have a cuspy halo. At much larger masses, constraints from kinematics 
and lensing in cD galaxies at the centers of relaxed clusters 
have shown that while the total mass profile is cuspy, the decomposed DM profiles are often shallower \citep{Sand02,Sand04,Newman13a,Newman13b}. 
Whether one theoretical mechanism can explain shallow density profiles over 6 orders of magnitude 
in halo mass or whether multiple 
mechanisms are conspiring together over different mass scales in unknown.
\subsection{Theoretical Explanations for the Existence of Cores}
\par Presuming that the inferences of cores in many galaxies is correct, 
the most satisfying resolution to the ``core-cusp'' problem would be to identify one physical mechanism
responsible for DM cores, and that this mechanism would meet all the observational constraints while 
competitor theories would not. Since many of the theoretical mechanisms are still being developed, such a 
falsification is presently impossible. We will briefly mention three types of proposed mechanisms. First, 
there may be structural features in real galaxies that are not present in the N-body simulations that can 
transfer energy and angular momentum to DM at small radii. \cite{ElZa01} make simulations where clumps of DM 
fall inward due to dynamical friction and deposit energy to the more numerous pool of DM particles at small radii. 
\cite{Tonini06} similarly make models where baryons and DM exchange angular momentum in the early stages 
of galaxy formation with a prediction that the inner disk will have predominantly tangential orbits. 
A string of works have modeled resonances between DM halos and bars which may be missed by usual N-body simulations 
that have the effect of creating DM cores \citep{Weinb02,Holle05,Weinb07a,Weinb07b}. That said, the importance of 
this mechanism is disputed by \cite{Sellw03} and others \citep{McMil05,Sellw08,Dubin09}. 
\par Second, non-CDM models produce much less structure on small scales. 
The two most prominent non-CDM candidates are warm dark matter \citep{Hoga00,Avila01,Abaza01,Kapli05,Cembr05,Strig07} and 
self-interacting DM \citep{Sperg00,Kapli00,Rocha13,Peter13}. 
While warm dark matter initially 
seemed promising as an explanation for cores, it now seems to be ruled out \citep[e.g.][]{Maccio12a}. 
In particular for self-interacting DM models, 
it appears that cross-section-to-mass values of $\sigma/\text{m}\sim0.1~\text{cm}^2~\text{g}^{-1}$ are compatible with all 
current constraints \citep{Rocha13} but perhaps too small to create substantial cores. 
\cite{Kuzi10} have made arguments that rotation curves in late-type galaxies do not provide compelling evidence 
for non-CDM models. Recent work by \cite{Kapli13} has found that the 
natural variation in fractional disk mass can explain much of the scatter in core densities and sizes and still leave room 
for non-CDM models, particularly self-interacting ones. One potentially falsifiable prediction from that simulation is how  
disk scale and core size may correlate. 
\par Third, baryons may be responsible for a feedback mechanism that 
dynamically heats the central DM. While early 
simulations of this effect, such as by the instantaneous removal of the disk potential in N-body simulations \citep{Nava96b}, 
failed to produce a long-lived core, more recent simulations have found greater success. High spatial resolution 
hydrodynamical simulations with updated prescriptions 
for supernova feedback have been used to model the effects in dwarf 
galaxies \citep{Gove10,Gover12,Maccio12b,DiCin14}. \cite{Pontz12} have 
presented analytic approximations that also irreversibly transfer energy to DM particles by repeatedly 
changing the baryonic potential on kpc scales. In both cases, the DM profile is seen to become shallower ($0.2<\gamma<0.8$) 
than the initial cusp over $10^8 M_{\odot}<M_{*}<10^{10} M_{\odot}$, which encompasses the range in our study. Interestingly, 
at the lower masses relevant to dSphs, this mechanism becomes ineffective at forming cores. These models have been 
compared to the THINGS HI observations \citep{Walt08,deBl08,Oh08,Oh11a,Oh11b} and produce similar density profiles. Another group has simulated 
feedback in dwarf and disk galaxies from SN and the radiation pressure 
from massive stars and find the latter to be most important \citep{Truji14}. 
Their simulations produce some cases of constant and rising star formation 
histories and match well the constraints of star formation history 
across redshift. The DM halos in their simulations are only weakly impacted 
by the feedback, with the strongest cases being their fiducial model 
dwRP\_1 at $\gamma=$0.7. Finally, it must be remembered that baryons 
could cause contraction rather than expansion of DM particles. 
The adiabatic contraction models of \cite{Blum86} have long been explored, yet the observational data are 
ambiguous regarding whether this process happens in real galaxies \citep{Dutt05,Dutt07,Dutt09,Thoma11}.  
\subsection{Observational History of DM Density Profile Measurements}
\par Several observational studies have been the impetus behind the previously discussed theoretical models. Following the original two 
``core-cusp'' papers \citep{Flore94,Moore94}, more HI data strengthened the case for cores \citep{deBl96,deBl97}. Rotation curves of 
ionized gas from optical longslit spectroscopy were obtained \citep{vdB01,McGau01,deBl01,deBl01b,deBl02,Swat03} as a 
better spatial resolution complement to the HI data. Some work with Fabry-Perot interferometers also permitted 
high resolution kinematic data over two spatial dimensions \citep{Blais01,Blais04}. 
This step allowed limited resolution and beam-smearing to be investigated 
as a source of systematic error that could be artificially manifesting cores. \cite{Blais04} found the beam-smearing to be 
significantly biasing the DM profile in NGC~5055, but others \citep{deBl02,March02,Genti04} concluded that beam-smearing was 
not a significant systematic for most observations. 
The debate moved on to other systematic 
sources of error, such as dynamical centers, slit position angle uncertainties, asymmetries, and non-circular motions. From these data, 
\cite{deBl03} argued that these possible systematics could not explain away the evidence for cores while others 
\citep{vdB01,Swat03,Rhee04} 
took the opposite view. \cite{Genti05} investigated non-circular motions from halo triaxiality in DDO~47 and found it to be 
an unimportant effect. 
\cite{Spek05} showed that longslit data alone cannot generally provide meaningful constraints on DM profiles. 
\par The next round of data used integral field spectrographs to deliver larger samples of 
two-dimensional kinematic maps at high resolution \citep{Swat03b,Simo03,Simo05,Kuzi06,Kuzi08}. These data are more immune to certain systematics such as 
slit misalignment, more able to quantify non-circular motion over a range of azimuthal angles, and much more constraining of 
DM parameters in the presence of degeneracies with stellar mass-to-light values and velocity anisotropies. \cite{Kuzi09} 
focused on constraining the systematics with dynamical models for their recent data. The broad result from these 
papers was that DM halos with a range of profiles exist, with cores present in some fraction of late-type dwarfs. 
Yet despite the improvement afforded by optical kinematic maps, the exact distribution 
of logarithmic DM slopes is still controversial \citep{Simo05,Oh11a}. 
\par A related topic of the baryonic mass-to-light ratio in such late-type dwarfs was addressed by 
the DiskMass Survey team \citep{Bers10a} and by \cite{Herr09}. These works used measurements of the 
vertical dispersion along with inferences for the disk scale lengths and the dynamical relation for an isothermal sheet to 
weigh the baryonic disks. The major conclusion is that the disks are substantially submaximal \citep{Bers11}. There is less 
agreement on the ability of these data to measure DM profiles. \cite{Herr09} claim their five quite 
massive galaxies prefer DM cores, while 
\cite{West11} give a thorough analysis of one galaxy in their sample, UGC~463, and conclude that the data cannot constrain the profile. 
\cite{Herr09} use planetary nebulae as the kinematic tracers while the DiskMass Survey uses both gaseous and stellar kinematics. 
\par Meanwhile, the THINGS team has worked to develop methods to better isolate velocity fields from non-circular motions \citep{Oh08} and 
gather a large sample of HI observations at homogeneous sensitivity and resolution. That team presented velocity fields and mass 
models for 19 galaxies in \cite{deBl08} and a more stringently selected subset of seven low-mass 
galaxies in \cite{Oh11a}. The 
mean value of logarithmic DM slopes for the dwarf galaxy subset was $0.29\pm0.07$. 
\subsection{Outline of Included Work}
\par In this paper we measure DM halo properties from kinematic models 
and try to find correlations to understand 
the physics that are causing deviation from $\Lambda$CDM 
predictions at small scales. 
We present our data and reduction methods in \S \ref{sec:obs}. In \S \ref{sec_kin}, we describe our methods 
to extract stellar and gaseous kinematics from the reduced data. The kinematic template construction details 
are given in Appendix \ref{app:temp}. Appendix \ref{app:mgedm} gives some numerical methods details. 
We describe our dynamics model constraints in \S \ref{sec:modelstot} along with 
tests for the effects of orbital anisotropy error (Appendix \ref{app:betatest}) 
and convergence (Appendix \ref{app:MCMCtest}). 
In \S \ref{sec:interp} we 
investigate the consistency of our results both internally and to external data sets. 
\S \ref{sec:dis} gives our discussion of the results, our analysis of 
stellar population parameters from Lick indices (with details in Appendix \ref{app:Lick}), 
and our attempts to find correlations between the DM logarithmic slope and 
other galaxy properties. Finally, we state our conclusions in \S \ref{sec:conc}. 
\par The topic of DM density profiles has suffered from imprecise language in much of the literature. 
For example, some studies refer to a density profile with $\gamma=0.5$ (or 
any value of $\gamma < 1$) as a ``core'', while others would describe the 
same profile as a ``shallow cusp'' since the density continues to increase 
toward smaller radii. Since all our measurements find intermediate 
values, we will not emphasize these discrete terms here and instead focus 
on the numerical value of $\gamma$. 
\par Throughout this paper, all magnitudes are quoted on the AB system. Mass-to-light values 
($\Upsilon_*$), as usual, are all relative to solar values of mass and 
luminosity. To convert from observed fluxes at a fixed distance into solar luminosity units, we 
need the absolute magnitude of the sun in the filters of interest. We adopt 
M$_{\odot,r}$=4.64 and M$_{\odot,R}$=4.61 \citep{Blant07}. The dimensionless Hubble constant is assumed to be 
0.7 throughout \citep{Komat11}.
\section{Observations and Reductions}
\label{sec:obs}
\subsection{Sample Selection}
\label{sec_samp}
\par Our observations are limited in scope and sample size by the sparse photon counts per resolution element achievable 
in a reasonable time. The observations are not complete to any strict selection criteria, but several 
parameters were considered when selecting targets. We observed galaxies at 3 Mpc$<$D$<$20 Mpc so that 
a modest number of tiled exposures could cover the targets and to allow the fibers to 
subsample the $\sim$1 kpc scale typically found for cores. We chose galaxies at moderate inclination, 40\arcdeg$<i<$75\arcdeg, 
so that the luminous scale lengths could be measured, rotation could be measured, and all components to the 
stellar velocity ellipsoid (SVE) would be present in projection. Our prejudice was to target the lowest 
surface brightness galaxies possible with the instrument, given that they should be more dark matter 
dominated, but we found that low surface brightness galaxies ($\mu_{0,B}>22.5$ mag arcsec$^{-2}$) 
required unfeasible exposures. Some properties of our sample are given in Table \ref{tab:proplog}.
In particular, we have listed in the total HI mass as derived from the m$_{21c}$ measurements in HyperLeda 
\citep{Patu03} via the standard relation for z=0 galaxies from RC3 \citep{deVa91}: 
\begin{equation}
\label{eq:HI}
\log M_{HI} = -0.4 \times m_{21} + 2 \times \log D +12.3364.
\end{equation} 
The original HI data sources can be found on the HyperLeda website. There are roughly ten 
measurements for each of our target galaxies that go into the HyperLeda values. 
We have taken distance estimators from the EDD Distance table in the Extragalactic Distance Database \citep{Tully09}. 
The table gives a preferred distances, based on one of a number of methods, as listed in the footnote. 

\begin{deluxetable*}{lccccccccccc}
\tabletypesize{\scriptsize}
\tablecaption{VIRUS-W Sample Properties\label{tab:proplog}}
\tablewidth{0pt}
\tablehead{
\colhead{Galaxy} & \colhead{$\alpha_0$} & \colhead{$\delta_0$} & \colhead{Distance} & \colhead{V$_{sys}$} &
\colhead{R$_d$} & \colhead{i} & \colhead{PA} & \colhead{$\log M_{HI}/M_{\odot}$} & \colhead{$\log L/L_{\odot}$} & \colhead{$\sigma_{g}$}\\
\colhead{} & \colhead{} & \colhead{} & \colhead{(Mpc)} & \colhead{(heliocentric)} & 
\colhead{(\arcsec)} & \colhead{(\arcdeg)} & \colhead{(\arcdeg)} & \colhead{} & \colhead{} & \colhead{}\\
\colhead{} & \colhead{} & \colhead{} & \colhead{} & \colhead{(km s$^{-1}$)} &
\colhead{} & \colhead{} & \colhead{} & \colhead{} & \colhead{} & \colhead{(km s$^{-1}$)}}
\startdata
NGC959 & 02:32:23.85 & +35:29:40.5 & 9.86\tablenotemark{1} & 595.0 & 20.0 & 55.2 & 67.9 & 8.33 & 8.93 & 15.0 \\
UGC2259 & 02:47:55.41 & +37:32:18.8 & 9.86\tablenotemark{1} & 581.5 & 22.9 & 40.9 & 156.2 & 8.58 & 8.36 & 10.3 \\
NGC2552 & 08:19:20.00 & +50:00:31.8 & 11.4\tablenotemark{1} & 523.3 & 35.2 & 52.7 & 60.54 & 8.81 & 9.10 & 15.4 \\
NGC2976 & 09:47:15.31 & +67:55:00.1 & 3.58\tablenotemark{2} & 2.0 & 46.6 & 61.9 & -38.1 & 8.04 & 8.98 & 17.5 \\
NGC5204 & 13:29:36.58 & +58:25:13.2 & 3.25\tablenotemark{2} & 200.0 & 30.3 & 46.8 & 176.8 & 8.30 & 8.37 & 14.7 \\
NGC5949 & 15:28:00.70 & +64:45:47.4 & 14.3\tablenotemark{1} & 440.0 & 19.1 & 62.0 & 144.1 & 8.36 & 9.41 & 11.5 \\
UGC11707 & 21:14:31.73 & +26:44:05.9 & 15.0\tablenotemark{3} & 898.0 & 30.5 & 72.7 & 54.5 & 9.15 & 9.04 & 10.2
\enddata
\tablecomments{Distance method used; 
1:Cosmicflow-1 group, 2:CMDs/TGRB, 3:Numerical Action kinematic model. All distances from \cite{Tully09}. 
The galaxy centers, disk scale lengths, R$_d$, inclinations, position angles, and stellar luminosities quoted here have been fit 
by us photometrically. The total HI masses have been taken from the literature. The systemic velocities and average 
gas dispersions, $\sigma_g$, have been fit by us spectroscopically.}
\end{deluxetable*}
\par We also preferred target galaxies for which DM profiles from gas kinematics 
had been measured in the literature. We describe their literature values and how 
they relate to our derivations in \S \ref{sec_N2976comp}. 
\par In addition to the sample of seven galaxies with high quality data, we attempted 
observations on four fainter sources. The S/N for these four is too low to 
achieve useful kinematic constraints, so we do not present them here. In order to 
reach a minimum S/N=10 per pixel, which is our desired threshold 
for kinematic extraction, 
we had to bin the galaxies into roughly ten bins per galaxy. The four dropped 
from the sample are UGC191 (observed for 18.7 hours), UGC3371 (observed for 22.1 
hours), UGC11557 (observed for 23.0 hours), and UGC12732 (observed for 15.1 hours). 
Our sample contains most of the well-behaved nearby galaxies in the northern 
hemisphere with high enough surface brightness to have their stellar 
kinematics measured with current instruments, telescopes, and manageable 
exposure times. Larger samples can yet be gathered with heavy exposure 
time investments. Some of us are leading a large observational project 
which includes longer allocations than available in this paper. These data
are the best prospect to further build a stellar kinematics sample of 
this sort with any instrumentation either built or in development. Since
we have shown in this work that DM density slopes have a very small
bias when determined through gaseous kinematics as compared to stellar, 
larger samples of high quality gaseous data with 2D spatial coverage 
are also desireable. Some of us are taking such data.
\subsection{Photometry}
\label{sec:phot}
\par Photometry is a necessary source of information on the contribution from luminous stellar matter to the potential. 
It is also used, once deprojected, to compute the weights along the line-of-sight through the 
potential to form the line-of-sight velocity distribution (LOSVD). Quality photometry is a necessary complement to the more expensive spectroscopic data when making 
resolved mass models. Several works have investigated the utility of different filters in this application. 
The goal of filter selection is to find colors that can reliably trace $\Upsilon_*$ with minimal sensitivity to 
complex star formation histories, dust, and nebular emission line contamination. These requirements naturally 
lead to red and near-infrared filters. While one goal is to have a photometric constraint on the absolute value of 
$\Upsilon_*$, it can alternatively be fit in the dynamical modeling. A more important consideration is to have a filter 
that is minimally sensitive to stellar population ages and dust so that gradients in $\Upsilon_*$ can 
be ignored in the dynamical modelings. Beginning with \cite{Bell01}, model values of $\Upsilon_*$ from 
simple stellar populations were shown to correlate with tabulated optical and near-infrared colors. Later work by 
\cite{Port04} and \cite{Zibe09} has updated such models and studied further systematics. 
As discussed in \cite{Bers10a}, the 
near-infrared models 
suffer some disagreement from differences in the treatment of thermally pulsating active asymptotic giant 
branch stars (TP-AGB), and the total uncertainty in the absolute $\Upsilon_*$ under the 
assumption that all the 
modeling uncertainties are independent is a factor of $\sim$4, even in the most optimal filter choices. Our 
filter choices are informed by these studies, but actual selection was based on availability and 
signal-to-noise. 

\subsubsection{New and Archival Data}
\par We have gathered optical photometry for the seven galaxies with high quality spectroscopic data. Table 
\ref{tab:photlog} gives the specific sources and properties. Most of the photometry comes from the 
Sloan Digital Sky Survey DR10 mosaic tool. Only UGC~02259 did not have publicly available imaging in our sample. 
We observed UGC~02259 with the S2KB CCD on the 0.9m WIYN telescope. The conditions were not photometric. 
Standard calibrations were taken. The data were reduced with the \texttt{CCDRED} package in IRAF \citep{Tody86}. 
The three related software packages of SExtractor \citep{Berti96}, SCAMP, and SWarp \citep{Berti02} 
were used to create the final image. 
SExtractor was used to detect sources, while SCAMP and SWarp were used to fit an astrometric 
solution and resample the exposures to a common image. Finally, we measured a rather uncertain photometric solution 
to the image by matching sources to the NOMAD catalog \citep{Zacha05}. NOMAD is a compilation catalog, and 
we have used the R-band magnitudes drawn originally from the USNO-B1 survey \citep{Monet03}. The absolute 
photometry is uncertain because our observing conditions were uncertain by several tenths of 
magnitudes and the source of the flux standards in this case used photometric plates. The 
photometric accuracy of USNO-B1 is quoted by \cite{Monet03} as 0.3 mag, and a recent recalibration by 
\cite{Madse13} claims an accuracy of 0.1 mag. We remind the reader that our main concern with photometry 
is relative accuracy as the absolute normalization affects only the IMF.
\begin{deluxetable*}{lcccr}
\tabletypesize{\scriptsize}
\tablecaption{Photometric Data Log\label{tab:photlog}}
\tablewidth{0pt}
\tablehead{
\colhead{Galaxy} & \colhead{Source} & \colhead{Acquisition Date} & \colhead{Filter} & \colhead{Exposure time (s)}}
\startdata
NGC0959  & 2 & 01-15-02 & R & 360 \\
UGC02259 & 4 & 11-04-11 & R & 1800 \\
NGC2552  & 1 & 04-25-00 & r & 53.9 \\
NGC2976  & 1 & 11-20-03 & r & 53.9 \\
NGC5204  & 1 & 04-15-01 & r & 53.9 \\
NGC5949  & 1 & 06-15-04 & r & 53.9 \\
UGC11707 & 3 & 05-14-96 & R & 600
\enddata
\tablecomments{Imaging source for the luminosity profile measurements; 
1:SDSS image, 2:Image from \cite{Taylo05}, 3:Image from \cite{Swat02}, 4:new image taken with WIYN 0.9m.}
\end{deluxetable*}
\par A general functional form to parametrize galactic photometry has been introduced by \cite{Monn92} and 
\cite{Emse94} which is 
very useful in the high signal-to-noise, high spatial resolution regime that accompanies nearby galaxies. It is also 
the parametrization necessary to execute the Jeans Anisotropic Modeling software (discussed 
further in \S \ref{sec_starmodels}). \cite{Capp02} has 
given an algorithm to generically represent data with MGEs. We adopt the formalism 
presented in \cite{Emse94} and reviewed in \cite{Capp08} where primed and unprimed variables represent 
observed and intrinsic (deprojected) values. The surface brightness is the sum of N two-dimensional, normal 
distributions: $L'_j$ is the luminosity of j-th term in units of $L_\odot$ in the chosen filter, $I'_j$ is 
the luminosity surface density in units of $L_\odot$ pc$^{-2}$, 
$\sigma'_j$ is the width along the major axis in units of 
arcsec, and $q'_j$ is the axial ratio. Some care 
must be taken as the smallest observed axial ratio sets a minimum inclination below which 
deprojection is not possible. In \S \ref{sec:modelstot} we fit inclination in combination with other parameters through Bayesian methods, 
and the inclination's prior probability is set to zero below this minimum. 
\par We used M.~Cappellari's publicly available IDL routine, named \texttt{FIND\_GALAXY}, 
to estimate priors to the galaxy centroids, inclinations, and PAs. \texttt{FIND\_GALAXY} works by 
collecting all pixels above a certain threshold and using the luminosity weighted moments. 
We used the top 10\% of pixels per galaxy, as ranked by flux, in this estimation. 
\par
\begin{figure*}
\centering
\includegraphics [scale=0.9,angle=0]{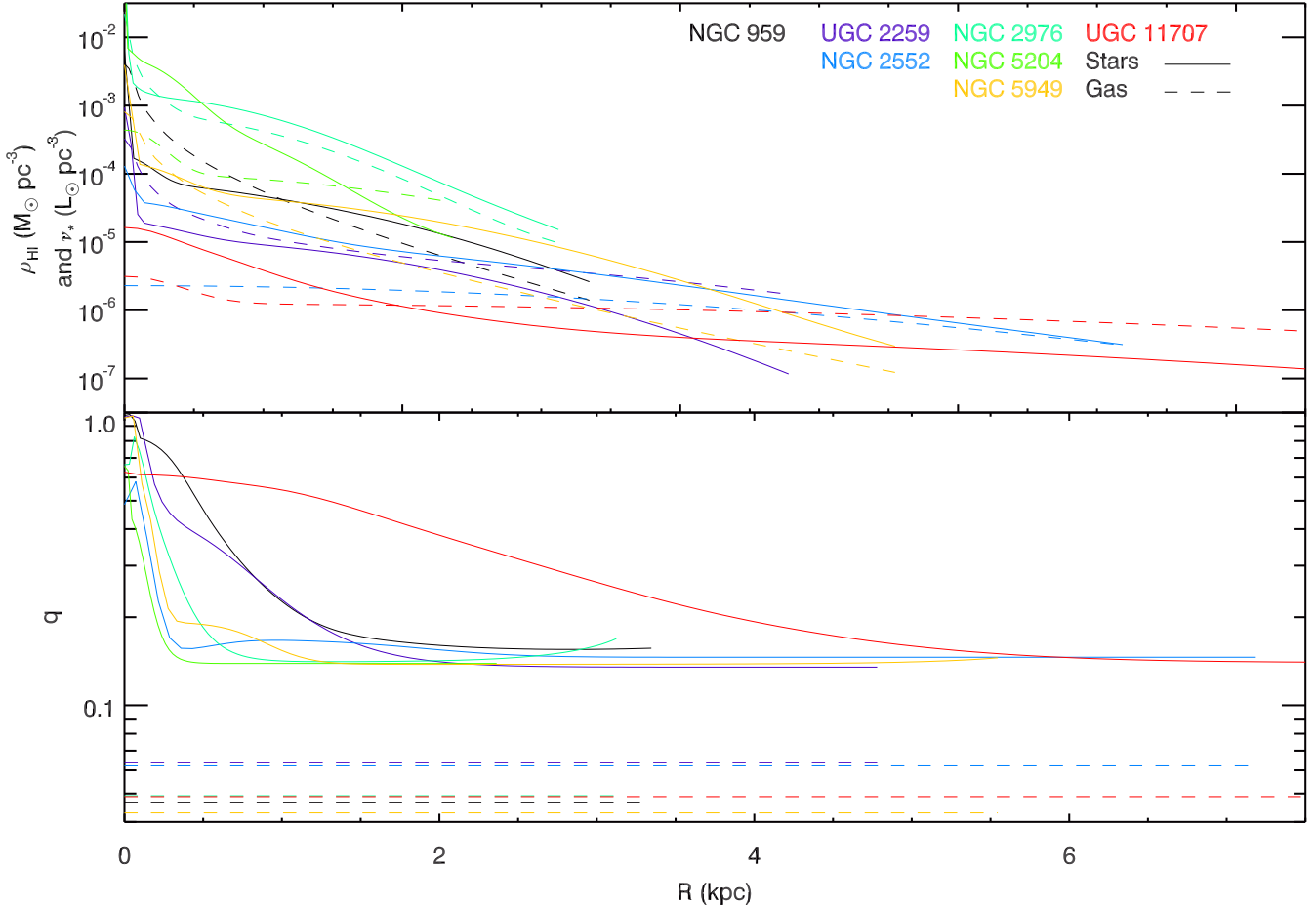}
\caption{The MGE fits to the stellar luminosities and HI masses. The 
bottom panel shows the intrinsic axial ratios for the nominal inclinations. 
Minimum axial ratios of 0.05 and 0.14 are enforced for the gas and stars, 
respectively.}
\label{fig:MGEdisp}
\end{figure*}

 We have fit all the photometry with the MGE software available from M.~Cappellari's webpage. The 
galaxy distance must be known to convert into absolute luminosity, and we adopted the values listed in 
Table \ref{tab:proplog}. The galaxy centroids, sky background levels, and masks for unrelated 
objects were determined manually. The axial ratio measured by \texttt{FIND\_GALAXY} determined the 
a priori inclination as listed in Table \ref{tab:proplog}. The standard relation of:
\begin{equation}
\label{eq:ellip}
\cos^2i=\frac{q'^2-q_{min}^2}{1-q_{min}^2},
\end{equation} 
relates the axial ratio to the inclination with q$_{min}$ being the intrinsic axial ratio of thin disk galaxies. 
For the stellar fits, we have enforced q$_{min}=$0.14, as motivated by many 
thin-disk galaxies when observed edge-on \citep[e.g.][]{Kreg02}. 
The residuals in any one angular and radial bin, named a sector, defined during the 
ellipse fitting generally display 15\% maximum deviation per galaxy and have root-mean-square residuals of 5\% per sector. 
The model luminosity profiles and axial ratios are shown in Figure \ref{fig:MGEdisp}. 
We list the total luminosities in Table \ref{tab:proplog}. 

Each galaxy is seen to be rounder at its center, with quite flat,
disk-like axis ratios at larger radii.  These transitions are robustly
seen in the photometry, whether numerically fit or even by eye.  In
four cases, NGC~2552, NGC~2976, NGC~5204, and NGC~5949, the rounder
structure is a compact nuclear star cluster that has negligible
influence on the global kinematics of the galaxy.  In the remaining
three systems, NGC~959, UGC~2259, and UGC~11707 there are larger
spheroidal structures that may be either bulges or, more likely,
pseudo-bulges.  Some of these transitions occur at the same radii
where we seek to measure the DM density slopes, so the DM density slope
and stellar axial ratio could be covariant at some level. We make a very
conservative estimate of the possible bias to the DM rotation curve as
follows. We have calculated the rotation curve of the stellar mass
from MGE fit terms for both NGC~2976 and NGC~959, and made a second
estimate where the axis ratio over the whole radial range is fixed to
the flattest value seen at large radii. In the process, we have
adjusted the surface density to maintain a constant mass for each MGE
component. For NGC~2976, the flatter profile reaches a higher
rotational velocity by 1.3 km~s$^{-1}$ at 150~pc, but converges to the
axially-changing case past 200~pc. This difference is easily within
the uncertainties on the velocities and therefore has no effect on the
DM estimates.  In the more extreme case of NGC~959, the flatter
distribution produces a rotation curve that is higher by
3--3.5~km~s$^{-1}$ between 250--1100~pc.  While still smaller than our
observational errors, this level of bias affects the DM powerlaw slope
estimates at the $\Delta\gamma=0.1$ level, comparable to the
uncertainties we derive on $\gamma$.  We note that a uniformly flat
stellar mass distribution is in conflict with our photometry and the
best estimates of the axis ratios are given by Figure
\ref{fig:MGEdisp}, but small biases resulting from errors in the
stellar axis ratios are possible.

\subsection{Archival HI Data}
\label{sec_HImass}
\par Neutral hydrogen can form a significant fraction of the mass 
at some radii in late-type dwarfs. We gather resolved HI measurements 
from the literature. In order to account for the mass of helium and 
metals, we have applied an additional factor to reach $M_{gas}=
1.4\times M_{HI}$. We generated MGE terms to represent the HI mass. 
NGC~959 and NGC~5949 do not have published, resolved 
HI data. Fortunately, they both have rather low values of $M_{HI}/L_{*,R}$, 
meaning that the detailed distributions are unlikely to be important 
to the gravitational potential. We have distributed their total 
HI masses as exponentials with scale lengths equal to the 
stellar scale lengths. For these two, the HI mass is effectively 
absorbed into $\Upsilon_*$. For UGC~2259, we have used the radial profile 
of Figure 5 in \cite{Cari88}. NGC~2552, NGC~5204, and UGC~11707 have their 
HI maps presented in \cite{Swat03}, and the authors have shared the images. 
The standard flux-mass equation, with $M_{HI}$ in units of $M_{\odot}$, 
D as distances in Mpc, and S as the HI flux integral in units of 
mJy km s$^{-1}$, is used as:
\begin{equation}
\label{eq:HImass}
M_{HI}=236\times D^2 \times S.
\end{equation} 
NGC~2976 has been measured in \cite{Stil02a}, and the HI radial profile 
as presented in \cite{Simo03} has been used to fit the MGE terms. 
The MGE fits have been forced to represent thin disks by constraining $q_j=0.05$, 
for the a priori inclination in all cases. The value of $q_j$ 
is comparable to the values of the average vertical scale lengths 
\citep{Baget11} and scale lengths (derived from the table of HI
surface densities in \cite{Leroy08}) for several THINGS dwarf galaxies. From 
those works, NGC~2976, IC~2574, and NGC~4214 appear marginally thinner 
(q$\sim$0.03) and NGC~4449 is marginally thicker (q$\sim$0.10). 
The model mass and axis ratios are shown 
in Figure \ref{fig:MGEdisp}. We have made no attempt at modeling molecular 
gas masses. In late-type dwarf galaxies where CO has been observed, the 
molecular gas is a small component to the total mass \citep[e.g.][]{Simo03,Simo05}.

\subsection{Integral Field Spectroscopy}
\label{sec_obs}
Data were taken with the Visible Integral-field Replicable Unit Spectrograph, Wendelstein model (VIRUS-W) 
\citep{Fabr08,Fabr12} mounted on the 2.7m Harlan J.~Smith telescope at the McDonald Observatory over several 
observing runs. VIRUS-W is a fiber-fed, integral-field unit (IFU) spectrograph. We used the 
high resolution mode which covers 4850\AA$<\lambda<$5470\AA\ 
and delivered $R=8300$ (35 km s$^{-1}$) by our measurement of arc lamp lines. 
We have also made fits to the instrumental resolution by comparing a small number of 
stars and higher resolution templates and found consistency with the lamp-based measurements. 
The instrumental resolution measurements are not totally immune to illumination issues 
as the fibers do not fully scramble signal radially before passing light to the spectrograph. However, 
the stellar continuum emission in these galaxies is quite regular and even the gas emission is 
often smoothly varying over several fiber radii. 
On the 2.7m telescope, the fibers have a diameter of 3\farcs1 and the 
field-of-view, 105\arcsec$\times$55\arcsec\ is 
one of the largest available for medium resolution IFUs.
Observing information is logged in Table \ref{tab:obslog}.  
In Table \ref{tab:proplog} we also list the median dispersion observed in the 
gas emission lines. This measurement is described in \S \ref{sec_kin} and its application to estimating a 
circular velocity is discussed in \S \ref{sec_gasmodels}. The instrumental dispersion 
has been measured off arc lamp lines and ranges from 15.4--18.0 km s$^{-1}$ for 
different fibers. The quoted gas dispersions have been corrected to an intrinsic value 
by subtracting off the fiber-specific instrumental dispersion in quadrature. 

\begin{deluxetable*}{lcccccc}
\tabletypesize{\scriptsize}
\tablecaption{VIRUS-W Data Log\label{tab:obslog}}
\tablewidth{0pt}
\tablehead{
\colhead{Galaxy} & \colhead{Dates} & \colhead{On-source} &
\colhead{$\left<\mu\right>_{e}$} & \colhead{Number of} & \colhead{S/N over}\\
\colhead{} & \colhead{(UTC)} & \colhead{Exposure}
& \colhead{(r-band mag} & \colhead{kinematic} & \colhead{4900-5400\AA}\\
\colhead{} & \colhead{} & \colhead{Time (hrs)} &
\colhead{arcsec$^{-2}$)} & \colhead{bins} & \colhead{(pixel$^{-1}$ bin$^{-1}$)}}
\startdata
NGC0959 & 23/8/11-27/8/11 & 10.0 & 21.9 & 92 & 18.9 \\
\ldots & 6/2/13-8/2/13 & & & & \\
UGC02259 & 22/12/11-28/12/11 & 7.0 & 23.5 & 43 & 11.2 \\
\ldots & 7/2/13 & & & & \\
NGC2552 & 22/12/11-26/12/11 & 8.0 & 23.0 & 63 & 16.0 \\
\ldots & 5/2/13-8/2/13 & & & & \\
NGC2976 & 16/5/12-20/5/12 & 6.0 & 21.4 & 251 & 16.7 \\
NGC5204 & 16/5/12-19/5/12 & 10.5 & 21.8 & 140 & 18.0 \\
\ldots & 5/2/13-8/2/13 & & & & \\
NGC5949 & 20/5/12 & 3.0 & 21.4 & 120 & 14.7 \\
UGC11707 & 23/8/11-1/9/11 & 16.0 & 23.5 & 37 & 17.5
\enddata
\end{deluxetable*}

\par The IFU data were reduced with a pipeline we originally developed for the Hobby-Eberly Telescope Dark 
Energy eXperiment \citep[HETDEX;][]{Hill04,Hill08}. The pipeline is named \texttt{vaccine} and is more 
fully described in \cite{Adam11}. In brief, the pipeline makes the usual CCD reduction steps 
of bias-subtraction and flat-fielding. Next, it traces the fibers' centers and measures their 
cross-dispersion profile. A wavelength solution is fit to arc lamp lines. Finally, a model for the 
background sky spectrum is made by fitting a bspline \citep{Die93,Kels03} to all the fibers lacking continuum sources. 
Careful attention was paid to the quality and robustness of the sky spectrum model driven by the 
faint emission-line source needs of the HETDEX pilot survey. The bspline fit is particularly well 
suited to this task as it is stable to outlier datapoints (such as unflagged cosmic rays) and 
can capture curvature better than a linear interpolation, particularly as we have hundreds of fibers that 
have sampled the background spectra at slightly offset wavelengths. 

\subsection{Kinematic Extraction of Stars and Gas}
\label{sec_kin}
\par First, the stellar data are binned to achieve sufficient S/N for 
kinematic extraction. The Voronoi binning scheme of \cite{Capp03} is used 
with a target of S/N=12 per pixel per bin. No binning is used for the gas 
extraction.   
\par We extract kinematics by finding a LOSVD kernel that convolves with a 
set of stellar template spectra to best match the data. The code we use for this task determines the optimal 
template weights and makes a maximum penalized likelihood estimate, in pixel space, of the LOSVD \citep{Gebh00}. 
The LOSVD was initialized as a normally distributed function around the systemic velocity with a 
standard deviation of 25 km s$^{-1}$. A regularization parameter of $\alpha=1$ was used, and experimentation 
showed that our results were not sensitive to regularizations higher or lower by a factor of three. This 
code is capable of making non-parametric LOSVD estimates, but for this work we restricted the fits to 
Gaussian LOSVDs. This is the same method we used in our previous work \citep{Adams12}, although we have changed 
the handling of the 
stellar templates. In some circumstances, nebular gas emission may interfere with this process. One may either 
fit regions with gas lines or mask them. Since the nebular gas lines (primarily H$\beta$ and [\ion{O}{3}]) do not 
lie in regions with prominent absorption lines and therefore do not coincide with much LOSVD information, we have 
simply masked 20\AA\ around [\ion{O}{3}]4959 and [\ion{O}{3}]5007 in the templates. H$\beta$ lies outside our fitting region. 
We have additionally masked 
5195--5205\AA\ in the observed wavelength range due to a variable \ion{N}{1} airglow line. The region used in the LOSVD fit 
was 4900--5350\AA. Most of the LOSVD information in this range is coming from three Lick Fe indices \citep{Wort97} and 
the Mg$_{\textrm{b}}$ triplet. Once the single best fit was found, Monte Carlo simulations through 100 iterations per bin 
were run to determine the errors on the central velocity and the dispersion.
\par Template mismatch can bias the kinematic measurements. This is true even for 
old, single population systems if the templates have the wrong metallicity and 
especially true for our young, multiple population systems if an improper age or 
sets of ages are used. If a template is selected 
that has a different equivalent width (EW) than the data, the fitting process can trade off too-large dispersions with 
too-strong template lines as a way to distribute the residuals and achieve a statistically better fit. 
Some checks, and even optional corrections, do exist in the \cite{Gebh00} 
LOSVD code. One can measure the optimal EW offset and stretch between the data and the best fitting model and 
choose to fit them as additional parameters. However, it is best to begin with a template set that is as representative 
of the data as possible. For this reason, we use a small number of composite templates that are fully described in Appendix 
\ref{app:temp}. The templates are based on empirical ELODIE spectra 
\citep{Prug01} and stacked to represent a 13.5 Gyr old population and 
then three populations with a 50\% mass fraction from a 13.5 Gyr 
population and variously a 1 Gyr, 250 Myr, and 50 Myr population. 
\par Compared to the stars, the extraction of gas velocities is much simpler. We make a single constrained fit to the 
H$\beta$, [\ion{O}{3}]4959, and [\ion{O}{3}]5007 lines. The line widths and radial velocity components are all 
constrained to hold the same value and are parametrized as Gaussians. The ratio of [\ion{O}{3}]5007 and [\ion{O}{3}]4959 is fixed 
to be 2.88 based on atomic physics. We have looked for significantly different velocities between H$\beta$ and the nebular 
lines and found none. One subtlety is the treatment of underlying Balmer absorption. The effect is minor as the 
absorption is usually much broader and only slightly offset from the Balmer emission, but we make the proper correction anyhow. 
We find the stellar kinematic bin to which each fiber belongs. From the stellar template weights and LOSVD values, we 
form and subtract the best fitting stellar model. Then, we fit the emission lines on the continuum-subtracted data with 
full error propagation. The velocity maps for the gaseous-traced line-of-sight velocities 
are shown in Figure \ref{fig:datmodmap}.
\section{Constraints from Dynamical Models}
\label{sec:modelstot}
\subsection{Model Parameterizations}
\label{sec_modelparam}
\par We have estimated the dynamics model parameters through Bayesian statistical 
methods. The parameters are quoted in Table \ref{tab:fits_param}. 
In our previous work, we used frequentist parameter estimation with a 
smaller number of parameters. In the present work, we use 
Markov Chain Monte Carlo (MCMC) methods to measure parameter constraints and 
covariances with the python-based software \texttt{emcee} \citep{Fore13}. 
Parameters can be estimated either way, but there are several advantages to the 
Bayesian approach. First, 
the frequentist approach requires stepping through one or a small number of 
parameters while optimizing the $\chi^2$ metric over 
the other, nuisance parameters. As is common to general N-dimensional 
optimization problems, it is difficult to ensure that an absolute, rather than local, minimum has been found. 
Second, the N-dimensional optimization process is very time 
consuming. We were able to run optimizations for single parameters and crude grids 
of parameter covariances, but run-time limits precluded finer grids. Finally, 
some of the parameters have useful priors, such as the galaxy geometric parameters, and 
Bayesian analysis can incorporate that additional information. 
\setlength{\tabcolsep}{2.5pt}
\begin{deluxetable*}{c|cr@{$\pm$}lr@{$\pm$}lr@{$\pm$}lr@{$\pm$}lr@{$\pm$}lr@{$\pm$}lr@{$\pm$}lr@{$\pm$}lr@{$\pm$}lr@{$\pm$}lr@{$\pm$}lr@{$\pm$}l}

\tabletypesize{\scriptsize}
\tablecaption{Model parameter constraints\label{tab:fits_param}}
\tablewidth{0pt}
\tablehead{
& \colhead{Galaxy} & \multicolumn{2}{c}{$\log$ M$_{200}$} & \multicolumn{2}{c}{c} 
& \multicolumn{2}{c}{$\gamma$} & \multicolumn{2}{c}{$\beta_{z}$} & \multicolumn{2}{c}{$\Upsilon_*$} & \multicolumn{2}{c}{$i$ (\arcdeg)} 
& \multicolumn{2}{c}{PA(\arcdeg)} & \multicolumn{2}{c}{v$_{sys}$} & \multicolumn{2}{c}{$\Delta\alpha_0$} 
& \multicolumn{2}{c}{$\Delta\delta_0$} & \multicolumn{2}{c}{$\sigma_{sys}$}\\
& & \multicolumn{2}{c}{/M$_{\odot}$} & \multicolumn{2}{c}{} 
& \multicolumn{2}{c}{} & \multicolumn{2}{c}{} & \multicolumn{2}{c}{} & \multicolumn{2}{c}{}
& \multicolumn{2}{c}{} & \multicolumn{2}{c}{(km s$^{-1}$)} & \multicolumn{2}{c}{(\arcsec)}
& \multicolumn{2}{c}{(\arcsec)} & \multicolumn{2}{c}{(km s$^{-1}$)}
}
\startdata
\parbox[t]{2mm}{\multirow{7}{*}{\rotatebox[origin=c]{90}{Gas-traced}}}
& NGC0959  & 11.06  & 0.23 & 16.7 & 2.0 & 0.88 & 0.15 & \multicolumn{2}{c}{\ldots} & 1.10 & 0.15 & 55.2 & 0.1 & 62.8  & 1.9 & 597.0 & 0.8 & 0.8  & 0.4 & -0.1 & 0.5 & 5.6 & 0.7 \\
& UGC02259 & 11.42  & 0.14 & 18.2 & 2.6 & 0.72 & 0.09 & \multicolumn{2}{c}{\ldots} & 1.07 & 0.27 & 41.0 & 0.2 & 159.9 & 1.5 & 581.6 & 1.1 & -0.8 & 1.0 & 1.0  & 1.2  & 5.5 & 0.5 \\
& NGC2552  & 11.33  & 0.11 & 18.1 & 2.0 & 0.38 & 0.11 & \multicolumn{2}{c}{\ldots} & 1.01 & 0.19 & 52.9 & 0.1 & 57.8  & 0.9 & 521.9 & 0.7 & 0.9  & 0.8 & 1.4  & 0.9 & 5.7 & 0.3 \\
& NGC2976  & 11.94  & 0.51 & 20.6 & 3.3 & 0.30 & 0.18 & \multicolumn{2}{c}{\ldots} & 0.83 & 0.22 & 62.0 & 0.2 & -34.4 & 1.6 & 5.3   & 1.1 & -1.3 & 1.5 & -1.5 & 1.3 & 6.9 & 0.4 \\
& NGC5204  & 11.36  & 0.16 & 18.7 & 2.1 & 0.85 & 0.06 & \multicolumn{2}{c}{\ldots} & 1.08 & 0.13 & 46.8 & 0.1 & 171.0 & 2.2 & 201.6 & 0.9 & 0.2  & 1.0 & -0.8 & 1.0 & 6.6 & 0.7 \\
& NGC5949  & 11.82  & 0.42 & 17.5 & 1.9 & 0.53 & 0.14 & \multicolumn{2}{c}{\ldots} & 1.16 & 0.34 & 62.0 & 0.1 & 148.5 & 2.0 & 442.4 & 1.7 & 1.2  & 0.6 & 0.2  & 0.7 & 5.4 & 0.8 \\
& UGC11707 & 11.49  & 0.18 & 15.1 & 1.6 & 0.41 & 0.11 & \multicolumn{2}{c}{\ldots} & 1.11 & 0.23 & 72.7 & 0.1 & 56.8  & 1.0 & 899.4 & 1.0 & 1.4  & 0.7 & 0.2  & 0.8 & 7.5 & 0.4 \\
\hline
\parbox[t]{2mm}{\multirow{7}{*}{\rotatebox[origin=c]{90}{Stellar-traced}}}
& NGC0959  & 11.64 & 0.32 & 18.5 & 2.4 & 0.73 & 0.10 & -0.05 & 0.20 & 1.08 & 0.27 & 55.3 & 0.1 & 65.4  & 2.2 & 596.5 & 2.0 & -0.6 & 2.0 & 0.4  & 1.0 & 3.0 & 1.1 \\
& UGC02259 & 11.62 & 0.61 & 16.7 & 5.7 & 0.77 & 0.21 &  0.28 & 0.39 & 1.10 & 0.44 & 41.1 & 0.2 & 157.6 & 3.3 & 578.7 & 2.9 & 0.9  & 2.3 & -0.6 & 3.0 & 3.3 & 1.2 \\
& NGC2552  & 11.23 & 0.38 & 15.8 & 3.6 & 0.53 & 0.21 &  0.35 & 0.29 & 1.24 & 0.55 & 52.8 & 0.2 & 56.9  & 3.1 & 521.9 & 2.7 & -1.3 & 2.5 & -0.3 & 2.4 & 3.6 & 0.7 \\
& NGC2976  & 11.56 & 0.46 & 17.7 & 2.5 & 0.53 & 0.14 & 0.49  & 0.07 & 0.93 & 0.21 & 62.0 & 0.1 & -33.6 & 3.0 & 1.5   & 1.3 & 0.6  & 2.3 & 0.3 & 1.9 & 2.3 & 0.6 \\
& NGC5204  & 11.76 & 0.51 & 18.3 & 3.3 & 0.77 & 0.19 & 0.65  & 0.19 & 1.30 & 0.42 & 47.0 & 0.1 & 172.6 & 3.5 & 201.4 & 2.2 & -0.9 & 2.4 & -1.6 & 2.7 & 2.9 & 1.6 \\
& NGC5949  & 11.46 & 0.22 & 17.5 & 1.9 & 0.72 & 0.11 & 0.11  & 0.17 & 1.20 & 0.28 & 62.1 & 0.1 & 146.3 & 1.3 & 441.3 & 1.5 & -0.5 & 0.7 & -0.1 & 0.7 & 2.9 & 1.1 \\
& UGC11707 & 11.13 & 0.37 & 17.3 & 4.9 & 0.65 & 0.26 & 0.34  & 0.26 & 1.07 & 0.44 & 72.8 & 0.2 & 53.1  & 4.0 & 896.6 & 3.8 & 1.0  & 4.3 & 0.5  & 2.9  & 3.6 & 1.2
\enddata
\end{deluxetable*}
\par Our gas-based models have ten terms: the logarithm of the virial mass, log(M$_{200}$), the concentration c,
and the inner density logarithmic slope of a generalized NFW
function (gNFW), the stellar mass-to-light ratio $\Upsilon_*$, the 
galaxy inclination, the position angle, the systemic velocity, 
the offset in right ascension and and declination of the dynamical 
center from the a prior value, and a systematic uncertainty. 
Three of these ten parameters describe the DM halo:
log(M$_{200}$), c, and $\gamma$. The original NFW density distribution 
\citep{Nava96a} contains a central cusp with $\gamma=1$. Note that in some of the 
observational literature, this inner density logarithmic slope is instead 
named $\alpha$ and the opposite sign is sometimes assigned to the definition. A very general 
halo profile form was introduced by \cite{Hern90} and 
explored by \cite{Zhao96} and \cite{Wyit01}, where the inner power law, 
the outer power law, and the sharpness of the transition are all 
variables. It was shown by \cite{Klyp01} that these parameters are too 
degenerate for realistic data to constrain, and a common choice is to 
leave only the inner power law slope free for a gNFW function. 
The density of the gNFW function is:
\begin{equation}
\label{eq:gNFW}
\rho(r)=\frac{\delta_c \rho_{crit}}{(r/r_s)^{\gamma}[1+(r/r_s)]^{3-\gamma}}, 
\end{equation} 
where $\delta_c$ is the overdensity factor and 
$\rho_{crit}$ is the critical density of the universe of:
\begin{equation}
\label{eq:crit}
\rho_{crit}\equiv\frac{3H^2(z)}{8 \pi G}, 
\end{equation} 
with Hubble parameter $H(z)$ and gravitational constant G. A single integral, 
which can be reduced to an incomplete gamma function for spherical halos, 
is given in \cite{Dutt05} and corrected in Equation 9 of \cite{Barn12}, 
to relate $\delta_c$ to c and $\gamma$. We have not fit an Einasto profile 
\citep{Einas65,Navar04,Navar10} as it does not differ significantly from the gNFW 
function at our resolution. 
The stellar-based parameters encompass all the gas-based parameters and 
further contain a stellar velocity ellipsoid anisotropy term, $\beta_z$, defined as 
\begin{equation}
\label{eq:betaz}
\beta_z\equiv1-\frac{\bar{v_z^2}}{\bar{v_R^2}}.
\end{equation} 
\par We provide fits under an alternative DM functional form commonly used in the literature:
the Burkert profile \citep{Burke95} as  
\begin{equation}
\label{eq:Burk}
\rho(r)=\frac{\rho_{b}}{(1+r/r_b)(1+(r/r_b)^2)}.
\end{equation}  
This form enforces a DM core. Burkert profiles have been used before, such as to
correlate core sizes with other observables in the study of
non-CDM models, and therefore its DM parameters will be quoted for
comparison to such studies. To make a 
direct comparison, we have fixed the  
systematic uncertainties to the values derived for the gNFW fits 
from Table \ref{tab:fits_param} and run our MCMC pipeline. For most galaxies under 
study the quality of the Burkert fit, as judged by   
$\chi^2$ with one more degree of freedom relative to the gNFW function
and shown in Table \ref{tab:chisq}, is slightly poorer. However, we are unable to 
rule out the Burkert form from the statistics, and the preference of a galaxy for one 
profile other another is not always consistent for the two tracers. This is not so surprising since 
model selection requires a higher statistical threshold ($\Delta\chi^2=\sqrt{2\nu}$ 
for 1-$\sigma$ significance where $\nu$ is the degrees of freedom) than 
parameter constraint ($\Delta\chi^2=1$ for 1-$\sigma$ significance). We give constraints on 
the two Burkert parameters in Table \ref{tab:Bfit}.

\begin{deluxetable*}{lrrrrcrr}
\tabletypesize{\scriptsize}
\tablecaption{Quality of fit for gNFW and Burkert models\label{tab:chisq}}
\tablewidth{0pt}
\tablehead{
& \multicolumn{3}{c}{Gas-traced models} & \phantom{a} & \multicolumn{3}{c}{Stellar-traced models} \\
\cline{2-4} \cline{6-8}\\
\colhead{Galaxy} & \colhead{N} & \colhead{gNFW $\chi^2$} & \colhead{Burkert $\chi^2$} & &
\colhead{N} & \colhead{gNFW $\chi^2$} & \colhead{Burkert $\chi^2$}}
\startdata
NGC0959  & 1152 &  935 & 1053 & &  92 &  42.6 &  47.2 \\
UGC02259 &  876 &  752 &  726 & &  43 &  27.8 &  37.9 \\
NGC2552  & 1848 & 1765 & 1836 & &  63 &  40.2 &  41.4 \\
NGC2976  & 1794 & 1608 & 1567 & & 251 & 184.6 & 180.9 \\
NGC5204  & 3768 & 3331 & 3330 & & 140 &  80.3 & 217.9 \\
NGC5949  &  234 &  172 &  177 & & 120 &  60.3 &  52.7 \\
UGC11707 & 1068 & 1011 & 1244 & &  37 &  26.3 &  17.6 
\enddata
\tablecomments{The number of datapoints are given by N. The degrees of freedom can be obtained with 
N and the number of fit terms, k. The values are k=9 for the gas-traced gNFW model and 10 for the 
stellar-traced gNFW model where we have fixed the systematic uncertainty. 
The Burkert models have k lower by one.}
\end{deluxetable*}

\begin{deluxetable*}{lr@{$\pm$}lr@{$\pm$}lcr@{$\pm$}lr@{$\pm$}l}
\tabletypesize{\scriptsize}
\tablecaption{Burkert DM parameter constraints\label{tab:Bfit}}
\tablewidth{0pt}
\tablehead{
& \multicolumn{4}{c}{Gas-traced models} & \phantom{a} & \multicolumn{4}{c}{Stellar-traced models} \\
\cline{2-5} \cline{7-10}\\
\colhead{Galaxy} & \multicolumn{2}{c}{$\rho_b$} & \multicolumn{2}{c}{$r_b$} &
& \multicolumn{2}{c}{$\rho_b$} & \multicolumn{2}{c}{r$_b$} \\
\colhead{} & \multicolumn{2}{c}{(M$_{\odot}$ pc$^{-3}$)} &
\multicolumn{2}{c}{(kpc)} &
& \multicolumn{2}{c}{(M$_{\odot}$ pc$^{-3}$)} &
\multicolumn{2}{c}{(kpc)}}
\startdata
NGC0959 & 0.20 & 0.04 & 1.5 & 0.2 & & 0.22 & 0.05 & 1.8 & 0.3 \\
UGC02259 & 0.21 & 0.03 & 1.7 & 0.1 & & 0.23 & 0.06 & 1.7 & 0.4 \\
NGC2552 & 0.10 & 0.02 & 2.3 & 0.4 & & 0.13 & 0.04 & 1.6 & 0.3 \\
NGC2976 & 0.15 & 0.02 & 2.3 & 0.6 & & 0.15 & 0.02 & 1.9 & 0.4 \\
NGC5204 & 0.25 & 0.05 & 1.7 & 0.3 & & 0.24 & 0.06 & 1.9 & 0.6 \\
NGC5949 & 0.18 & 0.04 & 1.9 & 0.3 & & 0.19 & 0.03 & 1.9 & 0.2 \\
UGC11707 & 0.10 & 0.03 & 2.5 & 0.5 & & 0.17 & 0.05 & 1.7 & 0.3
\enddata
\tablecomments{The measured constraints on the central densities, $\rho_b$ and the 
core radii, r$_b$, in the Burkert function form for DM halos.}
\end{deluxetable*}

\par The fourth parameter is the stellar mass-to-light ratio 
in the photometric band specific to the MGE terms, $\Upsilon_*$. The two 
bands we use here are $r$ and $R$. The fifth parameter is the 
systemic velocity of the galaxy. The next four parameters are geometric: 
the inclination $i$, the position angle PA, 
and the offsets from the nominal photometric centers, $\Delta\alpha_0$ and 
$\Delta\delta_0$ measured in arcsec. 
Lastly, the final parameter is a 
systematic kinematic uncertainty, $\sigma_{sys}$, that is added to the 
statistical uncertainty for the purpose of calculating model likelihoods. For the 
gas models this systematic uncertainty applies to the line-of-sight velocity 
and for the stellar models it applies to the second moment velocities. 
\par The limits on $\Upsilon_*$ are 
formed by considering some stellar 
populations synthesized with the isochrones of \cite{Bres12}. We have 
used a metallicity of [Fe/H]=-0.5, a Chabrier initial mass function (IMF), 
and considered single burst ages of 100 Myr, 3 Gyr, 13.5 Gyr, and a 
composite of 10\% 100 Myr and 90\% 3 Gyr by mass. These four sets, which 
ought to bound reasonable conditions in dwarf galaxies, correspond to 
$\Upsilon_{*,R}$=$\left\{0.20, 1.60, 6.03, 0.72\right\}$ and 
$\Upsilon_{*,r}$=$\left\{0.18, 1.26, 4.50, 0.70\right\}$. Salpeter IMF values will 
be $\approx$2$\times$ larger. The upper bounds are irrelevant as 
the best-fit parameters never go so high in the fit. In principle we 
need a lower bound to protect against an unphysical, massless stellar disk. 
None of the galaxies are purely young starbursts, and the composite value ought to 
represent a reasonable lower limit. However, given the uncertainty in the 
absolute model calibrations of $\Upsilon_*$, we will set 0.35 and 10.0 
as the limits for both bands. The zeropoint offset between filters is small 
enough here to ignore. We find that the data themselves require 
values of $\Upsilon_*$ that are sufficiently larger (Table \ref{tab:fits_param}) 
than the lower bound that 
the bound's exact value is unimportant. 
\par We assign priors to all the parameters as listed in Table \ref{tab:prior}. 
The \texttt{emcee} code operates by distributing ``walkers'' across the parameter 
space, moving the walkers around according to the relative likelihoods of the 
present the proposed parameters without saving the parameters in a ``burn-in'' phase, 
and finally recording a number of MCMC samples for each walker under the 
same type of movement rules. 
The walkers are initialized to randomized positions as described by the final two columns 
of Table \ref{tab:prior}. 
The systemic velocity and geometric parameters priors are centered around the values listed in 
Table \ref{tab:proplog}. The priors on the systemic velocities come from an optimally weighted 
average of our kinematic measurements. The priors on the geometric parameters come from our photometric 
fits as described in \S \ref{sec:phot}.
We experimented first with placing priors on M$_{200}$ and $c$ via 
\cite{Klyp11} and \cite{Behr13}, but we found very similar results by adopting the flat priors in Table \ref{tab:prior}. 
\par The inclinations must have a lower limit enforced to numerically 
allow deprojection of the MGE terms as discussed in \S \ref{sec:phot}. Several other 
parameters have their priors restricted to physically reasonable ranges. All priors are either 
constant or normal functions. 
\begin{deluxetable*}{lrrrrcrr}
\tabletypesize{\scriptsize}
\tablecaption{Bayesian Priors and MCMC Initialization\label{tab:prior}}
\tablewidth{0pt}
\tablehead{
& \multicolumn{4}{c}{Prior} & \phantom{a} & \multicolumn{2}{c}{Initialization}\\
\cline{2-5} \cline{7-8}\\
\colhead{Parameter} & \colhead{Form} & \colhead{Domain} &
\colhead{Peak value} & \colhead{Standard deviation} & \colhead{} & \colhead{Central value} & 
\colhead{Standard deviation}}
\startdata
M$_{200}$/M$_{\odot}$ & Uniform & 10$^8$--10$^{12.5}$ & \ldots & \ldots & & 10$^{11}$ & 1.8$\times$10$^{11}$ \\
c & Uniform & 2.0--40.0 & \ldots & \ldots & & 17.0 & 3.0 \\
$\gamma$ & Uniform & 0.0--2.0 & \ldots & \ldots & & 0.6 & 0.3 \\
$\beta_z$ & Uniform & -$\infty$--1.0 & \ldots & \ldots & & 0.3 & 0.3 \\
$\Upsilon_{*}$ & Uniform & 0.35--10.0 & \ldots & \ldots & & 1.0 & 0.3 \\
$i$/\arcdeg & Normal & $i_{min}$--90.0 & $i_0$ (Table \ref{tab:proplog}) & 4.0 & & $i_0$ & 0.2 \\
PA/\arcdeg & Normal & 0--360.0 & PA$_0$ (Table \ref{tab:proplog}) & 4.0 & & PA$_0$ & 5.0 \\
v$_{sys}$/km s$^{-1}$ & Normal & -$\infty$--$\infty$ & v$_{sys,0}$ (Table \ref{tab:proplog}) & 10.0 & & v$_{sys,0}$ & 5.0 \\
$\Delta\alpha_0$/\arcsec & Normal & -$\infty$--$\infty$ & 0.0 & 3.0 & & 0.0 & 3.0 \\
$\Delta\delta_0$/\arcsec & Normal & -$\infty$--$\infty$ & 0.0 & 3.0 & & 0.0 & 3.0 \\
$\sigma_{sys}$/km s$^{-1}$ & Normal & 0--$\infty$ & 5.0 & 5.0 & & 5.0 & 5.0 \\
$\rho_{b}$/M$_{\odot}$ pc$^{-3}$ & Uniform & 0--$\infty$ & \ldots & \ldots & & 0.1 & 0.02 \\
$r_{b}$/kpc & Uniform & 0--10 & \ldots & \ldots & & 1.0 & 0.1
\enddata
\tablecomments{Column 1 lists the model parameters, column 2 lists the form utilized for the 
prior probability distributions, column 3 lists the domain over which the priors have non-zero 
values, columns 4 and 5 list the normal parameters for the relevant cases, and columns 6 and 7 
list the normal parameters by which all the parameters are initialized.}
\end{deluxetable*}
\par
\begin{figure*}
\centering
\subfigure{\includegraphics[scale=0.58,angle=0]{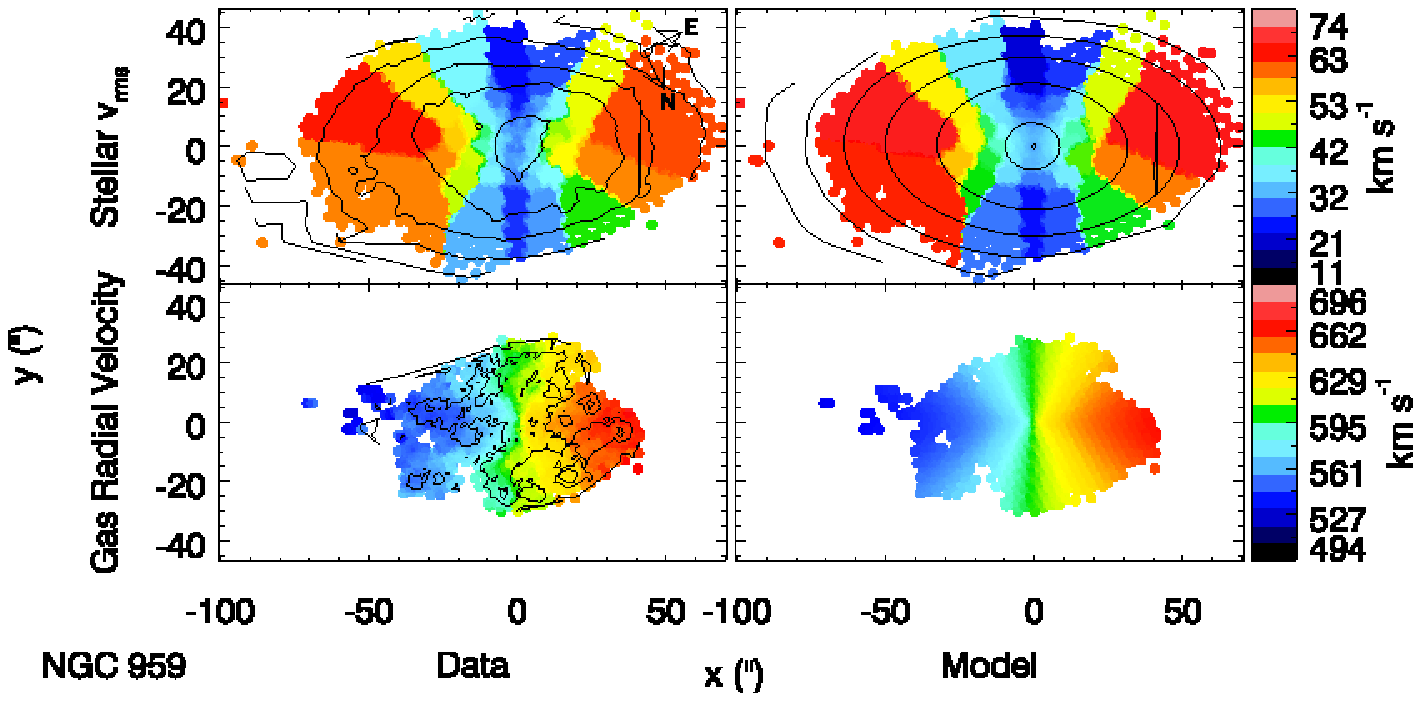}}
\subfigure{\includegraphics[scale=0.58,angle=0]{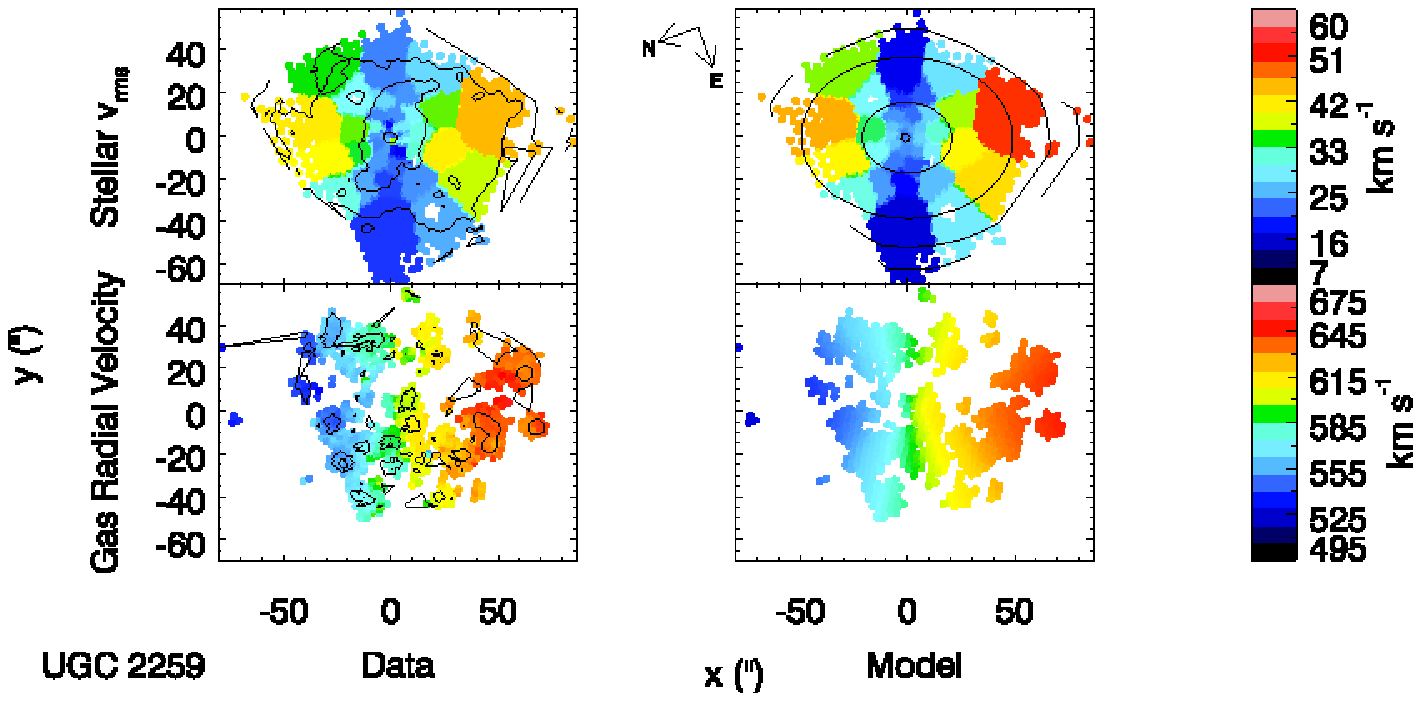}}\\
\subfigure{\includegraphics[scale=0.58,angle=0]{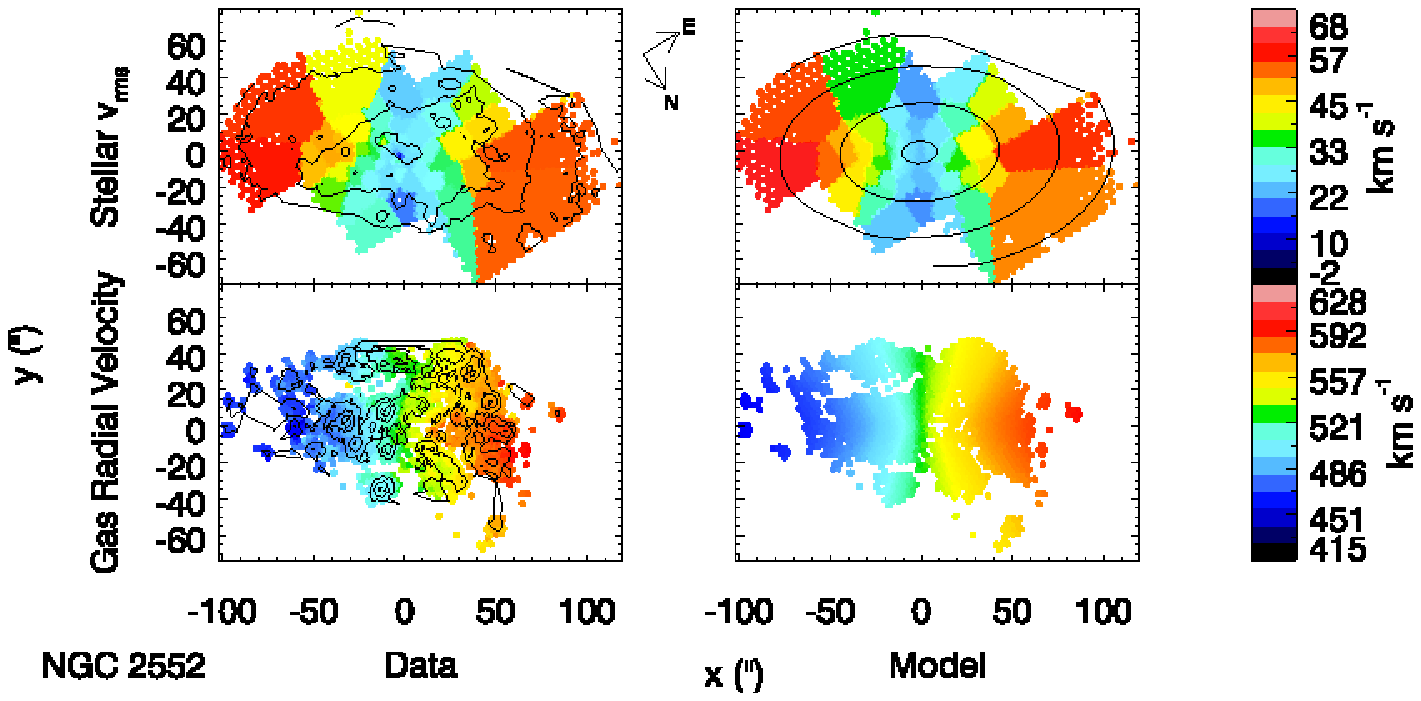}}
\subfigure{\includegraphics[scale=0.58,angle=0]{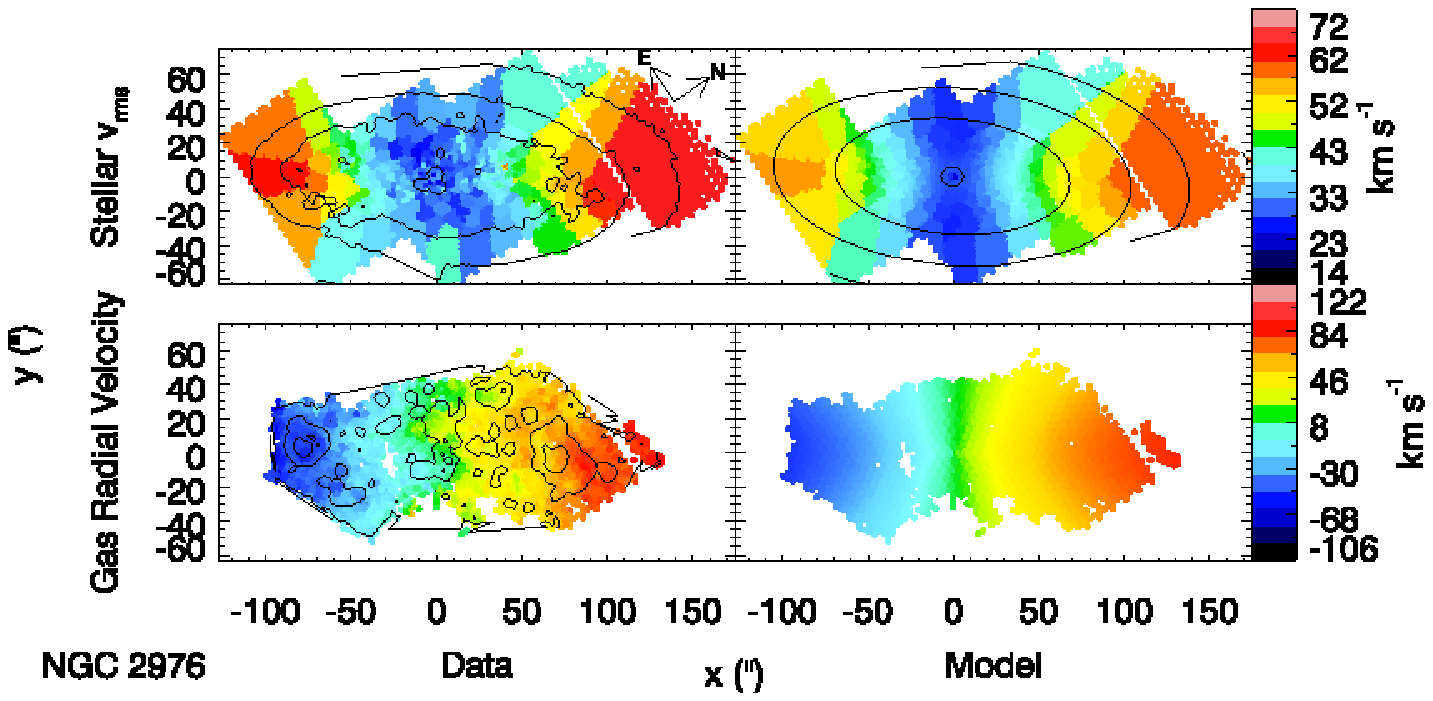}}\\
\subfigure{\includegraphics[scale=0.58,angle=0]{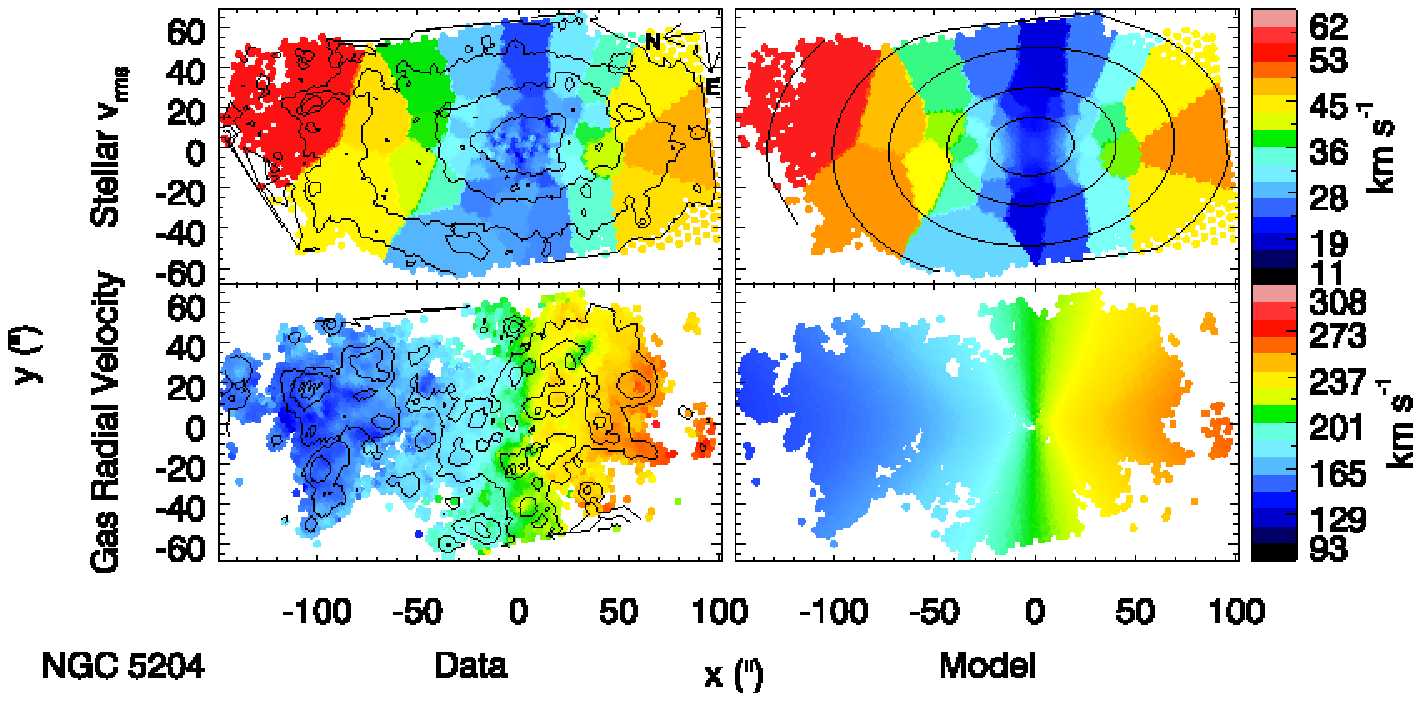}}
\subfigure{\includegraphics[scale=0.58,angle=0]{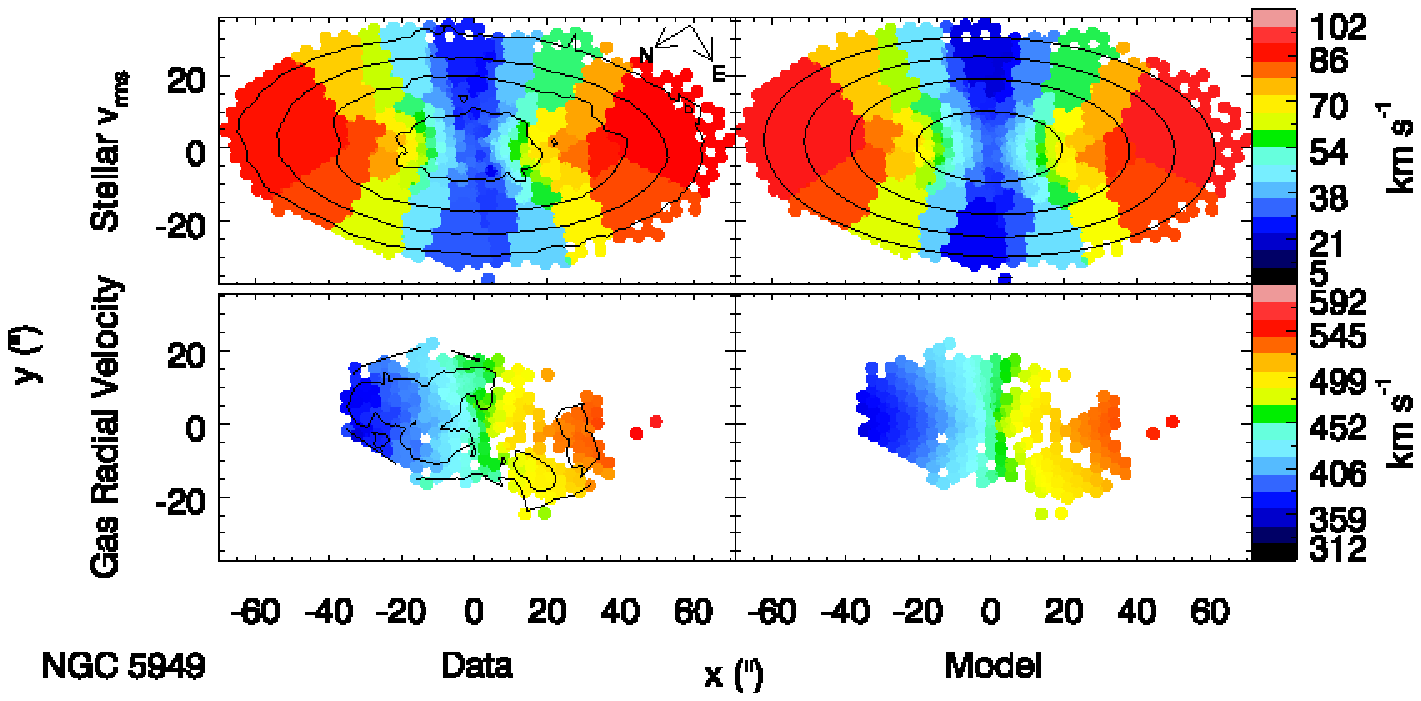}}\\
\subfigure{\includegraphics[scale=0.58,angle=0]{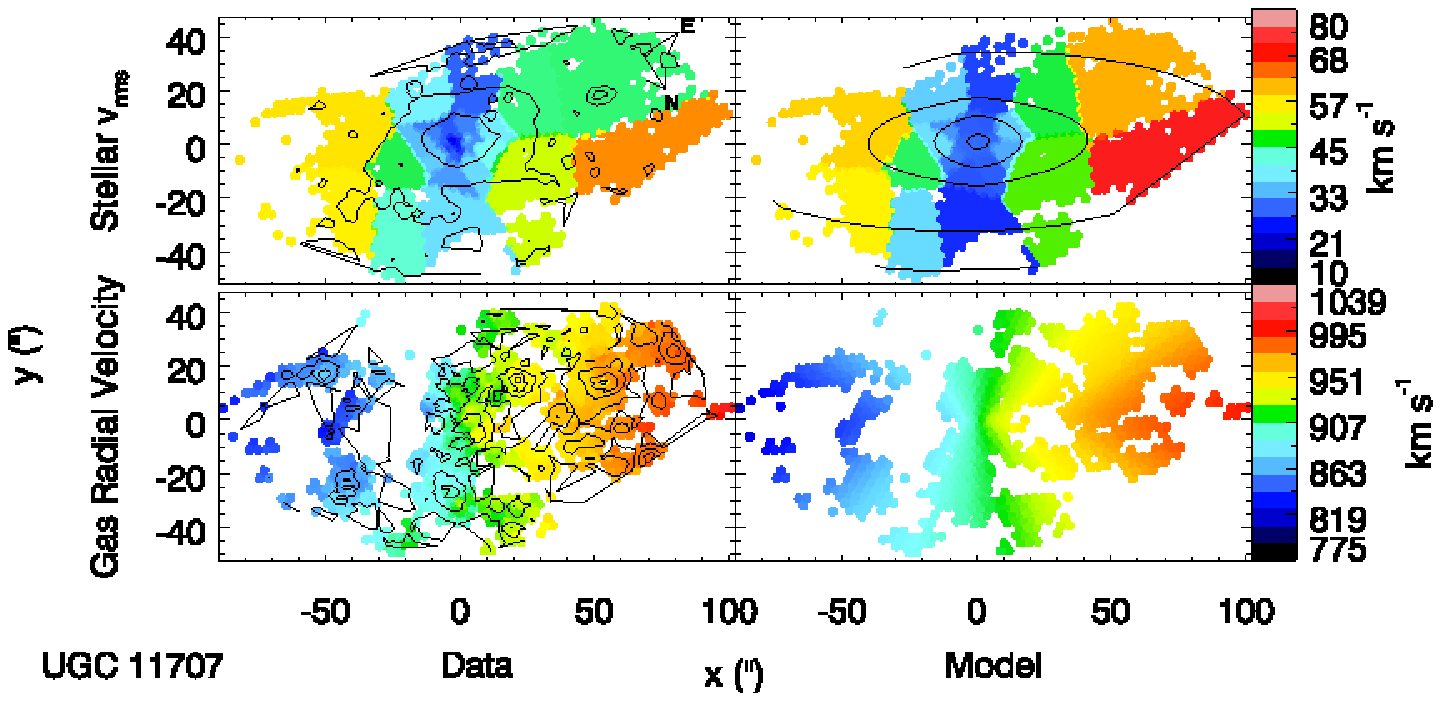}}
\caption{The gas-traced and stellar-traced data and 
the best fitting gNFW models. \textit{\textbf{Top}} Stellar kinematic 
fields in the v$_{rms}$ parameters. \textit{\textbf{Bottom}} Gas 
line-of-sight velocity kinematic maps. \textit{\textbf{Left}} Data. 
\textit{\textbf{Right}} The best fitting models. The contours 
show logarithmically scaled surface brightnesses for the continuum, on the 
stellar maps, 
and the emission lines, on the gas maps. The data-side contours show the 
actual surface brightnesses and the model-side stellar contour shows the 
MGE model. Stellar contours are spaced by one magnitude and gas contours 
by two magnitudes.}
\label{fig:datmodmap}
\end{figure*}

 Some further details are described in Appendices \ref{app:mgedm}--\ref{app:MCMCtest}. There, 
the details of the MGE computations, some simulations are run to test the 
recoverability of model parameters, and convergence tests in the length of MCMC chains and 
the number of parameter space ``walkers'' are described. The kinematic maps and the best fit 
gNFW models are shown in Figure \ref{fig:datmodmap}. The binning can 
be seen through neighboring fibers having constant velocities in the stellar maps. 
\subsection{Stellar-based Dynamical Models}
\label{sec_starmodels}
\par We use the Jeans Anisotropic Multi-Gaussian-Expansion (JAM) models of \cite{Capp08} to create the 
observables that correspond to a particular mass model. The JAM models assume that the velocity ellipsoid 
can be well represented by a cylindrical coordinate system and that the mass distribution is axisymmetric. 
Of course, an additional assumption to such models is that the system is virialized. The JAM software 
solves the Jeans equations. The Jeans equations are formed by taking moments of the collisionless Boltzmann 
equation. The Jeans equations have a closed form if the shape and orientation of the 
SVE are fixed. To close the equations, JAM uses the variable $\beta_z$ as a free parameter. JAM as provided by 
M.~Cappellari is written in \texttt{IDL}. Our earlier work \citep{Adams12} used this version. However, we 
needed both a faster version and one that could be easily called by the python-based \texttt{emcee} software. Therefore, 
we rewrote JAM in FORTRAN but without making any substantive changes to the algorithm. At necessary 
steps within the JAM algorithm, we used the \texttt{rlft3} routine from Numerical Recipes \citep{Press92} 
for 2D convolution and the \texttt{dqxgs} routine \citep{Fava91} in the SLATEC/QUADPACK math library for numerical integration. 
We ran numerous test cases between the IDL and FORTRAN implementations and always found agreement to better than 
1\% in velocities for any given mass model. The residuals are likely due to the differences in integration and 
convolution routines, but they are far below the level of interest for comparison to realistic data. JAM has also been 
ported into C by \cite{Watki13} and is publically available. 
\subsection{Gas-based Dynamical Models}
\label{sec_gasmodels}
\par
\begin{figure}
\centering
\includegraphics [scale=0.85,angle=0]{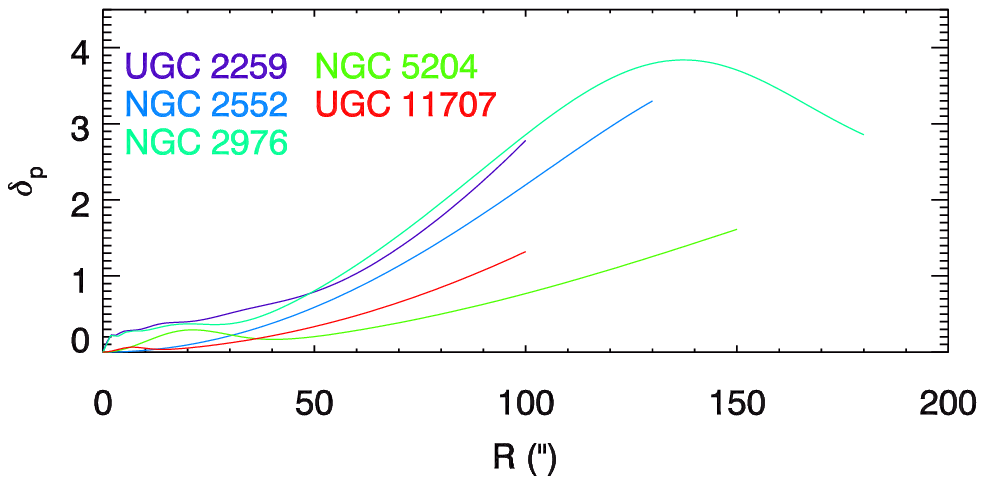}
\caption{The pressure support coefficient due to neutral gas 
density gradients. Only if this term took on large values at 
small radii would the correction to the circular velocity become 
significant. For the five well-measured cases, the HI is more 
diffuse than the stellar light, and this term is suppressed at 
small radii. NGC~959 and NGC~5949 lack resolved HI measurements and are 
therefore not shown.}
\label{fig:dP}
\end{figure}

 The tangential velocity of gas may not be a pure probe of the potential's 
circular velocity. In a similar fashion to asymmetric drift for stellar tracers, 
orbiting gas may experience additional pressure. \cite{Dalc10} have derived formulae 
for this situation. The equations are valid if $\sigma_g$ is dominated by turbulence, 
and the gas has an isotropic velocity ellipsoid. 
Gradients in the surface density of the gas, $\Sigma_g$, and 
the gas dispersion, $\sigma_g$, compose a pressure coefficient, $\delta_P$, as 
\begin{align}
\label{eq:gpres}
\delta_P=-\left(\frac{d \ln \sigma^2_g}{d \ln R} + \frac{d \ln \Sigma_g}{d \ln R}\right) \\
V^2_c=V^2_{g,\theta}+\delta_P \sigma^2_g.
\end{align}
In principle, the finite size of our fibers adds a beam-smearing component to our 
measured dispersions. However, the largest gradients across a fiber for the 
observed velocity fields are $\sim$ 5 km s$^{-1}$, and the contributed broadening 
with the round fibers will be less. Since we measure 15$\lessapprox \sigma_g \lessapprox$ 
20 km s$^{-1}$, the beam smearing is insignificant. We find negligible dispersion 
gradients in all our galaxies; no galaxies show a drop in $\sigma_g$ by more 
than 5 km s$^{-1}$ from center-to-edge. The only significant pressure term may be from 
the radially declining gas density. We must assume that the ionized gas distribution 
is similar to the neutral gas's for this computation. 
Figure \ref{fig:dP} shows the pressure support coefficient evaluated for each 
galaxy. We lack actual resolved HI measurements in NGC~959 and NGC~5949. 
For the five measured cases, the pressure support 
coefficient does not reach a value of unity until R$>$50\arcsec. At those 
radii, the rotational velocity is much larger than the gas dispersion so pressure 
support is not important anywhere. 
As noted by \cite{Dalc10}, the effect of this term is to steepen rotation 
curves, but perhaps not a level that solves the cusp-core problem. 
Therefore, we conclude that pressure support in the gas is not significant so 
that we can assume purely circular motion in a thin disk.

\par We have followed the routines of M.~Cappellari, adapted into FORTRAN, to calculate circular velocity for an 
arbitrary MGE model. Assuming purely circular motion in a thin disk, we 
can project these circular velocities for each position, given a position angle and inclination, to create a 
model that can be directly compared to the gas data. The gas-traced models have no use for the stellar velocity 
anisotropy, but all the other input parameters are shared with the stellar-traced models. 
\par The models previously presented have assumed axisymmetry. Dropping 
this assumption adds significant complexity. In order to model 
triaxial structures through stellar kinematics, the best option is to 
use triaxial Schwarzschild codes, for which \cite{vdBo08} is the current standard. 
The code has primarily been used to assess the robustness of supermassive 
black hole mass estimates in the presence of triaxial halos. The main limitation 
is that only static potentials can currently be modeled. The code could profitably 
be employed to more generally fit triaxial halo shapes in our sample, but it 
cannot fit structures with a pattern speed such as bars. We hypothesize that 
bars may be a significant dynamical perturber to the gas kinematics, but less so 
to the hotter stellar kinematics. We do not attempt to model triaxial structure 
through stellar kinematics herein.
\par Fortunately, the problem of triaxial structure affecting gaseous kinematics has 
been extensively studied. The general solution for an arbitrary number of harmonic 
terms is presented in \cite{Shoen97}. A general solution named 
\texttt{DiskFit}, accompanied by software, 
has been given in a set of papers \citep{Spek07,Sell10}. The basic idea is that a bisymmetric 
potential component, $m=2$, will contribute to the sky-projected harmonic 
modes of $m'=m\pm1$. This leads to Equation 5 in \cite{Spek07}
\begin{align}
\label{eq:ss07}
v_{rad}=v_{sys}+\sin i(v_{1,t}\cos\theta-&v_{2,t}\cos(2(\theta-\phi_b))\cos\theta \nonumber \\
-&v_{2,r}\sin(2(\theta-\phi_b))\sin\theta), 
\end{align}
where $\theta$ is relative to the major axis and $\phi_b$ is the angle 
between the projected major axis and the long axis of the triaxial structure, 
assumed to be a bar by \cite{Spek07}. 
When data sets show more complex structure than simple 
tangential motion, two simple choices present themselves. One may fit the 
motion as radial, first-order motion (inflow or outflow) or as the tangential and radial 
terms of a bisymmetric distortion. \cite{Spek07} used data on NGC~2976 to 
explain the kinematic twist with each model and showed that the derived 
circular velocity curve is sensitive to this choice. However, the gas 
kinematics can be fit nearly as well with either choice, and the interpretation 
is degenerate. Our stellar kinematic data on NGC~2976 show a powerful 
route toward breaking this degeneracy and favor the bisymmetric distortion fit 
to the gas. A primary goal of this study is to determine whether such 
bisymmetric distortions are consistently biasing the inferred potentials 
in late-type dwarf galaxies or if NGC~2976 is an outlier. 
\par
\begin{figure}
\centering
\includegraphics [scale=0.85,angle=0]{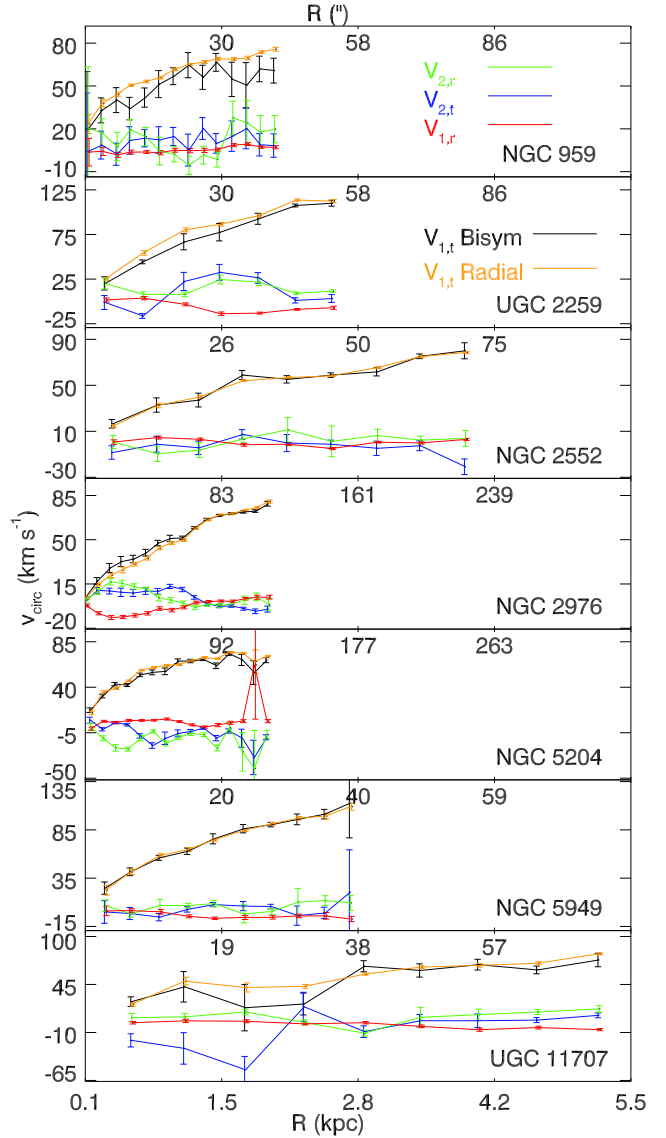}
\caption{DiskFit models for all seven galaxies. Both an $m=1$ model with a radial 
component, v$_{1,r}$, and a 
bisymmetric, $m=2$ model with radial, v$_{2,r}$, and 
tangential, v$_{2,t}$, components are fit. The bisymmetric 
model can represent a bar. Given the much larger uncertainties in the first order tangential 
velocity for the bisymmetric model, the two circular velocity curves are 
generally consistent. 
\textit{\textbf{NGC~2976}} The two fits are
similar to those in \cite{Spek07}. The bisymmetric model returns larger velocities at
smaller radii as a cuspier model would produce. The statistical quality of the two fits
are indistinguishable, but the stellar-traced mass models are more compatible with the
bisymmetric gas models. \textit{\textbf{UGC~11707}} Both models
fit a large tangential velocity to the bin at R=1.1 kpc. However, the fit is
unlikely to represent global, gravitational-driven motion. Due to dust and
patchiness in the gas distribution, the azimuthal coverage for this ring is poor and the
velocities are distinct from the overall kinematics of the galaxy.
A small pocket of high velocity gas is likely driving this unrealistic solution. Within the large
uncertainties, the two fits are consistent.}
\label{fig:disk1}
\end{figure}

 We have adapted a tilted ring code to perform the same operations as \texttt{DiskFit}. We 
fit both radial and bisymmetric models to each of 
our galaxies. The position angles, inclinations, and centers were left free in the fits. 
Errors were estimated with 100 bootstrap resamplings. The fits are shown in Figure \ref{fig:disk1}. We recover the result from \cite{Spek07} that NGC~2976 has a substantially 
different rotation curve under the two parametrizations, with the bisymmetric rotation 
curve having larger rotational velocities at smaller radii. 
UGC~11707 shows unrealistic behavior, but the problem is only 
present in the second smallest radial bin for both models. 
The gas velocity map of UGC~11707 shows a small pocket of high velocity gas at 12\arcsec\ 
off the major axis. The gas emission in this bin is quite patchy. 
The high velocity measurement is real and significant, but it is unlikely to truly be a signal of 
regular, global, gravitationally dominated motion. The remaining five galaxies show strong agreement 
in the radial and bisymmetric models. Some bar and viewing angles can still hide kinematic 
features, but we see good evidence in one of seven galaxies for bars that produce a 
significant bias in the rotation curves.

\subsection{MOND parameter estimation}
\par Modified Newtonian Dynamics (MOND) was first proposed as an alternative 
explanation to dark matter for the flat rotation curves of galaxies 
at large radii by \citet{Milgr83}. The one additional physical constant 
required by MOND is $a_0$, the small acceleration at which objects start to 
deviate from Newton's Laws. The modification is usually written as 
$\mu(a/a_0)\times a = GMr^{-2}$ with a common choice of 
$\mu(a/a_0)=(1+a_0/a)^{-1}$. Under this form, the acceleration at the 
radius where a dark matter fit has an enclosed DM mass fraction of one-half 
can be interpreted instead as the value for $a_0$. We have made such an 
estimate from our data for the gNFW fits. We find 
$a_0=$0.78$\pm$0.37$\times$10$^{-10}$ m s$^{-2}$ from the gas-based data and 
$a_0=$0.56$\pm$0.40$\times$10$^{-10}$ m s$^{-2}$ from the stellar-based data. 
For both, we find root-mean-squares of 0.57$\times$10$^{-10}$ m s$^{-2}$, 
meaning that our data appear to favor a non-constant value of a$_0$. The 
significance for a non-constant value, however, is modest as 5/7 and 3/7 of 
the gas-based and stellar-based measurements deviate from the central 
estimate by 1-$\sigma$ significance.

\section{Interpreting the Mass Models}
\label{sec:interp}
\subsection{Agreement Between Kinematic Tracers}
\label{sec_disdis}
\par
\begin{figure}
\centering
\includegraphics [scale=0.85,angle=0]{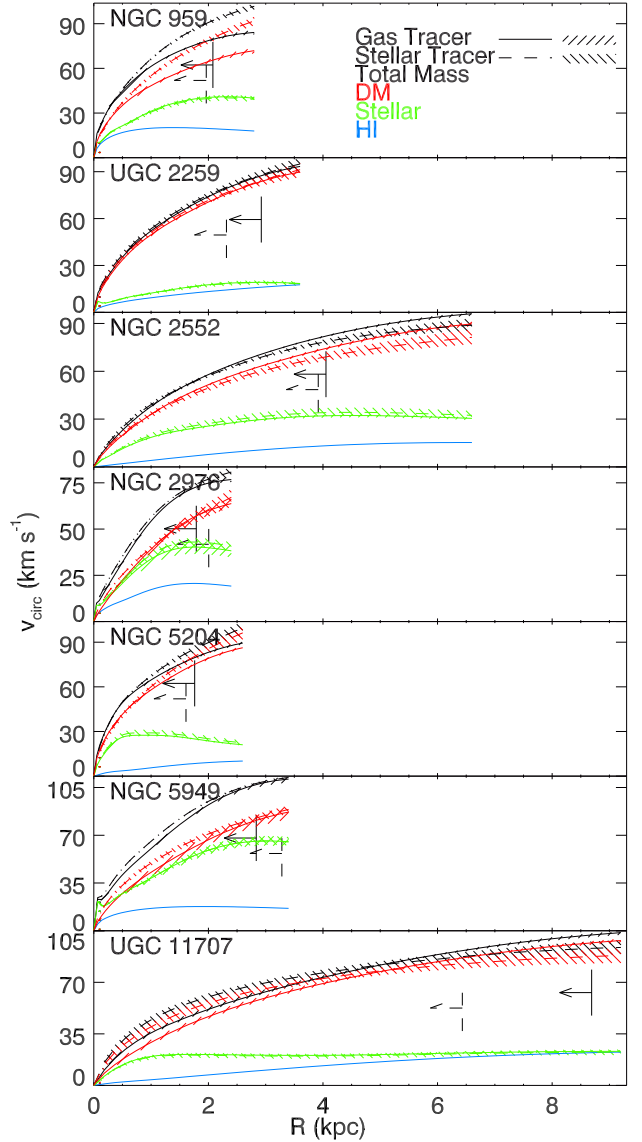}
\caption{The rotation curves for the best fit parameters from the gas-traced and stellar-traced data.
The median rotation curves for the various parts of the mass budget are shown as well as the 1-$\sigma$
confidence bands. Arrows are drawn to show the largest radius bin for each tracer. 
The circular velocity curves have been fit for each tracer as described in 
the text. For the gas data, the fit is made to the observed rotational 
velocity field. For the stellar data, the fit is made to the quadrature 
sum of rotational velocity and dispersion. 
\textit{\textbf{NGC~959}} 
The two tracers do not agree in their large-radii normalizations. The disagreement is robustly contained in the data, and
is likely due to our models not containing necessary complexity, such as from non-spherical DM halos. 
\textit{\textbf{UGC~2259}} The gas-traced model appears cuspier. 
\textit{\textbf{NGC~2552}} The large-radii normalization again disagrees
between the two models.
\textit{\textbf{NGC~2976}} A subtle but significant disagreement exists in the shape of the rotation curves. The 
disagreement is mainly in normalization with $\Upsilon_*$ fit differently between the two models. If a fixed value is used, 
the disagreement is primarily in a cuspier shape to the stellar-traced model. 
\textit{\textbf{NGC~5204}} The two
tracers show excellent agreement in their mass models. 
\textit{\textbf{NGC~5949}} The two
tracers show reasonable agreement in their mass models, but with the 
stellar-based model being modestly more cuspy. 
\textit{\textbf{UGC~11707}} The two
tracers again show modest disagreement at small radii, but the large 
error bars may explain the difference.} 
\label{fig:rotplot1}
\end{figure}

 We find that the posterior mass models inferred from the gas and stars 
are not always consistent. However, the density profiles are usually 
consistent. 
We show the inferred rotation curves for both kinematic tracers in Figure \ref{fig:rotplot1}. The curves, 
their decomposition into various mass components, and the 1-$\sigma$ ranges around the total rotation curves are 
created by averaging rotation curves from 1000 of the MCMC samples. NGC~2552 shows clear 
disagreement with the gas-traced asymptotic velocity being larger than the stellar-traced one, while the 
situation is reversed for NGC~959. 
We have tried evaluating the stellar-traced models against the 
gas data and vice versa, and the fits for these two galaxies are truly poor representations of the swapped data sets. 
In our current framework, we 
have no set of parameters that can bring the potentials into agreement. We speculate that the two potentials could be 
brought into better agreement by adding non-spherical structure to the DM halo, as the gas and stars will react differently. 
The simplest approach would be to make oblate or prolate DM halos in JAM. A constant ellipticity to the DM halo 
would shift the normalization of the rotation curves \citep{Simo05}. More complicated structure, such as 
ellipticity varying as a function of vertical height, could change the rotation curve shape, although a 
significant shape change could only come about by very contrived alignments, vertical scalings, and 
strong ellipticity gradients. Since the gas is the thinnest component, a constant ellipticity shift to the 
DM will change the gas-traced normalization more strongly than the stellar-traced one. This general problem is beyond the 
scope of this work. NGC~2976 shows modest, but significant, disagreement in the same sense of the radial and bisymmetric DiskFit 
models and as we found in \cite{Adams12}. The stellar-traced rotation curve has larger amplitude at smaller radii. The other four 
galaxies, UGC~2259, NGC~5204, NGC~5949, and UGC~11707 show fairly good agreement.
\par
\begin{figure*}
\centering
\subfigure{\includegraphics [scale=0.58,angle=0]{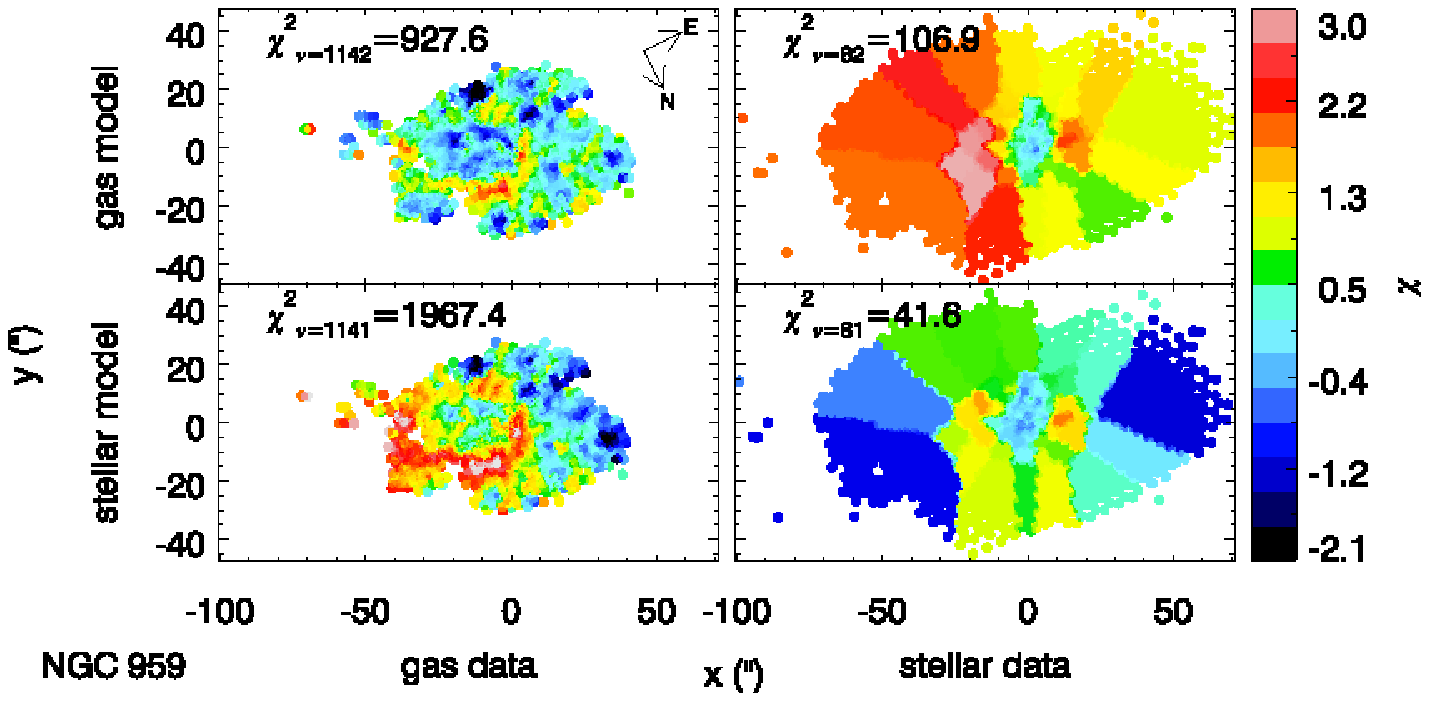}}
\subfigure{\includegraphics [scale=0.58,angle=0]{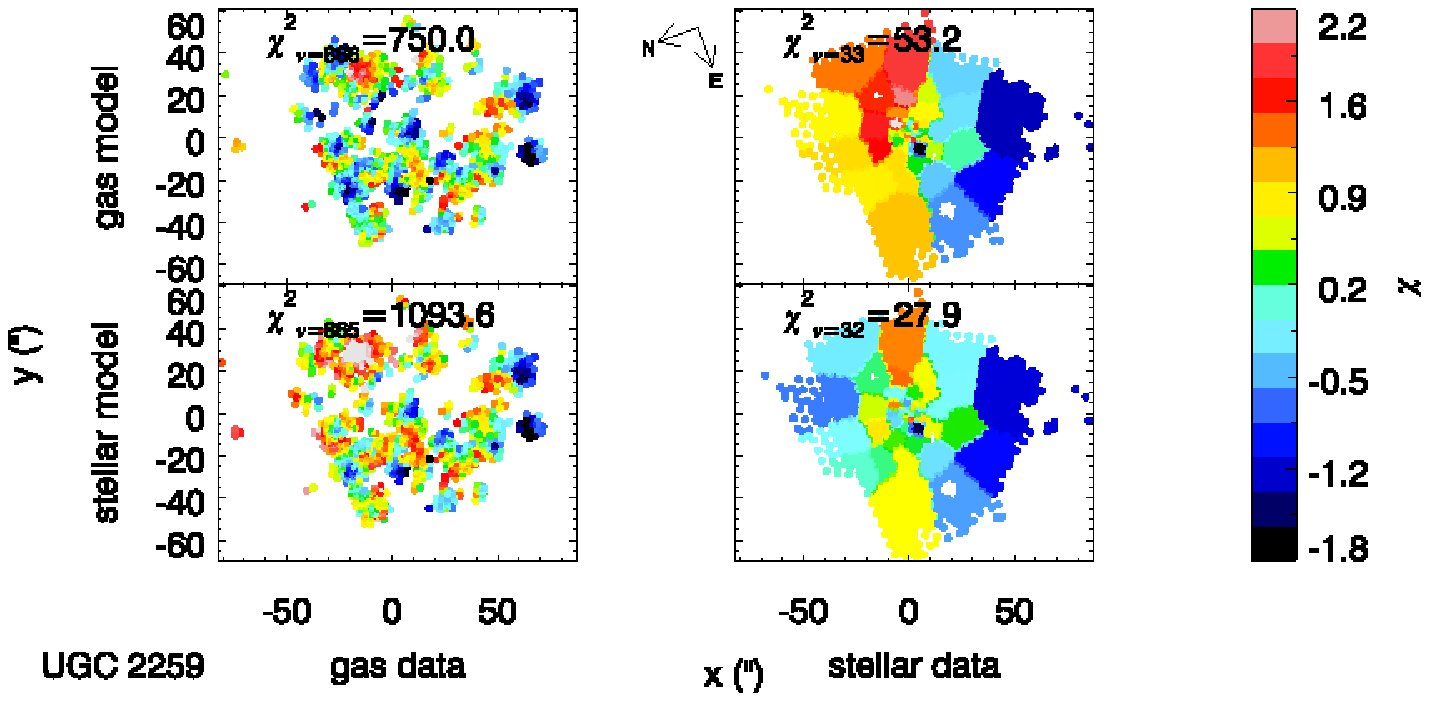}}\\
\subfigure{\includegraphics [scale=0.58,angle=0]{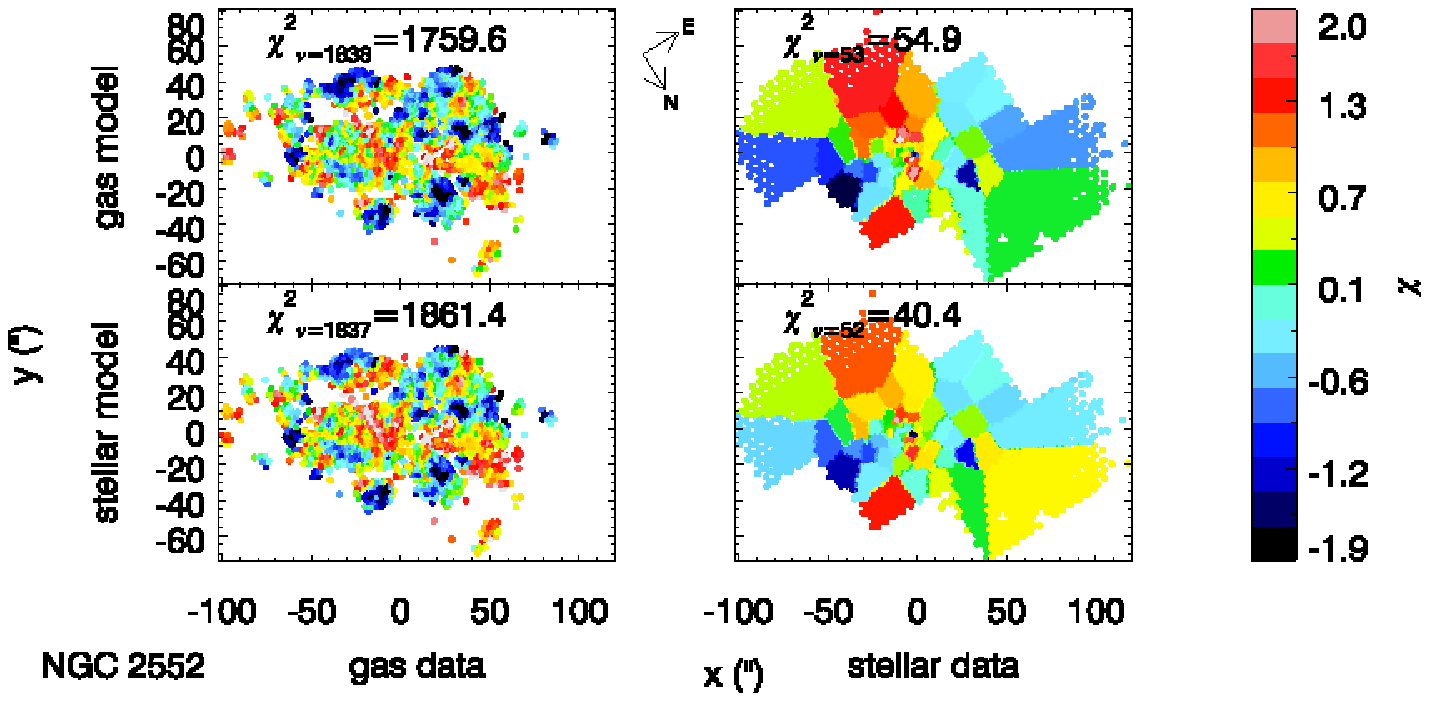}}
\subfigure{\includegraphics [scale=0.58,angle=0]{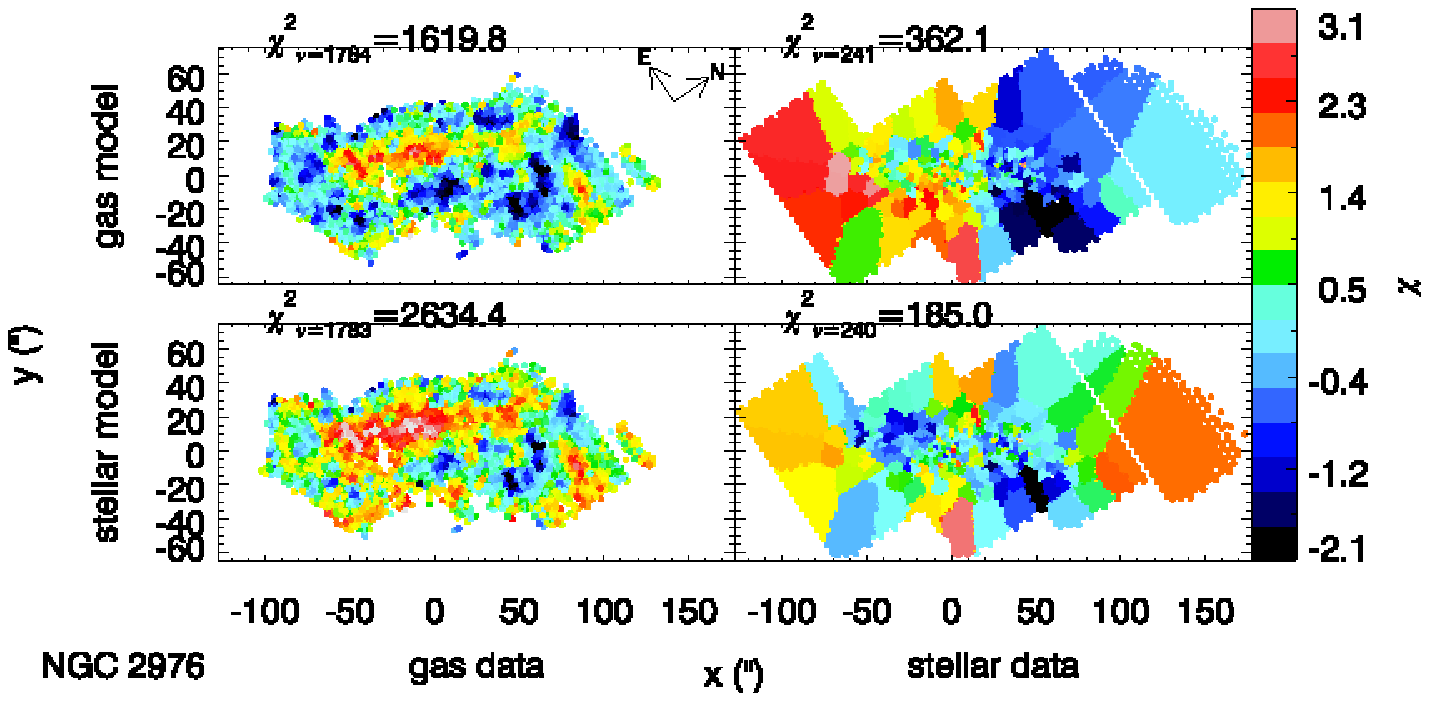}}\\
\subfigure{\includegraphics [scale=0.58,angle=0]{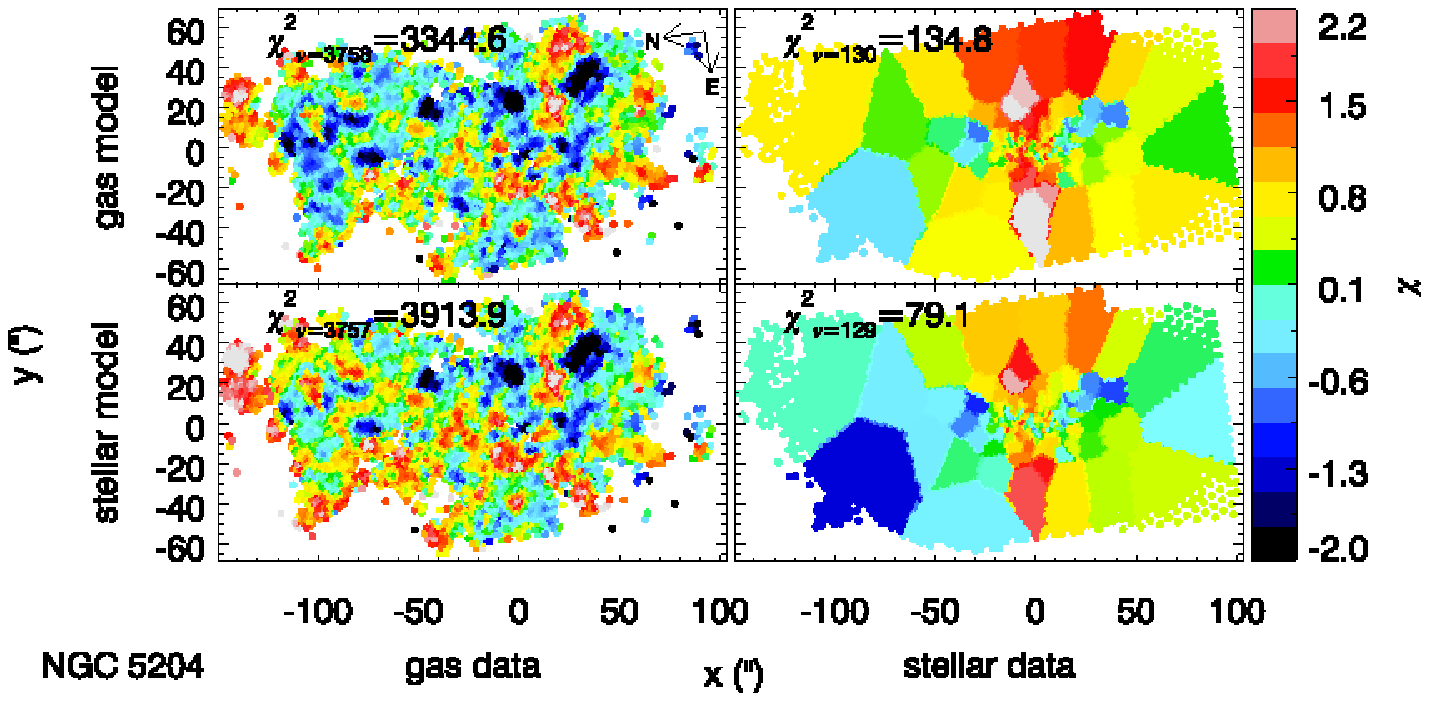}}
\subfigure{\includegraphics [scale=0.58,angle=0]{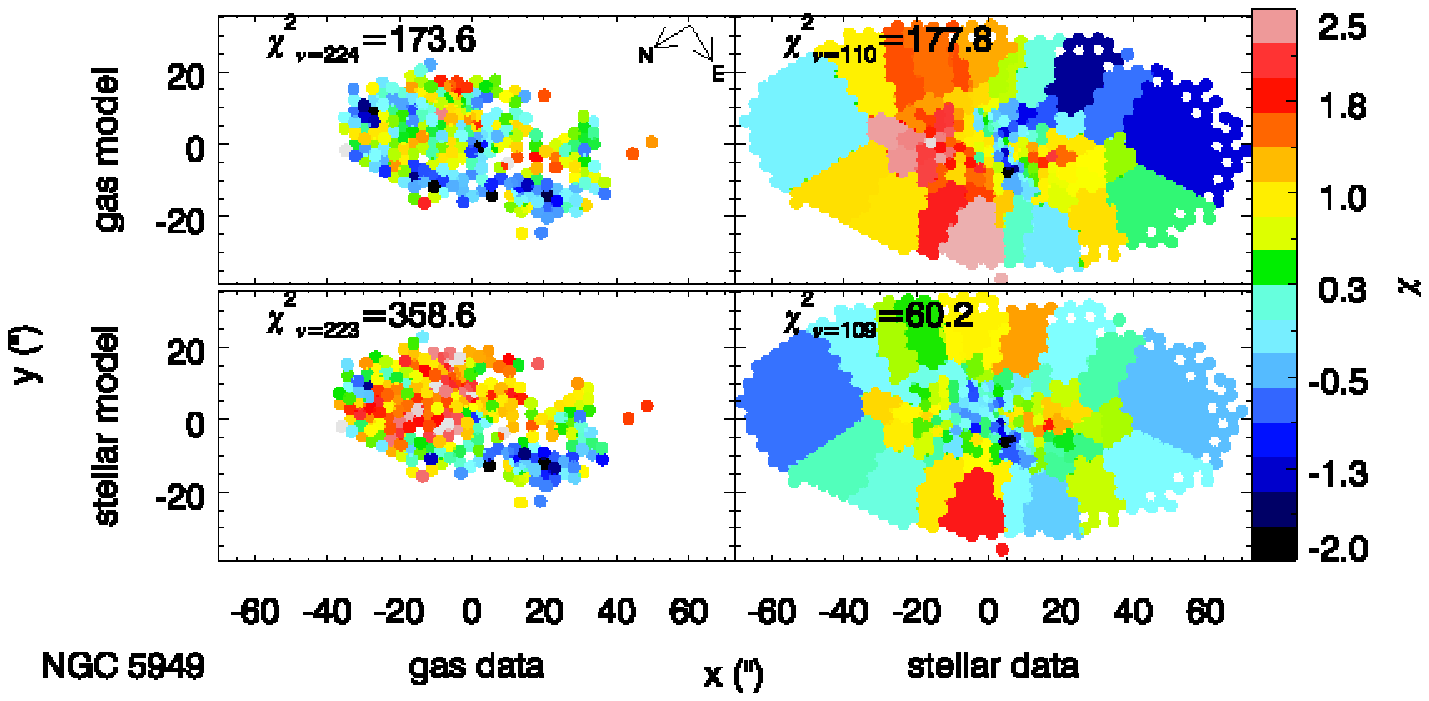}}\\
\subfigure{\includegraphics [scale=0.58,angle=0]{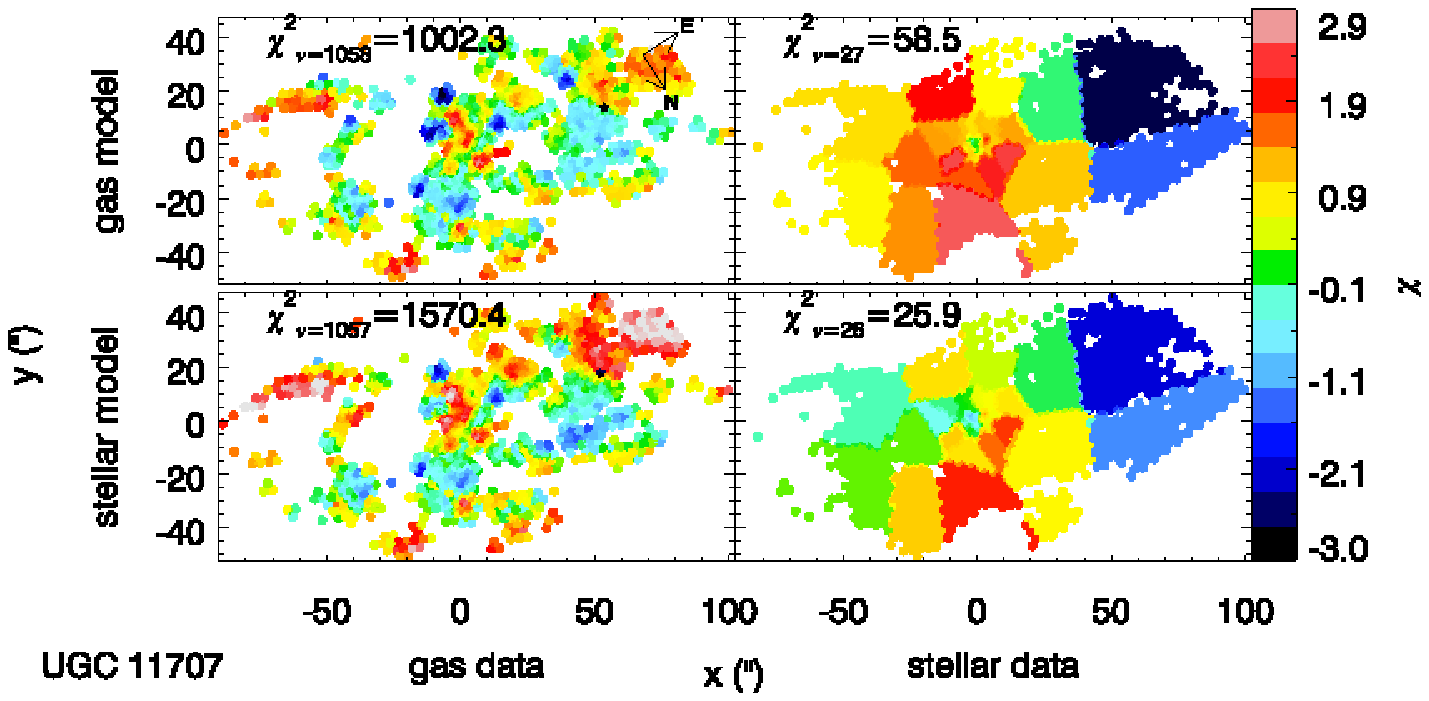}}
\caption{Quality-of-fit $\chi$ maps. \textit{\textbf{Left}} The 
line-of-sight gas velocities compared to models. \textit{\textbf{Right}} The 
stellar second moment velocities compared to models. 
\textit{\textbf{Top}} Model made with the best-fitting parameters based on the gas data. 
\textit{\textbf{Bottom}} Model made with the best-fitting parameters based on the stellar data. 
The total $\chi^2$ values and degrees of freedom are written in every panel. 
These plots indicate how strongly the two tracers disagree and show the regions that are 
causing the disagreement, when present. The major conflict in \textit{\textbf{NGC~959}} is that the 
stellar model robustly has a larger asymptotic velocity than the gas. The 
two models for \textit{\textbf{UGC~2259}} are quite similar and without
structured residuals in the swapped models. For \textit{\textbf{NGC~2552}}, 
the two models are quite similar, with most of the
structured residuals in the swapped models attributable to
slightly different kinematic centers and PAs. The two models for 
\textit{\textbf{NGC~2976}} show modest disagreement, with most of the
structured residuals in the swapped models attributable to
slightly different kinematic centers and $\gamma$. The two models 
for \textit{\textbf{NGC~5204}} are quite similar. For \textit{\textbf{NGC~5949}}, the 
two models show modest disagreement, with most of the
structured residuals in the swapped models attributable to
differences in $\gamma$. Finally, the two models for \textit{\textbf{UGC~11707}} 
show modest disagreement. The main
cause is a difference in $\gamma$, but even it is
compatible within the errors between the two tracers.}
\label{fig:chisq1}
\end{figure*}

 The data regions driving each fit can be seen in 
Figure \ref{fig:chisq1}. There, the residuals to the 
gas and stellar kinematic maps are shown along with the residuals from 
the parameter set selected by the alternate tracer. 

\par
\begin{figure}
\centering
\includegraphics [scale=0.85,angle=0]{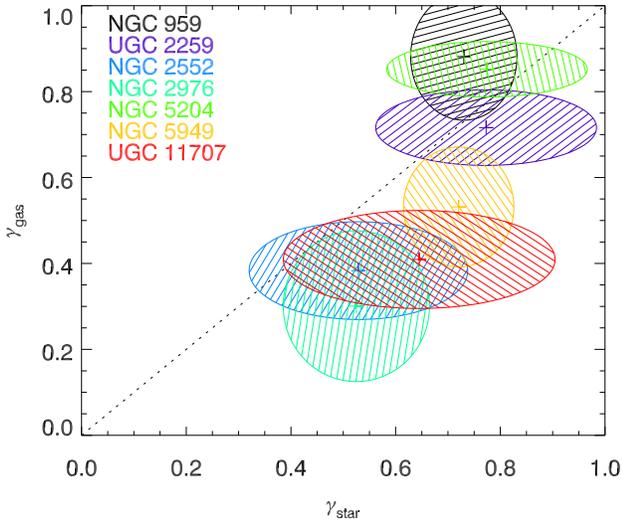}
\caption{The logarithmic DM slopes, $\gamma$, as derived from the gas- and stellar-traced models. The 
2D error ellipses represent 1-$\sigma$ uncertainties. The constraints on 
$\gamma$ are generally consistent between the two data sets. The line shown is the one-to-one relation expected if 
both tracers are unbiased.}
\label{fig:gam2}
\end{figure}

 The logarithmic DM slopes derived from the gas and stellar tracers are compared in 
Figure \ref{fig:gam2}. The data agree fairly well with the simple one-to-one relation. 
The most cored halos show a small bias toward being steeper from the 
stellar-traced models, but there is no doubt that several galaxies 
are incompatible with NFW profiles. NGC~2976 agrees with the one-to-to relation within the 
uncertainties, in contrast to our previous work. The gas-traced model 
has become steeper, because the model has swept over several PAs and not modeled 
non-circular terms, and the stellar-traced model has become shallower because the 
stellar photometry and mass has increased as discussed in \S \ref{sec_N2976comp}. 
Two galaxies are outliers beyond 1-$\sigma$ significance, which is expected from 
normal statistics alone. NGC~5949 looks more 
cuspy from the stellar-traced models. 
NGC~5949, however, does not have evidence for a 
bar from the gas kinematics field and the discrepancy is harder to 
explain. It may be that more complex triaxial structure is causing this 
subtle (barely 1-$\sigma$) difference between the two tracers, or the 
model disagreement may be due to random chance. 
\subsection{Comparison with Literature Density Profile Measurements}
\label{sec_N2976comp}
\par In \cite{Adams12} (hereafter A12) we presented lower quality data on NGC~2976. In 
addition to acquiring higher resolution, higher S/N data, we 
have also improved the analysis methods by adopting a Bayesian
approach.  We first address the question of whether the
A12 data and analysis are consistent with our new results
on NGC~2976, and then broaden this comparison to consider the other
galaxies with published density profile slopes in the literature.
\par We find that 
the qualitative result with the stars in NGC~2976 reporting a steeper DM halo than 
the gas remains in the A12 data, but the quantitative limits on the logarithmic slope 
of the density profile change. The most substantial change is that the SDSS photometry 
on NGC~2976 report double the luminosity as the SINGS image originally used. The SDSS 
photometry is in better agreement with that from \cite{Simo03}. 
From Figure 13 of A12, it can be seen that because of the
degeneracy between the stellar mass-to-light ratio and the density
profile slope, such a shift in the stellar 
mass decreases the most likely value of the logarithmic slope to $\sim$0.5. We ran 
the new MCMC pipeline with the A12 data and MGE components and recovered the 
same best-fit value as in Figure 13 of A12. Next, we ran the A12 data with 
the new pipeline but with the updated stellar luminosity. The central estimates 
for $\gamma$ with the A12 data and A12 analysis, the A12 data and new analysis, and new 
data and new analysis, respectively, are: 0.9, 0.62, and 0.53. The 
facts that the A12 data barely resolved the dispersions, were binned differently, 
and lacked the thorough parameter search from the MCMC pipeline do not 
make a significant difference given the strong consistency between the A12 and 
new constraints when using a common MGE model. In the case of a zero mass, or very low mass, 
disk for NGC~2976, the disagreement between the stellar and gaseous kinematics remains significant. 
The better agreement in $\gamma$ between the two tracers for NGC~2976
achieved in the present analysis (Figure \ref{fig:gam2}) is 
due to the two models settling on different values of $\Upsilon_*$ (Table \ref{tab:fits_param}). 
\par
\begin{figure}
\centering
\includegraphics [scale=0.85,angle=0]{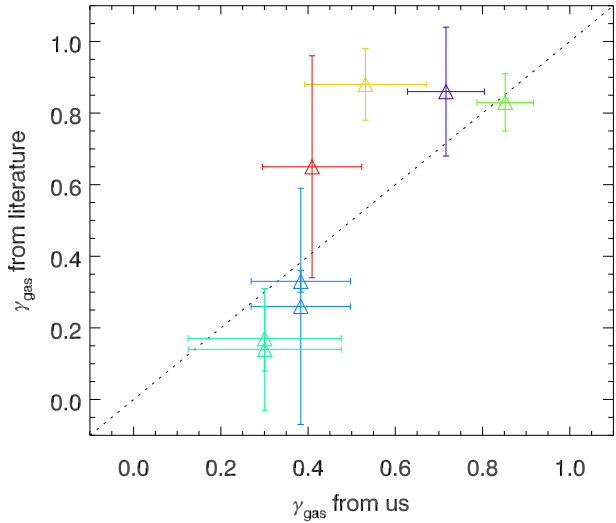}
\caption{The logarithmic DM slopes, $\gamma$, as fit from gas kinematics in this work and 
from all available literature values. The only measurement significantly discordant from a one-to-one relation 
is the NGC~5949 measurement from \cite{Simo05} being cuspier. Our stellar-based measurement is 
closer to the \cite{Simo05} value.}
\label{fig:gamlit}
\end{figure}

 Five of the remaining six galaxies have measurements of $\gamma$
available in the literature. We have plotted the literature values and our 
measurements in Figure \ref{fig:gamlit}. 
\par For the sixth galaxy, NGC~959, no numerical value of
$\gamma$ has been published. NGC~959 was fit with pseudo-isothermal (cored) 
and NFW profiles in \cite{Kuzi08} under a zero disk mass assumption. There was a 
very modest statistical preference for the cored fit, with $\chi^2_r$ values of 
1.2 and 1.7, respectively.  
\par UGC~2259 was measured by 
\cite{deBl96} and only fit with a pseudo-isothermal function. They found 
a core size of 0.4--1.2 kpc depending on the value of $\Upsilon_*$. \cite{Swat03} 
measured UGC~2259 to have $\chi^2_r$ values of 1.8 and 2.3 for 
pseudo-isothermal and NFW functions respectively. They also fit a gNFW function with 
$\gamma=0.86\pm0.18$. For
UGC~2259, we measure $\gamma = 0.72 \pm 0.09$ from the gas and $\gamma
= 0.77 \pm 0.21$ from the stellar kinematics, in excellent agreement
with \cite{Swat03}. 
\par NGC~2552 was fit by \cite{deBl01} to have 
$\gamma=0.33\pm0.03$. \cite{Swat03}
measured NGC~2552 to have $\chi^2_r$ values of 1.5 and 3.4 for
pseudo-isothermal and NFW functions. They also fit a gNFW function with 
$\gamma=0.26\pm0.33$. \cite{Kuzi06} find moderately poor pseudo-isothermal fits ($\chi^2_r=3.8$) but 
very poor NFW fits ($\chi^2_r=40$). \cite{Kuzi08} reinforce these findings for 
additional values of $\Upsilon_*$. \cite{Span08} find that the pseudo-isothermal ($\chi^2_r=2.5$) and 
NFW ($\chi^2_r=2.9$) functions fit the data to a similar level. We find slightly
steeper profiles using both the gas ($\gamma = 0.38 \pm 0.11$) and
the stars ($\gamma = 0.53 \pm 0.21$), but the results are consistent
within the uncertainties.
\par NGC~2976 was fit in 
\cite{Simo03} to have $\gamma=0.17\pm0.09$ for a minimal disk or 
$\gamma=0.01\pm0.13$ for a maximal disk. \cite{deBl08} made fits with 
$\Upsilon_*$ freely determined and found $\chi^2_r=0.5$ for a 
pseudo-isothermal fit and $\chi^2_r=1.7$ for a NFW fit. Both fits had their parameters run into physical 
boundaries, with the pseudo-isothermal core radius going to infinity and the NFW concentration going to zero. As 
discussed above, we find a value in NGC~2976 
steeper than that presented in \cite{Simo03}, especially for our stellar-traced value. Our present value is 
shallower than that presented in \cite{Adams12}.  
\par NGC~5204 
was measured by \cite{Swat03} to have $\chi^2_r$ values of 1.0 and 1.7 for
pseudo-isothermal and NFW functions. They also fit a gNFW function with 
$\gamma=0.83\pm0.16$. \cite{Span08} find that the pseudo-isothermal 
($\chi^2_r=0.6$) model fits moderately better than the NFW ($\chi^2_r=2.1$) function. Our gas-traced 
measurement ($\gamma = 0.85 \pm 0.06$) is spot-on
with the \cite{Swat03} value and our
stellar-traced value ($\gamma = 0.77 \pm 0.19$) also agrees within the
uncertainties.
\par \cite{Simo05} make a disk-subtracted fit to NGC~5949 and find $\gamma=0.88$. 
NGC~5949 is fit by \cite{Span08} to a 
similarly likely level with pseudo-isothermal ($\chi^2_r=1.0$) and
NFW ($\chi^2_r=1.4$) functions. Our measurements for NGC~5949 ($\gamma = 0.53 \pm
0.14$ from the gas and $\gamma = 0.72 \pm 0.11$ from the stars)
are shallower than the value determined by
\cite{Simo05}, although not by a large margin when the combined
uncertainties are taken into account.
\par \cite{Swat03} measured UGC~11707 to have $\chi^2_r$ values of 1.5 and 0.5 for
pseudo-isothermal and NFW functions. They also fit a gNFW function with 
$\gamma=0.65\pm0.31$. \cite{Span08} finds a moderately better pseudo-isothermal ($\chi^2_r=2.6$) 
than NFW ($\chi^2_r=3.8$) function fit for UGC~11707. We measure $\gamma = 0.41 \pm 0.11$ for the gas kinematics and
$\gamma = 0.65 \pm 0.26$ for the stellar kinematics of UGC~11707, in
agreement with the value from \cite{Swat03} given the large uncertainty.
\par
\begin{figure}
\centering
\includegraphics [scale=0.85,angle=0]{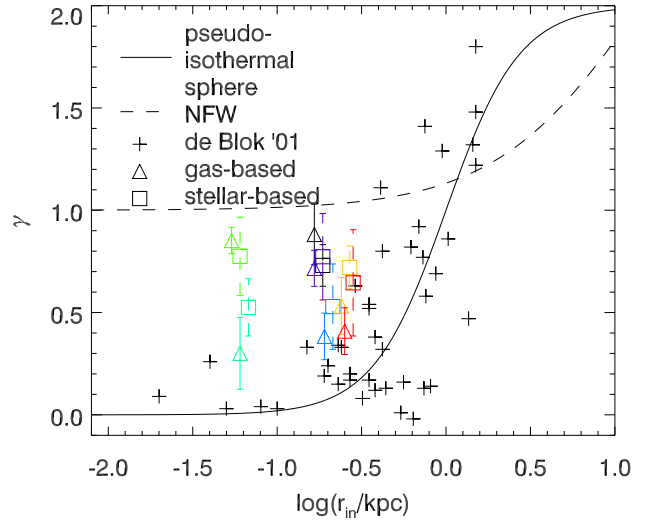}
\caption{Demonstration that the spatial resolution of the data is sufficient 
to discriminate between models. The crosses are from the 
compilation of data in Figure 1 of \citet{deBl01}. When the spatial resolution 
is poor, as indicated by the innermost radial data point, a cupsy measurement 
can be obtained even if the true density profile contains a core, 
like the pseudo-isothermal 
sphere. Our data are shown with the color coding per galaxy as in 
Figure \ref{fig:gam2}. The innermost radius is determined by adding the 
seeing to the fiber radius in quadrature. All of our 
data have high enough spatial resolution that 
fully cored profiles would have been detected if present. The solid curve 
assumes a 
core radius of 1 kpc for the pseudo-isothermal sphere, 
and the dashed curve assumes c=10 and V$_{200}$=100 km s$^{-1}$ 
for the NFW functional form.}
\label{fig:gamrin}
\end{figure}

 Finally, we consider logarithmic density slopes compared to the 
spatial resolution of the data. It has been shown before \citep{deBl01} that 
data taken with insufficient spatial resolution can bias the measurement of 
fully cored systems to higher values of $\gamma$. Figure 1 of 
\citet{deBl01} looked at this issue through data compiled from three 
earlier studies. In Figure \ref{fig:gamrin}, we add our data to this plot. 
One can see that our data lie far below the turnoff where biases occur for 
reasonable core sizes. Our results are therefore not dependent on spatial 
resolution. Additionally, our data occupy a space not populated by 
the \citet{deBl01} study. We genuinely find systems intermediate between 
cores and cusps rather than fully cored halos. 

\section{Discussion}
\label{sec:dis}
\subsection{The Distribution of Logarithmic Density Profile Slopes}

Because theoretical models generally do not attempt to reproduce the
detailed properties of individual galaxies, the best way to compare
observational results with theoretical predictions is to consider the
distribution of dark matter density profile slopes.  The previously
published values are $\gamma = 0.73 \pm 0.44$ from five galaxies
analyzed by \citet{Simo05} and $\gamma = 0.29 \pm 0.07$ determined for
seven THINGS dwarf galaxies by \citet{Oh11a}.  Our sample of seven
galaxies yields $\gamma = 0.67 \pm 0.10$ for the stellar kinematics
and $\gamma = 0.58 \pm 0.24$ for the gas.  Interestingly, our mean
slope using either tracer is significantly steeper than the THINGS
\ion{H}{1} measurements, but we find a narrower distribution than
\cite{Simo05}.  In fact, the uncertainty on the mean value of $\gamma$
we derive includes both intrinsic and measurement contributions, so
that the intrinsic scatter from galaxy to galaxy appears to approach
zero.  These data are thus consistent with a universal profile, but
not with the slope predicted by dark matter-only N-body simulations in
$\Lambda$CDM.  Of course, our sample is still small enough that we
could have failed to include intrinsically cored or cuspy galaxies simply by
chance.


\par To obtain the best available estimate of the distribution of
density profile slopes, we combine our sample with the highest-quality
literature measurements, including only objects for which
high-resolution two-dimensional velocity fields are available. We 
do not include results from \cite{Kuzi08,Kuzi09} because they do not derive 
logarithmic density slopes. 
\citet{Simo05} observed three galaxies that do not overlap with our
sample: NGC~4605 ($\gamma = 0.78$), NGC~5963 ($\gamma = 1.20$), and
NGC~6689 ($\gamma = 0.79$).  The other comparable data set is that of
\cite{Oh11a}.  We note several important differences between our
methodology and theirs.  First, the THINGS dwarfs have lower stellar
masses than our targets, but both are in the range over which SN
feedback models are predicted to be effective at changing the DM
profiles.  Second, rather than fitting a density profile to the full
rotation curve or velocity field, \cite{Oh11a} measure the density
profile slope by fitting a power law only to the innermost $\sim5$
points of the rotation curve ($r \lesssim 1$~kpc), which are the ones
potentially most subject to systematic uncertainties and assumptions
about the baryonic mass distribution. 
Most of the THINGS sample also exhibits serious kinematic peculiarities
that may bias the inferred density profiles. For example, five of the seven galaxies 
(NGC~2366, Holmberg~I, Holmberg~II, DDO~53, DDO~154) have 
significant kinematic asymmetries between the approaching and receding sides of the
galaxy, two have substantial non-circular motions (IC~2574, NGC~2366) , and one has a 
declining rotation curve (M81dwB). 
If we include only the best behaved THINGS galaxies, Ho~II and DDO~154, with 
the \cite{Simo05} sample and our new results, that
leaves us with a total sample of 12 objects.  This combined set produces a mean slope of
$\gamma = 0.63$ and a scatter of $\sim0.28$, essentially independent
of whether we use gaseous or stellar kinematics.  We suggest that
these values represent the best current constraints on the
observational distribution of dark matter density profiles on galaxy
scales. These values are strikingly similar
to those measured for the dark matter profiles of galaxy clusters by 
\cite{Newman13b}. 
\subsection{Systematic Uncertainties}
\par Our data have reduced sensitivity to some sources of systematic error and 
are equally sensitive to other systematic errors as previous data sets. By using 
stellar kinematics, we are potentially adding an additional systematic through the 
vertical orbital anisotropy. The tests presented in Appendix \ref{app:MCMCtest} were designed to 
test for this problem through simulations, and the level of systematic bias 
in $\gamma$ is found to be $<0.1$. The properties of high spatial resolution and 
two dimensional kinematics in our data retire a swath of systematic risks relating to 
locating the kinematic center and major axis alignment. While binning in our stellar 
kinematic extraction could plausibly be a risk, we find again from the simulations 
that this introduces no bias. Gas-based analysis is usually performed under the 
assumption of infinitely thin, planar, and circular motion. Our stellar-based 
analysis is predicated on the less restrictive assumptions of a steady-state, 
axisymmetric potential. While we find cases that can be resolved 
by breaking the usual gas modeling assumptions, the biases in $\gamma$ are small and 
comparable to the statistical uncertainties. Three sources of systematic 
error to which we, and other methods, are still potentially susceptible are 
triaxial DM halo structure \citep{Valen07,Kowal13,Valen14}, M/L$_*$ variations with radius, and 
more complex orbital anisotropy than can be fit with a single $\beta_z$ term. A fourth, minor concern is that the assumption in the JAM models of 
a cylindrically aligned dispersion ellipsoid is invalid. 
This is unlikely to be important because regions near the disk plane 
will be cylindrically aligned and the extra-planar regions will only 
become influential at very high inclinations. 
We note that the main effect 
of a constant ellipticity triaxial halo is to change the rotation curve normalization 
\citep{Simo05} and the indications from color and Lick index gradients are that 
M/L$_*$ gradients are small or absent in our galaxies. We do not identify any source 
of systematic error that can plausibly affect our measurements of $\gamma$ at a 
significant level.  

\subsection{Inner DM Surface Density}
\par Many authors have found that the inner DM surface density ($\rho_b\times r_b$) 
takes on a constant value for all galaxies and galaxy clusters measured. \citet{Donat09}
determine this constant to be 141$^{+82}_{-52}$ M$_{\odot}$ pc$^{-2}$.
We find an average of $\rho_b\times r_b=$321 M$_{\odot}$ pc$^{-2}$ from the gas and
$\rho_b\times r_b=$341 M$_{\odot}$ pc$^{-2}$ from the
stars. One galaxy, NGC 2552, lies removed from these averages by 1.5$\sigma$ in both tracers,
and no others lie beyond 1$\sigma$. Within the limits of our errors and small sample,
the inner DM surface density is constant but significantly different than previous 
estimates. Additionally, the inner DM surface densities of 
M31 dSphs appear offset from those of the MW \citep{Walke10} 
and the DM surface density may not be a constant.
\subsection{Correlations with DM Halo Profile}
\label{sec_corr}
\par
\begin{figure}
\centering
\includegraphics [scale=0.85,angle=0]{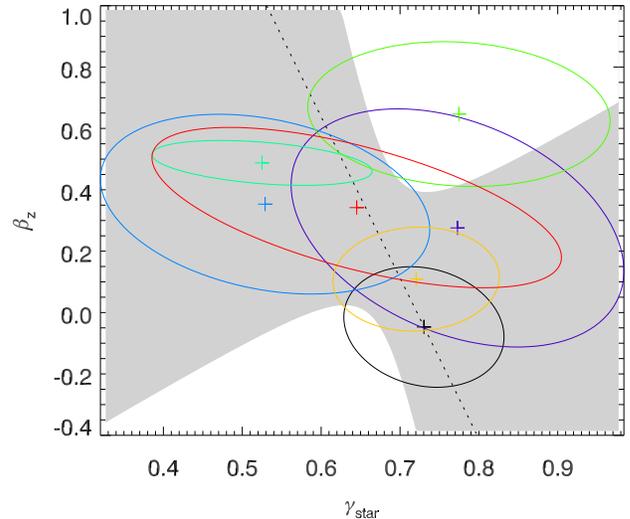}
\caption{The stellar-based logarithmic DM slopes and vertical orbital anisotropies. A maximum likelihood 
linear fit to the stellar-based values is largely unconstrained, but shown by the line and 
shaded 1-$\sigma$ confidence interval. The
color coding per galaxy is the same as in Figure \ref{fig:gam2}.}
\label{fig:gambeta}
\end{figure}

\begin{figure}
\centering
\includegraphics [scale=0.85,angle=0]{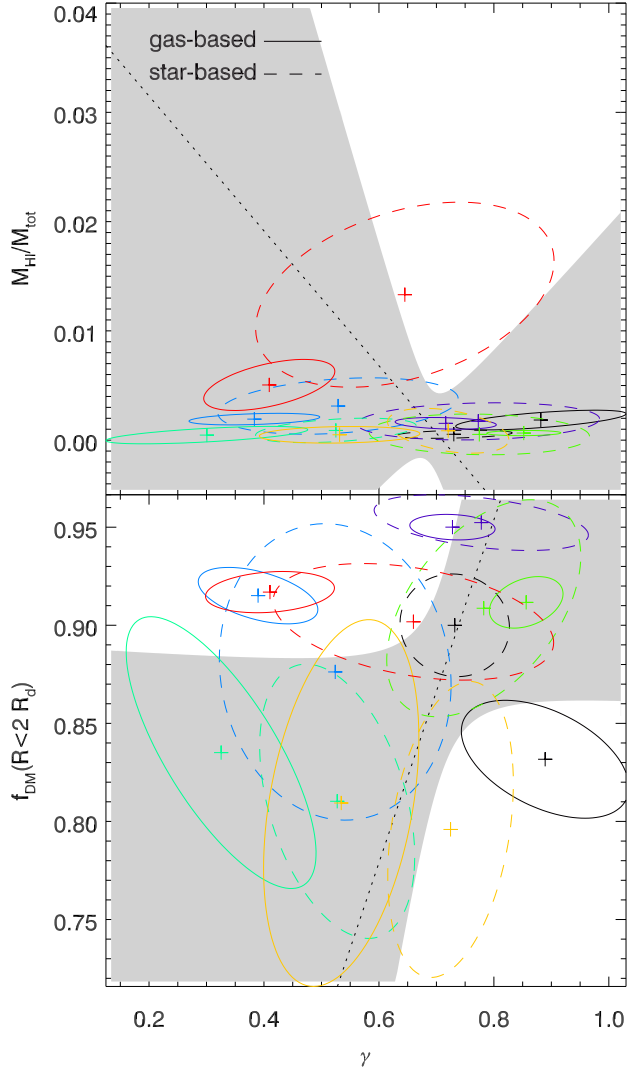}
\caption{Investigation of correlations with the logarithmic DM slope, $\gamma$. The
color coding per galaxy is the same as in Figure \ref{fig:gam2}. The maximum likelihood 
linear fit to the stellar-based values is shown. Stellar-based values are dotted and gas-based values use solid lines. 
\textit{\textbf{Top}} The ratio of HI to stellar mass. The data do not make a strong constraint and only 
exclude the strongest of correlations. The best fit is anti-correlated. SN feedback 
models would presumably predict a positive correlation between these values, if anything. More data to strengthen the 
anti-correlation and robust model predictions are necessary to investigate this preliminary evidence. \textit{\textbf{Bottom}} 
The DM mass fraction within a sphere of twice the stellar scale radius. The data exclude an anti-correlation. 
}
\label{fig:gamHI}
\end{figure}

\begin{figure}
\centering
\includegraphics [scale=0.85,angle=0]{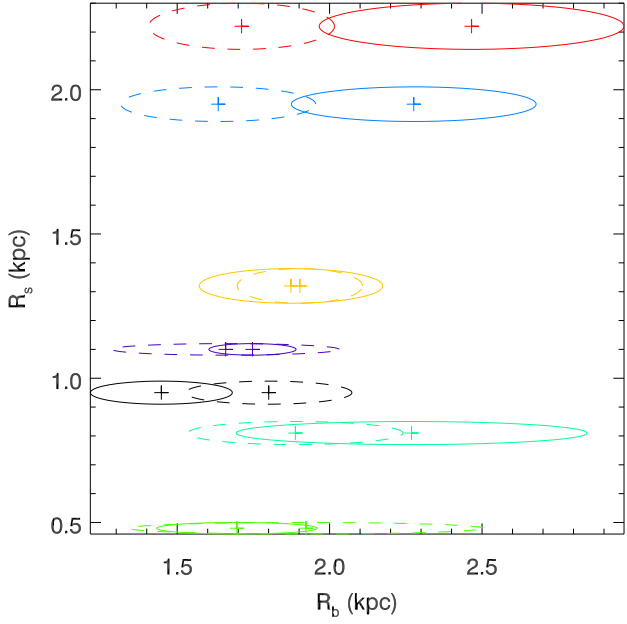}
\caption{The stellar scale lengths, R$_s$, and the halo core sizes, R$_b$, fit 
with a Burkert function. SIDM simulations are being created that are 
starting to make falsifiable predictions in these variables \citep{Kapli13}. 
One trend in the simulations is that higher baryon fraction, along with 
more massive disks and larger stellar scale lengths, leads to smaller 
halo cores. Stellar-based values are dotted
and gas-based values use solid lines. The maximum likelihood linear fit for the stellar-based
measurements is for 
an anti-correlation, but the correlation strength is weak and the uncertainty band fills 
nearly the entire plot. 
The color coding per galaxy is the same as in Figure \ref{fig:gam2}.}
\label{fig:szsz}
\end{figure}

 Although one might expect correlations with baryon content or stellar populations, 
we know of no secondary observables that are predicted to correlate with the 
logarithmic DM slope in the SN feedback or non-CDM models. Lacking firm theoretical guidance, 
we explore several parameters for clues to the underlying physics. We show the 
stellar-traced logarithmic DM slopes compared to the orbital anisotropies in Figure 
\ref{fig:gambeta}. NGC~2976 and NGC~5204 have significantly flattened stellar velocity 
ellipsoids as seen by previous studies (Appendix \ref{app:betatest}), but two others are 
intermediate and the remaining three galaxies are significantly more isotropic. 
There is an anti-correlation between 
$\gamma_{star}$ and $\beta_z$, however the large error bars leave only a modest 
significance to the correlation. We next plot the logarithmic DM slope 
against the relative mass in HI and stars in Figure \ref{fig:gamHI}. The SN feedback 
models make repeated starbursts at high redshifts. We cannot assess whether 
galaxies having experienced such disruptive events should still be 
expected to be gas-poor or whether subsequent accretion would wash out the signal. 
However, a strong trend between cored-ness and gas-deficit would perhaps be 
an indicator for the SN mechanism. The strength of correlation between $\gamma$ and the gas content is 
quite weak, and we can only exclude a very strong positive correlation. We have also calculated the HI gas mass internal to 
2 $R_d$ relative to the total mass as a way to investigate the central 
HI content that may be more sensitive to SN feedback. The results look like a scaled 
version of the total HI fraction.
We calculate the DM, stellar, and HI masses within two scale scale lengths. One 
non-analytic integral \citep[Equation 5 of ][]{Wyit01} was required to compute the 
gNFW enclosed mass. The DM fraction is compared to the logarithmic DM slopes in 
Figure \ref{fig:gamHI}. There is a weak correlation, but one also consistent with the 
uncertainties and a null hypothesis. This correlation may also naturally arise without 
a feedback mechanism because more-cored DM halos have lower central masses. We also 
look at the Burkert core sizes compared to the stellar disk scales in Figure \ref{fig:szsz}. 
The self-interacting DM models of \cite{Kapli13} predict correlations between disk scale length 
and DM core size at high baryon fraction but anti-correlation between DM core size and disk mass. For our sample drawn from a range of disk masses and 
intermediate baryon fractions, the second trend is most important and would 
imply an anti-correlation in DM core size and stellar disk scale. 
However, we find neither 
a positive nor negative correlation in our data.
\par
\begin{figure}
\centering
\includegraphics [scale=0.85,angle=0]{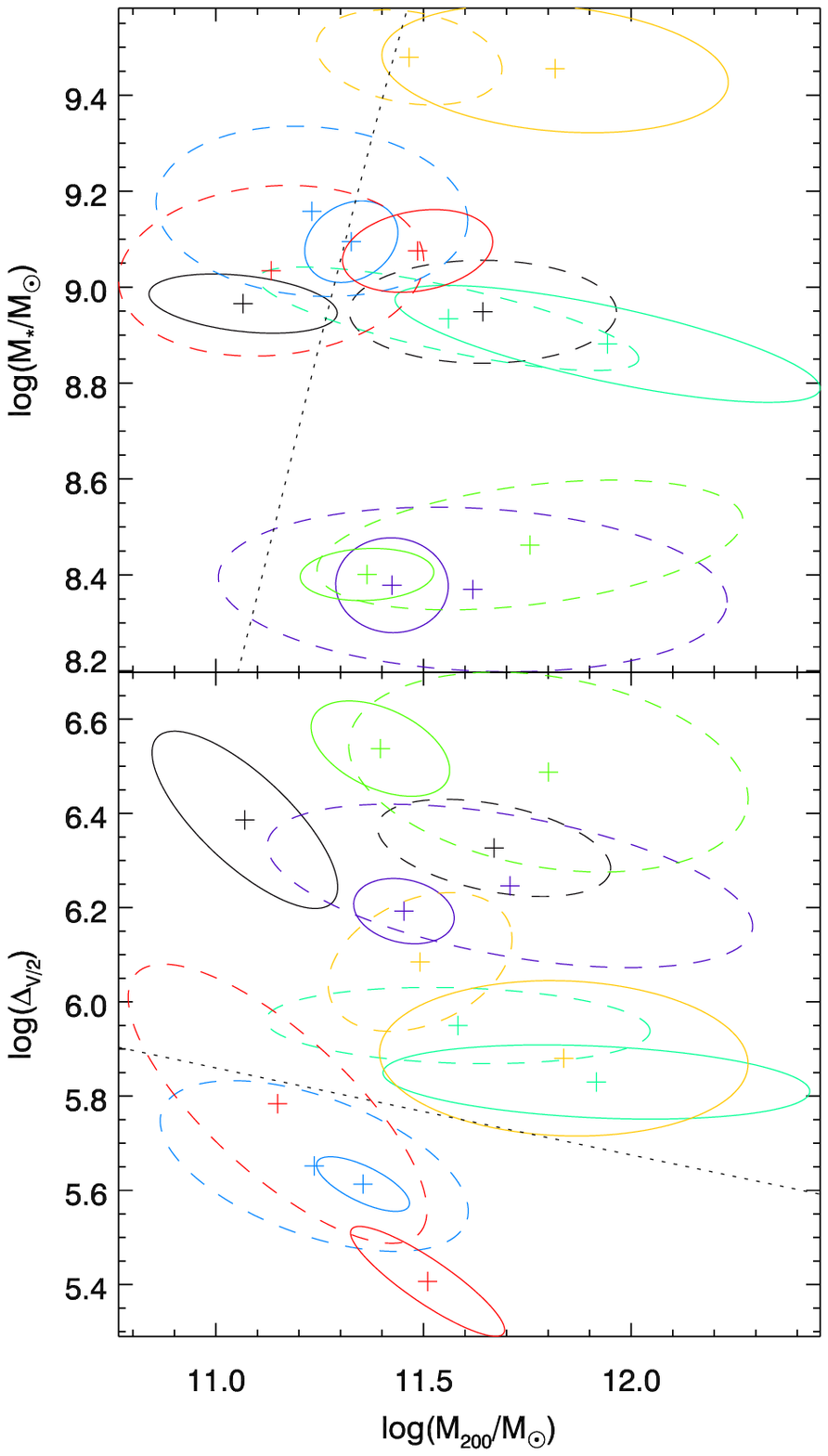}
\caption{Measured galaxy properties as a function of virial mass and 
compared to correlations seen in theoretical models. The
color coding per galaxy is the same as in Figure \ref{fig:gam2}. The lines shown are 
not fits to our data. Stellar-based values are dotted and gas-based values use solid lines. 
\textit{\textbf{Top}} The total DM and stellar masses are compared. The line is the relation found 
for $z=0$ subhalo abundance matching models in \cite{Behr13}. The data have reasonable 
agreement to the model, although with some scatter. \textit{\textbf{Bottom}} The measured virial masses and 
the non-parametric overdensity parameter from
\cite{Alam02} are compared. We calculate $\Delta_{V/2}$ for our data with the gNFW fits and by assigning 
V$_{\textrm{max}}$ to the circular velocity at our largest measured radius. Note that this is not how the 
theoretical values were estimated and may be a source of bias. To investigate that point, we also calculated 
V$_{\textrm{max}}$ from the full gNFW function and found $\Delta_{V/2}$ values closer to, 
but still not below, the line.  
The line is calculated for the NFW function and the M$_{200}$-c relation found
for central halos models in \cite{Klyp11}. Despite contrary findings in previous studies, 
we see good agreement in $\Delta_{V/2}$ with the $\Lambda$CDM predictions.}
\label{fig:MM}
\end{figure}

 Finally, we investigate other parameters that have a long history of being 
correlated in N-body and semi-analytic models. In \citet{Ferre12}, the halo masses of dwarf galaxies are measured and have       
smaller DM halos over the range $10^6<M_{*}/M_{\odot}<10^7$ than the semi-analytic
models of \citet{Guo11} predict. The two likely explanations for their data are that
vast numbers of dwarf galaxies are being missed by current surveys or that feedback is able
to dramatically reduce the DM halo masses in dwarfs. However, at the stellar mass values
studied in our work, the \citet{Ferre12} measurements indicate agreement with
semi-analytic modeling. We show the total stellar mass 
compared to the DM mass for our sample in Figure \ref{fig:MM}. The stellar mass to halo mass 
is a strong prediction of the data-tuned models of \cite{Behr13}. We find 
that on average, our measurements agree with the $z=0$ models. Another robust 
prediction from N-body models is the DM mass-concentration relation. Since 
our data do not adhere to the NFW functional form, it is not meaningful to 
compare our values of $c$ with those from simulations. \cite{Alam02} have 
presented a non-parametric overdensity parameter, $\Delta_{V/2}$, to address exactly 
this issue. The value simply represents the overdensity relative to the critical 
density at the radius where the circular velocity equals one-half the halo's maximum. This 
value can be derived numerically for an arbitrary density profile. We measure the value for 
our gNFW fits in Figure \ref{fig:MM} as compared to N-body model predictions. Note that we have 
assigned V$_{\textrm{max}}$ to the circular velocity at our largest measured radius for this plot. 
If we instead calculate V$_{\textrm{max}}$ from the full gNFW function, $\Delta_{V/2}$ becomes lower but 
several galaxies take on V$_{\textrm{max}}$ values significantly larger than their \ion{H}{1} linewidths 
would indicate. The two likely causes of this mismatch are that neither our
data nor the \ion{H}{1} linewidths are reaching the maximum velocity
in the halo or the virial mass cannot be robustly constrained by data 
from the central few kpc and the gNFW fits are mitigating the residuals 
at small radii by biasing high M$_{200}$ which is acceptable statistically as
the large radii bins have large uncertainties. The
mismatch we see between V$_{\textrm{max}}$ and V$_{\textrm{flat}}$ is
not the same as the core-cusp issue because our galaxies are not
underdense in $\Delta_{V/2}$ by either measure. The N-body predictions come from NFW fits and the M$_{200}$-c relation for central halos from 
the Bolshoi simulations \cite{Klyp11}. Surprisingly, our measurements lie at or above the 
predictions. $\Delta_{V/2}$ has been 
seen to lie below $\Lambda$CDM predictions in \cite{Alam02} and again in \cite{Simo05}, albeit 
within 1-$\sigma$ of the cosmic scatter. Contrary to many previous results, the galaxies in our 
sample do not seem to be underconcentrated relative to theoretical predictions. While the logarithmic DM slopes indicate a core/cusp 
problem, $\Delta_{V/2}$ is sensitive to larger radii regions and does not indicate a deficiency in 
DM mass. The mismatch we see between V$_{\textrm{max}}$ and V$_{\textrm{flat}}$ 
is not the same as the core-cusp issue because our galaxies are not 
underdense in $\Delta_{V/2}$ by either measure. 
Together, $\gamma$ and $\Delta_{V/2}$ limit the radial range of mass redistribution and 
energy input permitted by any feedback model. 
\par We have quantified possible correlations based on the stellar-traced kinematic models in 
Table \ref{tab:lineq}. The parametrization is limited to a linear model. Since we have 
errors in both dimensions and often correlated errors, we have used a general maximum 
likelihood solution for the fit parameters instead of the normal equations for linear least squares. 
We evaluate the likelihood for each model by finding the orthogonal residual to the line of each datapoint and 
the projection of the error ellipse along the same direction, which when divided, form the deviates. The 
sum of the squares of the deviates are then minimized and the covariance matrix is found through a 
non-linear least-squares solver. As suggested in \cite{Hogg10}, we have fit the terms $\theta$ and $b_{\perp}$ 
instead of a traditional slope and intercept. Doing so treats all slopes as equally likely a priori with 
implicit, flat priors on $\theta$ and $b_{\perp}$. The simple relation, 
\begin{equation}
\label{eq:lineq}
y=x \times \tan\theta + \frac{b_{\perp}}{\cos\theta}.
\end{equation} 
gives the linear fit for variables $x$ and $y$ in terms of $\theta$ and $b_{\perp}$. The 
dependent and independent variables and $b_{\perp}$ share the same units for each fit except 
for the [MgFe'] fit where $b_{\perp}$ has units of \AA. $\theta$ is given 
in radians. We have propagated the uncertainties through Taylor expansion to the dependent variable for all figures with a useful 
constraint. When no confidence interval is shown, the intervals fill nearly the entire plot range and are 
mentioned in the caption. 
\begin{deluxetable*}{llcccccl}
\tabletypesize{\scriptsize}
\tablecaption{Measured parameter correlations.\label{tab:lineq}}
\tablewidth{0pt}
\tablehead{
\colhead{Dependent} & \colhead{Independent} & \colhead{$\theta$} & \colhead{$b_{\perp}$} & \colhead{$\sigma^2_{\theta}$} & \colhead{$\sigma_{\theta b_{\perp}}$} & \colhead{$\sigma^2_{b_{\perp}}$} & \colhead{Figure} \\
\colhead{parameter} & \colhead{parameter} & \colhead{} & \colhead{} & \colhead{} & \colhead{} & \colhead{} & \colhead{number}
}
\startdata
$\beta_z$ & $\gamma$ & 1.76 & -0.71 & 0.057 & -0.0044 & 0.0033 & \ref{fig:gambeta} \\
$M_{HI}/M_{tot}$ & $\gamma$ & 3.08 & -0.044 & 0.014 & 0.0098 & 0.0068 & \ref{fig:gamHI} \\
$f_{DM}(R<2R_d)$ & $\gamma$ & 0.72 &  0.19 & 0.26 & -0.29 & 0.32 & \ref{fig:gamHI} \\
$R_s$ & $R_b$ & 1.69 & -1.95 & 0.051 & -0.058 & 0.078 & \ref{fig:szsz} \\
$[$MgFe'$]$ & $\gamma$ & 1.60 & -0.73 & 0.042 & -0.065 & 0.102 & \ref{fig:lickrad} \\
Mg$_{\textrm{b}}$/$\langle$Fe$\rangle$ & $\gamma$ & 1.47 & -0.53 & 0.010 & -0.016 & 0.029 & \ref{fig:lickrad}
\enddata
\end{deluxetable*}

\subsection{Lick indices and stellar populations}
\label{sec_lick}
\par Our data cover a set of Lick indices that are useful as basic age/metallicity/$\alpha$-abundance indicators 
in old stellar populations. The recent star formation in these galaxies complicates the straightforward application 
of simple stellar population models, but we present a simple method to isolate the older 
stellar populations in our spectra. We then employ a well-known algorithm that inverts indices into stellar population parameters. 
The absolute values and gradients of such population parameters may provide empirical clues into the physics shaping the dark matter halos. 

\par We have tried using the \texttt{EZ\_AGES} software \citep{Grav08} to invert line indices into stellar population parameters. 
Also provided in this package is \texttt{LICK\_EW}, which makes standardized EW measurements on the Lick index 
system for any observations taken with higher resolution than the original Lick system. We cover the indices 
H$\beta$ (although not the entire blue Lick sideband; see Appendix \ref{app:Lick}), Fe5015, Mg$_1$, Mg$_2$, Mg$_b$ (which is encompassed in the Mg$_2$ bandpass and therefore not independent), Fe5270, Fe5335, and Fe5406. Often, a more robust combination is 
quoted as $\langle$Fe$\rangle$, which is the average of Fe5270 and Fe5335. At the heart of \texttt{EZ\_AGES} 
are the stellar population models from \cite{Schi07}. The inversion is not simple because at low resolution there 
are no lines that sample a single element. One simple approximation is to use the combined indices of 
\cite{Thom03} of [MgFe']=$\sqrt{Mg_{\textrm{b}}\times(0.72\textrm{Fe}5270+0.28\textrm{Fe}5335)}$ as a [Mg/Fe]-independent index and 
Mg$_{\textrm{b}}$/$\langle$Fe$\rangle$ as a [Fe/H]-independent index. 
\par For comparing galaxies of different ages, it is better to use 
the \texttt{EZ\_AGES} solution. Our efforts to isolate the indices from old stellar populations and apply \texttt{EZ\_AGES} 
are documented in Appendix \ref{app:Lick}. We cannot find corrections that have valid solutions within \texttt{EZ\_AGES}. 
The indices, particularly H$\beta$ and $\langle$Fe$\rangle$, fall lower than all model grids. Instead, we simply report the 
two \cite{Thom03} indices, with correction factors to isolate the old stellar populations, and warn the reader that 
the measurements may yet be affected by younger star formation.
\par
\begin{figure}
\centering
\includegraphics [scale=0.85]{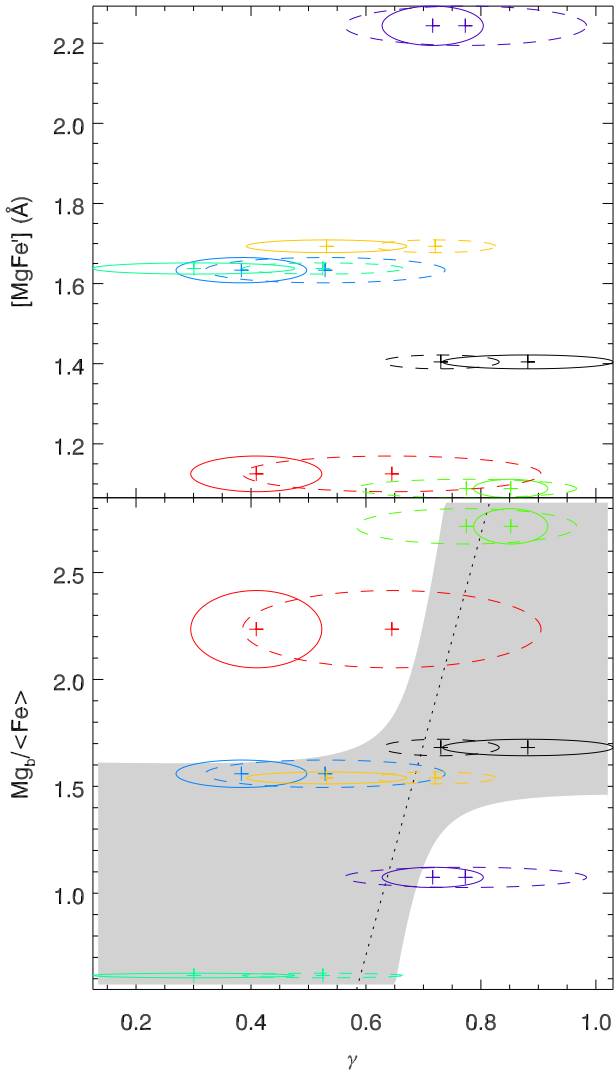}
\caption{The DM density logarithmic slopes compared to two linear combinations of 
corrected equivalent widths. The 
values have been corrected, by a ratio of template values, to isolate the old (3--13 Gyr) populations 
present for which Lick indices are commonly used. The Lick indices still 
fall off the \texttt{EZ\_AGES} grids after correction, but not by as much. These measurements may help validate or 
falsify recent SN feedback models, but model 
predictions for abundances do not yet exist. Stellar-based 
values are dotted and gas-based values use solid lines. The maximum likelihood
linear model fit of the stellar-based values is shown. The 
confidence interval for [MgFe'] covers nearly the entire plotted range and is not shown. 
The confidence interval for Mg$_{\textrm{b}}$/$\langle$Fe$\rangle$ is shown in gray. This is the opposite trend we expect for a 
SN feedback model. However, rigorous predictions by the modeling groups must be compared to our results 
before this becomes strong evidence. Further observations are also warranted to reach better statistics. Mg$_{\textrm{b}}$/$\langle$Fe$\rangle$ 
may also be affected by uncorrected age effects since the \texttt{EZ\_AGES} models cannot match the data. The
color coding per galaxy is the same as in Figure \ref{fig:gam2}. The maximum likelihood
linear fit to the stellar-based values is shown. Stellar-based values are dotted and gas-based values use solid lines.}
\label{fig:lickrad}
\end{figure}

 The correlations of [MgFe'] and Mg$_{\textrm{b}}$/$\langle$Fe$\rangle$ with $\gamma$ are shown in Figure \ref{fig:lickrad}. 
Neither shows strong correlation with $\gamma$, but a weak correlation for Mg$_{\textrm{b}}$/$\langle$Fe$\rangle$ does 
exist for $\gamma$ based on the stellar models. This is opposite to the trend we would expect from SN feedback. 
No correlation exists for the gas-measured values of $\gamma$. We present this result with the caveat that since 
\texttt{EZ\_AGES} models cannot fit our data, the Mg$_{\textrm{b}}$/$\langle$Fe$\rangle$ may also have contamination from unremoved age effects.  
\par We find one weak correlation of $\gamma$ with chemical structure, but the trend is in an 
unexpected direction and sensitive to the source of our $\gamma$ measurement. 
We also do not know how strong the relation between $\gamma$ and chemical abundance patterns 
is predicted to be from SN feedback models and whether are measurements are unexpected. 
In the case of a single starburst creating 
the DM cores, one may expect the most strongly cored galaxies to also be 
highest in Mg$_{\textrm{b}}$/$\langle$Fe$\rangle$. 
However, since the SN feedback models predict repeated bursts of star 
formation, it may be that the multiple generations of starbursts will wash out the $\alpha$-enhancement 
from any one short starburst. The answer will depend on how important the last burst is to the luminosity 
and abundance buildup. We advocate that quantitative abundance predictions be extracted 
from SN feedback simulations in future works. 

\section{Conclusions}
\label{sec:conc}
\par We have measured the kinematics in seven late-type dwarf galaxies with a 
wide-field, optical, integral field spectrograph. The 
gas radial velocities were fit with tilted ring models. The functional 
forms we explored were one with radial motion and a second for a bar-like 
potential, We have also measured 
radial velocities and dispersions for the stars and fit axisymmetric, anisotropic 
Jeans models to the stellar kinematics. To fully sample the large dimensional parameter 
spaces, we have made Bayesian parameter estimates with a Markov Chain Monte Carlo program. 
\par We find that generalized Navarro-Frenk-White (gNFW) density profiles 
adequately describe the 
kinematic fields. Burkert profiles can also fit most of the sample adequately. The 
values of the central logarithmic slope of the DM density profile, $\gamma$, we obtain 
always lie below the value for a pure NFW fit, 
although not significantly so in some cases. The stellar 
kinematics likely have different systematics than the gas, and may perhaps be less biased to 
unmodeled effects such as non-axisymmetric structure. However, we find that the bulk gNFW 
parameters between the gas-traced and stellar-traced models are usually in good 
agreement given the uncertainties. This is particularly true for $\gamma$. NGC~2976 
has been known to permit an alternate gas-traced solution if a bar is present \citep{Spek07}. 
NGC~2976 indeed looks most cored when the gas velocity field is fit with a 
radial component and more cuspy when either fit with the gas velocity field plus a bar 
component or fit with the stellar velocity field. The two patches of offset, 
peaked star formation in NGC~2976 may also signpost a weak bar. 
NGC~5949 shows 1-$\sigma$ evidence for a cuspier stellar-traced fit, but the 
two forms of gas-traced fits do not disagree in this case. The remaining five galaxies show 
no kinematic evidence for bars or biased values of $\gamma$. Even for the two 
galaxies with a potentially biased $\gamma$, the amplitude of the bias is too small to reconcile the 
results with a fully cuspy profile. The mean and standard deviation for the 
sample from the stellar fits are $\gamma=0.67\pm0.10$.  
\par Finally, we have searched for parameter correlations with $\gamma$ in the hopes of 
bolstering or rejecting some of the theoretical mechanisms that may core out DM halos. 
We find weak and likely unimportant correlations of $\gamma$ with vertical orbital anisotropy and 
DM fraction within 2 disk scale lengths. 
We measure $\Delta_{V/2}$, a previously used metric for the DM overdensity within 
the half-maximum velocity region, and we find that $\Delta_{V/2}$ is in line or even 
higher than recent 
N-body simulation predictions for cuspy halos. For our galaxies, 
the core-cusp problem manifests in $\gamma$ but not in $\Delta_{V/2}$. Since these two 
values are sensitive to masses at somewhat different scales, this result may limit the 
radial range to which DM mass is getting redistributed. We have looked for a correlation 
between the core sizes and the stellar disk sizes and found none. Some 
recent self-interacting DM simulations have predicted such a relation. Finally, we have 
used the available Lick indices to search for systematic differences in Fe or $\alpha$-abundance. 
Within the large observational uncertainties, we find no correlations of [MgFe'] with $\gamma$ 
but a weak correlation with a Lick index combination sensitive to $\alpha$-abundance. This 
is contrary to our expectations for SN feedback models with bursty 
star formation histories.
\par This work builds upon an earlier study where one galaxy was found to give 
significantly different measurements of $\gamma$ when measured in gas and stars. Our 
expanded sample size and analysis shows that some biases in traditional gas fits 
do exist, and that the biases can be explained by bar-like non-axisymmetric structure. However, 
the amplitude of the biases, when present, appear to be too small to be 
primarily responsible for the core-cusp problem. While a larger sample is still 
desirable and is being pursued by members of our group, these results suggest that 
some mechanism is indeed altering DM halos beyond the physics present in the 
ordinary $\Lambda$CDM N-body simulations. 
We have searched for correlations to explicate such a mechanism, 
and none are yet promising as directions for further research.

\acknowledgments
We thank the engineering staff of the McDonald Observatory for their work, particularly 
David Doss, Kevin Meyer, and John Kuehne. Thanks also to Chris Burns and Andrew Benson for help in 
using Carnegie's eero computing cluster. An
anonymous referee's comments substantially improved this paper's
content and presentation. 
This research has made use of the NASA/IPAC Extragalactic Database (NED) which is operated by the 
Jet Propulsion Laboratory, California Institute of Technology, under contract with the National 
Aeronautics and Space Administration, the SIMBAD database, operated at CDS, Strasbourg, France, 
NASA's Astrophysics Data System Bibliographic Services, and the HyperLeda database. Also, this 
research made use of SDSS. Funding for the Sloan Digital Sky Survey (SDSS) has been 
provided by the Alfred P. Sloan Foundation, the Participating Institutions, the 
National Aeronautics and Space Administration, the National Science Foundation, the 
U.S. Department of Energy, the Japanese Monbukagakusho, and the Max Planck Society. 
The SDSS Web site is http://www.sdss.org/. The SDSS is managed by the Astrophysical 
Research Consortium (ARC) for the Participating Institutions. The Participating 
Institutions are The University of Chicago, Fermilab, the Institute for Advanced 
Study, the Japan Participation Group, The Johns Hopkins University, the Korean 
Scientist Group, Los Alamos National Laboratory, the Max-Planck-Institute for 
Astronomy (MPIA), the Max-Planck-Institute for Astrophysics (MPA), New Mexico 
State University, University of Pittsburgh, University of Portsmouth, 
Princeton University, the United States Naval Observatory, and the University 
of Washington. 

{\it Facilities:} \facility{Smith (VIRUS-W)}.
\bibliography{dwarf_DM}

\begin{thebibliography}{187}
\expandafter\ifx\csname natexlab\endcsname\relax\def\natexlab#1{#1}\fi

\bibitem[{{Abazajian} {et~al.}(2001){Abazajian}, {Fuller}, \&
  {Patel}}]{Abaza01}
{Abazajian}, K., {Fuller}, G.~M., \& {Patel}, M. 2001, \prd, 64, 023501

\bibitem[{{Adams} {et~al.}(2011){Adams}, {Blanc}, {Hill}, {et~al.}}]{Adam11}
{Adams}, J.~J., {Blanc}, G.~A., {Hill}, G.~J., {et~al.} 2011, \apjs, 192, 5

\bibitem[{{Adams} {et~al.}(2012){Adams}, {Gebhardt}, {Blanc},
  {et~al.}}]{Adams12}
{Adams}, J.~J., {Gebhardt}, K., {Blanc}, G.~A., {et~al.} 2012, \apj, 745, 92

\bibitem[{{Alam} {et~al.}(2002){Alam}, {Bullock}, \& {Weinberg}}]{Alam02}
{Alam}, S.~M.~K., {Bullock}, J.~S., \& {Weinberg}, D.~H. 2002, \apj, 572, 34

\bibitem[{{Amorisco} {et~al.}(2013{\natexlab{a}}){Amorisco}, {Agnello}, \&
  {Evans}}]{Amori13}
{Amorisco}, N.~C., {Agnello}, A., \& {Evans}, N.~W. 2013{\natexlab{a}}, \mnras,
  429, L89

\bibitem[{{Amorisco} \& {Evans}(2012)}]{Amori12}
{Amorisco}, N.~C., \& {Evans}, N.~W. 2012, \mnras, 419, 184

\bibitem[{{Amorisco} {et~al.}(2013{\natexlab{b}}){Amorisco}, {Zavala}, \& {de
  Boer}}]{Amori13b}
{Amorisco}, N.~C., {Zavala}, J., \& {de Boer}, T.~J.~L. 2013{\natexlab{b}},
  ArXiv e-prints

\bibitem[{{Avila-Reese} {et~al.}(2001){Avila-Reese}, {Col{\'{\i}}n},
  {Valenzuela}, {D'Onghia}, \& {Firmani}}]{Avila01}
{Avila-Reese}, V., {Col{\'{\i}}n}, P., {Valenzuela}, O., {D'Onghia}, E., \&
  {Firmani}, C. 2001, \apj, 559, 516

\bibitem[{{Bagetakos} {et~al.}(2011){Bagetakos}, {Brinks}, {Walter},
  {et~al.}}]{Baget11}
{Bagetakos}, I., {Brinks}, E., {Walter}, F., {et~al.} 2011, \aj, 141, 23

\bibitem[{{Barnab{\`e}} {et~al.}(2012){Barnab{\`e}}, {Dutton}, {Marshall},
  {et~al.}}]{Barn12}
{Barnab{\`e}}, M., {Dutton}, A.~A., {Marshall}, P.~J., {et~al.} 2012, \mnras,
  423, 1073

\bibitem[{{Behroozi} {et~al.}(2013){Behroozi}, {Wechsler}, \&
  {Conroy}}]{Behr13}
{Behroozi}, P.~S., {Wechsler}, R.~H., \& {Conroy}, C. 2013, \apj, 770, 57

\bibitem[{{Bell} \& {de Jong}(2001)}]{Bell01}
{Bell}, E.~F., \& {de Jong}, R.~S. 2001, \apj, 550, 212

\bibitem[{{Bershady} {et~al.}(2011){Bershady}, {Martinsson}, {Verheijen},
  {Westfall}, {Andersen}, \& {Swaters}}]{Bers11}
{Bershady}, M.~A., {Martinsson}, T.~P.~K., {Verheijen}, M.~A.~W., {Westfall},
  K.~B., {Andersen}, D.~R., \& {Swaters}, R.~A. 2011, \apjl, 739, L47+

\bibitem[{{Bershady} {et~al.}(2010{\natexlab{a}}){Bershady}, {Verheijen},
  {Swaters}, {et~al.}}]{Bers10a}
{Bershady}, M.~A., {Verheijen}, M.~A.~W., {Swaters}, R.~A., {et~al.}
  2010{\natexlab{a}}, \apj, 716, 198

\bibitem[{{Bershady} {et~al.}(2010{\natexlab{b}}){Bershady}, {Verheijen},
  {Westfall}, {et~al.}}]{Bers10b}
{Bershady}, M.~A., {Verheijen}, M.~A.~W., {Westfall}, K.~B., {et~al.}
  2010{\natexlab{b}}, \apj, 716, 234

\bibitem[{{Bertin} \& {Arnouts}(1996)}]{Berti96}
{Bertin}, E., \& {Arnouts}, S. 1996, \aaps, 117, 393

\bibitem[{{Bertin} {et~al.}(2002){Bertin}, {Mellier}, {Radovich},
  {et~al.}}]{Berti02}
{Bertin}, E., {Mellier}, Y., {Radovich}, M., {et~al.} 2002, in Astronomical
  Society of the Pacific Conference Series, Vol. 281, Astronomical Data
  Analysis Software and Systems XI, ed. D.~A. {Bohlender}, D.~{Durand}, \&
  T.~H. {Handley}, 228

\bibitem[{{Binney} \& {Tremaine}(2008)}]{Binn08}
{Binney}, J., \& {Tremaine}, S. 2008, {Galactic Dynamics: Second Edition}
  (Princeton University Press)

\bibitem[{{Blais-Ouellette} {et~al.}(2001){Blais-Ouellette}, {Amram}, \&
  {Carignan}}]{Blais01}
{Blais-Ouellette}, S., {Amram}, P., \& {Carignan}, C. 2001, \aj, 121, 1952

\bibitem[{{Blais-Ouellette} {et~al.}(2004){Blais-Ouellette}, {Amram},
  {Carignan}, \& {Swaters}}]{Blais04}
{Blais-Ouellette}, S., {Amram}, P., {Carignan}, C., \& {Swaters}, R. 2004,
  \aap, 420, 147

\bibitem[{{Blanton} \& {Roweis}(2007)}]{Blant07}
{Blanton}, M.~R., \& {Roweis}, S. 2007, \aj, 133, 734

\bibitem[{{Blumenthal} {et~al.}(1986){Blumenthal}, {Faber}, {Flores}, \&
  {Primack}}]{Blum86}
{Blumenthal}, G.~R., {Faber}, S.~M., {Flores}, R., \& {Primack}, J.~R. 1986,
  \apj, 301, 27

\bibitem[{{Bosma}(1978)}]{Bosma78}
{Bosma}, A. 1978, PhD thesis, PhD Thesis, Groningen Univ., (1978)

\bibitem[{{Bosma}(1981{\natexlab{a}})}]{Bosma81a}
---. 1981{\natexlab{a}}, \aj, 86, 1791

\bibitem[{{Bosma}(1981{\natexlab{b}})}]{Bosma81b}
---. 1981{\natexlab{b}}, \aj, 86, 1825

\bibitem[{{Breddels} {et~al.}(2013){Breddels}, {Helmi}, {van den Bosch},
  {et~al.}}]{Bredd13}
{Breddels}, M.~A., {Helmi}, A., {van den Bosch}, R.~C.~E., {et~al.} 2013,
  \mnras, 433, 3173

\bibitem[{{Bressan} {et~al.}(2012){Bressan}, {Marigo}, {Girardi},
  {et~al.}}]{Bres12}
{Bressan}, A., {Marigo}, P., {Girardi}, L., {et~al.} 2012, \mnras, 427, 127

\bibitem[{{Bruzual} \& {Charlot}(2003)}]{BC03}
{Bruzual}, G., \& {Charlot}, S. 2003, \mnras, 344, 1000

\bibitem[{{Burkert}(1995)}]{Burke95}
{Burkert}, A. 1995, \apjl, 447, L25

\bibitem[{{Cappellari}(2002)}]{Capp02}
{Cappellari}, M. 2002, \mnras, 333, 400

\bibitem[{{Cappellari}(2008)}]{Capp08}
---. 2008, \mnras, 390, 71

\bibitem[{{Cappellari} \& {Copin}(2003)}]{Capp03}
{Cappellari}, M., \& {Copin}, Y. 2003, \mnras, 342, 345

\bibitem[{{Cappellari} {et~al.}(2007){Cappellari}, {Emsellem}, {Bacon},
  {et~al.}}]{Capp07}
{Cappellari}, M., {Emsellem}, E., {Bacon}, R., {et~al.} 2007, \mnras, 379, 418

\bibitem[{{Carignan} {et~al.}(1988){Carignan}, {Sancisi}, \& {van
  Albada}}]{Cari88}
{Carignan}, C., {Sancisi}, R., \& {van Albada}, T.~S. 1988, \aj, 95, 37

\bibitem[{{Cembranos} {et~al.}(2005){Cembranos}, {Feng}, {Rajaraman}, \&
  {Takayama}}]{Cembr05}
{Cembranos}, J.~A.~R., {Feng}, J.~L., {Rajaraman}, A., \& {Takayama}, F. 2005,
  Physical Review Letters, 95, 181301

\bibitem[{{Chabrier}(2003)}]{Chab03}
{Chabrier}, G. 2003, \pasp, 115, 763

\bibitem[{{Colucci} {et~al.}(2012){Colucci}, {Bernstein}, {Cameron}, \&
  {McWilliam}}]{Coluc12}
{Colucci}, J.~E., {Bernstein}, R.~A., {Cameron}, S.~A., \& {McWilliam}, A.
  2012, \apj, 746, 29

\bibitem[{{Dalcanton} \& {Stilp}(2010)}]{Dalc10}
{Dalcanton}, J.~J., \& {Stilp}, A.~M. 2010, \apj, 721, 547

\bibitem[{{de Blok} \& {Bosma}(2002)}]{deBl02}
{de Blok}, W.~J.~G., \& {Bosma}, A. 2002, \aap, 385, 816

\bibitem[{{de Blok} {et~al.}(2003){de Blok}, {Bosma}, \& {McGaugh}}]{deBl03}
{de Blok}, W.~J.~G., {Bosma}, A., \& {McGaugh}, S. 2003, \mnras, 340, 657

\bibitem[{{de Blok} \& {McGaugh}(1997)}]{deBl97}
{de Blok}, W.~J.~G., \& {McGaugh}, S.~S. 1997, \mnras, 290, 533

\bibitem[{{de Blok} {et~al.}(2001{\natexlab{a}}){de Blok}, {McGaugh}, {Bosma},
  \& {Rubin}}]{deBl01}
{de Blok}, W.~J.~G., {McGaugh}, S.~S., {Bosma}, A., \& {Rubin}, V.~C.
  2001{\natexlab{a}}, \apjl, 552, L23

\bibitem[{{de Blok} {et~al.}(2001{\natexlab{b}}){de Blok}, {McGaugh}, \&
  {Rubin}}]{deBl01b}
{de Blok}, W.~J.~G., {McGaugh}, S.~S., \& {Rubin}, V.~C. 2001{\natexlab{b}},
  \aj, 122, 2396

\bibitem[{{de Blok} {et~al.}(1996){de Blok}, {McGaugh}, \& {van der
  Hulst}}]{deBl96}
{de Blok}, W.~J.~G., {McGaugh}, S.~S., \& {van der Hulst}, J.~M. 1996, \mnras,
  283, 18

\bibitem[{{de Blok} {et~al.}(2008){de Blok}, {Walter}, {Brinks},
  {Trachternach}, {Oh}, \& {Kennicutt}}]{deBl08}
{de Blok}, W.~J.~G., {Walter}, F., {Brinks}, E., {Trachternach}, C., {Oh}, S.,
  \& {Kennicutt}, R.~C. 2008, \aj, 136, 2648

\bibitem[{{de Vaucouleurs} {et~al.}(1991){de Vaucouleurs}, {de Vaucouleurs},
  {Corwin}, {et~al.}}]{deVa91}
{de Vaucouleurs}, G., {de Vaucouleurs}, A., {Corwin}, Jr., H.~G., {et~al.}
  1991, {Third Reference Catalogue of Bright Galaxies}

\bibitem[{{Di Cintio} {et~al.}(2014){Di Cintio}, {Brook}, {Macci{\`o}},
  {et~al.}}]{DiCin14}
{Di Cintio}, A., {Brook}, C.~B., {Macci{\`o}}, A.~V., {et~al.} 2014, \mnras,
  437, 415

\bibitem[{{Dierckx}(1993)}]{Die93}
{Dierckx}, P. 1993, {Curve and surface fitting with splines} (Monographs on
  Numerical Analysis, Oxford: Clarendon, |c1993)

\bibitem[{{Donato} {et~al.}(2009){Donato}, {Gentile}, {Salucci},
  {et~al.}}]{Donat09}
{Donato}, F., {Gentile}, G., {Salucci}, P., {et~al.} 2009, \mnras, 397, 1169

\bibitem[{{Dubinski} {et~al.}(2009){Dubinski}, {Berentzen}, \&
  {Shlosman}}]{Dubin09}
{Dubinski}, J., {Berentzen}, I., \& {Shlosman}, I. 2009, \apj, 697, 293

\bibitem[{{Dutton} {et~al.}(2005){Dutton}, {Courteau}, {de Jong}, \&
  {Carignan}}]{Dutt05}
{Dutton}, A.~A., {Courteau}, S., {de Jong}, R., \& {Carignan}, C. 2005, \apj,
  619, 218

\bibitem[{{Dutton} \& {van den Bosch}(2009)}]{Dutt09}
{Dutton}, A.~A., \& {van den Bosch}, F.~C. 2009, \mnras, 396, 141

\bibitem[{{Dutton} {et~al.}(2007){Dutton}, {van den Bosch}, {Dekel}, \&
  {Courteau}}]{Dutt07}
{Dutton}, A.~A., {van den Bosch}, F.~C., {Dekel}, A., \& {Courteau}, S. 2007,
  \apj, 654, 27

\bibitem[{{Einasto}(1965)}]{Einas65}
{Einasto}, J. 1965, Trudy Astrofizicheskogo Instituta Alma-Ata, 5, 87

\bibitem[{{El-Zant} {et~al.}(2001){El-Zant}, {Shlosman}, \& {Hoffman}}]{ElZa01}
{El-Zant}, A., {Shlosman}, I., \& {Hoffman}, Y. 2001, \apj, 560, 636

\bibitem[{{Emsellem} {et~al.}(1994){Emsellem}, {Monnet}, \& {Bacon}}]{Emse94}
{Emsellem}, E., {Monnet}, G., \& {Bacon}, R. 1994, \aap, 285, 723

\bibitem[{{Fabricius} {et~al.}(2008){Fabricius}, {Barnes}, {Bender},
  {et~al.}}]{Fabr08}
{Fabricius}, M.~H., {Barnes}, S., {Bender}, R., {et~al.} 2008, in Presented at
  the Society of Photo-Optical Instrumentation Engineers (SPIE) Conference,
  Vol. 7014, Society of Photo-Optical Instrumentation Engineers (SPIE)
  Conference Series

\bibitem[{{Fabricius} {et~al.}(2012){Fabricius}, {Grupp}, {Bender},
  {et~al.}}]{Fabr12}
{Fabricius}, M.~H., {Grupp}, F., {Bender}, R., {et~al.} 2012, in Society of
  Photo-Optical Instrumentation Engineers (SPIE) Conference Series, Vol. 8446,
  Society of Photo-Optical Instrumentation Engineers (SPIE) Conference Series

\bibitem[{Favati {et~al.}(1991)Favati, Lotti, \& Romani}]{Fava91}
Favati, P., Lotti, G., \& Romani, F. 1991, ACM Trans. Math. Softw., 17, 218

\bibitem[{{Ferrero} {et~al.}(2012){Ferrero}, {Abadi}, {Navarro}, {Sales}, \&
  {Gurovich}}]{Ferre12}
{Ferrero}, I., {Abadi}, M.~G., {Navarro}, J.~F., {Sales}, L.~V., \& {Gurovich},
  S. 2012, \mnras, 425, 2817

\bibitem[{{Flores} \& {Primack}(1994)}]{Flore94}
{Flores}, R.~A., \& {Primack}, J.~R. 1994, \apjl, 427, L1

\bibitem[{{Foreman-Mackey} {et~al.}(2013){Foreman-Mackey}, {Hogg}, {Lang}, \&
  {Goodman}}]{Fore13}
{Foreman-Mackey}, D., {Hogg}, D.~W., {Lang}, D., \& {Goodman}, J. 2013, \pasp,
  125, 306

\bibitem[{{Freeman}(1970)}]{Freem70}
{Freeman}, K.~C. 1970, \apj, 160, 811

\bibitem[{{Gebhardt} {et~al.}(2000){Gebhardt}, {Richstone}, {Kormendy},
  {et~al.}}]{Gebh00}
{Gebhardt}, K., {Richstone}, D., {Kormendy}, J., {et~al.} 2000, \aj, 119, 1157

\bibitem[{{Gentile} {et~al.}(2005){Gentile}, {Burkert}, {Salucci}, {Klein}, \&
  {Walter}}]{Genti05}
{Gentile}, G., {Burkert}, A., {Salucci}, P., {Klein}, U., \& {Walter}, F. 2005,
  \apjl, 634, L145

\bibitem[{{Gentile} {et~al.}(2004){Gentile}, {Salucci}, {Klein}, {Vergani}, \&
  {Kalberla}}]{Genti04}
{Gentile}, G., {Salucci}, P., {Klein}, U., {Vergani}, D., \& {Kalberla}, P.
  2004, \mnras, 351, 903

\bibitem[{{Gerssen} \& {Shapiro Griffin}(2012)}]{Gers12}
{Gerssen}, J., \& {Shapiro Griffin}, K. 2012, \mnras, 423, 2726

\bibitem[{{Governato} {et~al.}(2010){Governato}, {Brook}, {Mayer}, {Brooks},
  {Rhee}, {Wadsley}, {Jonsson}, {Willman}, {Stinson}, {Quinn}, \&
  {Madau}}]{Gove10}
{Governato}, F., {Brook}, C., {Mayer}, L., {Brooks}, A., {Rhee}, G., {Wadsley},
  J., {Jonsson}, P., {Willman}, B., {Stinson}, G., {Quinn}, T., \& {Madau}, P.
  2010, \nat, 463, 203

\bibitem[{{Governato} {et~al.}(2012){Governato}, {Zolotov}, {Pontzen},
  {et~al.}}]{Gover12}
{Governato}, F., {Zolotov}, A., {Pontzen}, A., {et~al.} 2012, \mnras, 422, 1231

\bibitem[{{Graves} \& {Schiavon}(2008)}]{Grav08}
{Graves}, G.~J., \& {Schiavon}, R.~P. 2008, \apjs, 177, 446

\bibitem[{{Greene} {et~al.}(2012){Greene}, {Murphy}, {Comerford},
  {et~al.}}]{Gree12}
{Greene}, J.~E., {Murphy}, J.~D., {Comerford}, J.~M., {et~al.} 2012, \apj, 750,
  32

\bibitem[{{Greene} {et~al.}(2013){Greene}, {Murphy}, {Graves},
  {et~al.}}]{Gree13}
{Greene}, J.~E., {Murphy}, J.~D., {Graves}, G.~J., {et~al.} 2013, \apj, 776, 64

\bibitem[{{Guo} {et~al.}(2011){Guo}, {White}, {Boylan-Kolchin},
  {et~al.}}]{Guo11}
{Guo}, Q., {White}, S., {Boylan-Kolchin}, M., {et~al.} 2011, \mnras, 413, 101

\bibitem[{{Hernquist}(1990)}]{Hern90}
{Hernquist}, L. 1990, \apj, 356, 359

\bibitem[{{Herrmann} \& {Ciardullo}(2009)}]{Herr09}
{Herrmann}, K.~A., \& {Ciardullo}, R. 2009, \apj, 705, 1686

\bibitem[{{Hill} {et~al.}(2004){Hill}, {Gebhardt}, {Komatsu}, \&
  {MacQueen}}]{Hill04}
{Hill}, G.~J., {Gebhardt}, K., {Komatsu}, E., \& {MacQueen}, P.~J. 2004, in
  American Institute of Physics Conference Series, Vol. 743, The New Cosmology:
  Conference on Strings and Cosmology, ed. {R.~E.~Allen, D.~V.~Nanopoulos, \&
  C.~N.~Pope}, 224--233

\bibitem[{{Hill} {et~al.}(2008b){Hill}, {MacQueen}, {Smith}, {et~al.}}]{Hill08}
{Hill}, G.~J., {MacQueen}, P.~J., {Smith}, M.~P., {et~al.} 2008b, in Society of
  Photo-Optical Instrumentation Engineers (SPIE) Conference Series, Vol. 7014,
  231

\bibitem[{{Hogan} \& {Dalcanton}(2000)}]{Hoga00}
{Hogan}, C.~J., \& {Dalcanton}, J.~J. 2000, \prd, 62, 063511

\bibitem[{{Hogg} {et~al.}(2010){Hogg}, {Bovy}, \& {Lang}}]{Hogg10}
{Hogg}, D.~W., {Bovy}, J., \& {Lang}, D. 2010, ArXiv e-prints

\bibitem[{{Holley-Bockelmann} {et~al.}(2005){Holley-Bockelmann}, {Weinberg}, \&
  {Katz}}]{Holle05}
{Holley-Bockelmann}, K., {Weinberg}, M., \& {Katz}, N. 2005, \mnras, 363, 991

\bibitem[{{Jardel} \& {Gebhardt}(2012)}]{Jarde12}
{Jardel}, J.~R., \& {Gebhardt}, K. 2012, \apj, 746, 89

\bibitem[{{Jardel} {et~al.}(2013){Jardel}, {Gebhardt}, {Fabricius},
  {et~al.}}]{Jarde13}
{Jardel}, J.~R., {Gebhardt}, K., {Fabricius}, M.~H., {et~al.} 2013, \apj, 763,
  91

\bibitem[{{Johnson} {et~al.}(2006){Johnson}, {Ivans}, \& {Stetson}}]{Johns06}
{Johnson}, J.~A., {Ivans}, I.~I., \& {Stetson}, P.~B. 2006, \apj, 640, 801

\bibitem[{{Kaplinghat}(2005)}]{Kapli05}
{Kaplinghat}, M. 2005, \prd, 72, 063510

\bibitem[{{Kaplinghat} {et~al.}(2013){Kaplinghat}, {Keeley}, {Linden}, \&
  {Yu}}]{Kapli13}
{Kaplinghat}, M., {Keeley}, R.~E., {Linden}, T., \& {Yu}, H.-B. 2013, ArXiv
  e-prints

\bibitem[{{Kaplinghat} {et~al.}(2000){Kaplinghat}, {Knox}, \&
  {Turner}}]{Kapli00}
{Kaplinghat}, M., {Knox}, L., \& {Turner}, M.~S. 2000, Physical Review Letters,
  85, 3335

\bibitem[{{Katz} {et~al.}(1998){Katz}, {Soubiran}, {Cayrel}, {Adda}, \&
  {Cautain}}]{Katz98}
{Katz}, D., {Soubiran}, C., {Cayrel}, R., {Adda}, M., \& {Cautain}, R. 1998,
  \aap, 338, 151

\bibitem[{{Kelson}(2003)}]{Kels03}
{Kelson}, D.~D. 2003, \pasp, 115, 688

\bibitem[{{Klypin} {et~al.}(2001){Klypin}, {Kravtsov}, {Bullock}, \&
  {Primack}}]{Klyp01}
{Klypin}, A., {Kravtsov}, A.~V., {Bullock}, J.~S., \& {Primack}, J.~R. 2001,
  \apj, 554, 903

\bibitem[{{Klypin} {et~al.}(2011){Klypin}, {Trujillo-Gomez}, \&
  {Primack}}]{Klyp11}
{Klypin}, A.~A., {Trujillo-Gomez}, S., \& {Primack}, J. 2011, \apj, 740, 102

\bibitem[{{Komatsu} {et~al.}(2011){Komatsu}, {Smith}, {Dunkley},
  {et~al.}}]{Komat11}
{Komatsu}, E., {Smith}, K.~M., {Dunkley}, J., {et~al.} 2011, \apjs, 192, 18

\bibitem[{{Kowalczyk} {et~al.}(2013){Kowalczyk}, {{\L}okas}, {Kazantzidis}, \&
  {Mayer}}]{Kowal13}
{Kowalczyk}, K., {{\L}okas}, E.~L., {Kazantzidis}, S., \& {Mayer}, L. 2013,
  \mnras, 431, 2796

\bibitem[{{Kregel} {et~al.}(2002){Kregel}, {van der Kruit}, \& {de
  Grijs}}]{Kreg02}
{Kregel}, M., {van der Kruit}, P.~C., \& {de Grijs}, R. 2002, \mnras, 334, 646

\bibitem[{{Kuzio de Naray} {et~al.}(2010){Kuzio de Naray}, {Martinez},
  {Bullock}, \& {Kaplinghat}}]{Kuzi10}
{Kuzio de Naray}, R., {Martinez}, G.~D., {Bullock}, J.~S., \& {Kaplinghat}, M.
  2010, \apjl, 710, L161

\bibitem[{{Kuzio de Naray} {et~al.}(2008){Kuzio de Naray}, {McGaugh}, \& {de
  Blok}}]{Kuzi08}
{Kuzio de Naray}, R., {McGaugh}, S.~S., \& {de Blok}, W.~J.~G. 2008, \apj, 676,
  920

\bibitem[{{Kuzio de Naray} {et~al.}(2006){Kuzio de Naray}, {McGaugh}, {de
  Blok}, \& {Bosma}}]{Kuzi06}
{Kuzio de Naray}, R., {McGaugh}, S.~S., {de Blok}, W.~J.~G., \& {Bosma}, A.
  2006, \apjs, 165, 461

\bibitem[{{Kuzio de Naray} {et~al.}(2009){Kuzio de Naray}, {McGaugh}, \&
  {Mihos}}]{Kuzi09}
{Kuzio de Naray}, R., {McGaugh}, S.~S., \& {Mihos}, J.~C. 2009, \apj, 692, 1321

\bibitem[{{Leroy} {et~al.}(2008){Leroy}, {Walter}, {Brinks},
  {et~al.}}]{Leroy08}
{Leroy}, A.~K., {Walter}, F., {Brinks}, E., {et~al.} 2008, \aj, 136, 2782

\bibitem[{{Macci{\`o}} {et~al.}(2012{\natexlab{a}}){Macci{\`o}}, {Paduroiu},
  {Anderhalden}, {Schneider}, \& {Moore}}]{Maccio12a}
{Macci{\`o}}, A.~V., {Paduroiu}, S., {Anderhalden}, D., {Schneider}, A., \&
  {Moore}, B. 2012{\natexlab{a}}, \mnras, 424, 1105

\bibitem[{{Macci{\`o}} {et~al.}(2012{\natexlab{b}}){Macci{\`o}}, {Stinson},
  {Brook}, {et~al.}}]{Maccio12b}
{Macci{\`o}}, A.~V., {Stinson}, G., {Brook}, C.~B., {et~al.}
  2012{\natexlab{b}}, \apjl, 744, L9

\bibitem[{{Madsen} \& {Gaensler}(2013)}]{Madse13}
{Madsen}, G.~J., \& {Gaensler}, B.~M. 2013, ArXiv e-prints

\bibitem[{{Marchesini} {et~al.}(2002){Marchesini}, {D'Onghia}, {Chincarini},
  {et~al.}}]{March02}
{Marchesini}, D., {D'Onghia}, E., {Chincarini}, G., {et~al.} 2002, \apj, 575,
  801

\bibitem[{{Markwardt}(2009)}]{Mark09}
{Markwardt}, C.~B. 2009, in Astronomical Society of the Pacific Conference
  Series, Vol. 411, Astronomical Data Analysis Software and Systems XVIII, ed.
  D.~A. {Bohlender}, D.~{Durand}, \& P.~{Dowler}, 251

\bibitem[{{McGaugh} {et~al.}(2001){McGaugh}, {Rubin}, \& {de Blok}}]{McGau01}
{McGaugh}, S.~S., {Rubin}, V.~C., \& {de Blok}, W.~J.~G. 2001, \aj, 122, 2381

\bibitem[{{McMillan} \& {Dehnen}(2005)}]{McMil05}
{McMillan}, P.~J., \& {Dehnen}, W. 2005, \mnras, 363, 1205

\bibitem[{{Milgrom}(1983)}]{Milgr83}
{Milgrom}, M. 1983, \apj, 270, 365

\bibitem[{{Monet} {et~al.}(2003){Monet}, {Levine}, {Canzian},
  {et~al.}}]{Monet03}
{Monet}, D.~G., {Levine}, S.~E., {Canzian}, B., {et~al.} 2003, \aj, 125, 984

\bibitem[{{Monnet} {et~al.}(1992){Monnet}, {Bacon}, \& {Emsellem}}]{Monn92}
{Monnet}, G., {Bacon}, R., \& {Emsellem}, E. 1992, \aap, 253, 366

\bibitem[{{Moore}(1994)}]{Moore94}
{Moore}, B. 1994, \nat, 370, 629

\bibitem[{Mor{\'e}(1978)}]{More78}
Mor{\'e}, J.~J. 1978, in Lecture Notes in Mathematics, Vol. 630, Numerical
  Analysis, ed. G.~Watson (Springer Berlin Heidelberg), 105--116

\bibitem[{{Mucciarelli} {et~al.}(2008){Mucciarelli}, {Carretta}, {Origlia}, \&
  {Ferraro}}]{Mucci08}
{Mucciarelli}, A., {Carretta}, E., {Origlia}, L., \& {Ferraro}, F.~R. 2008,
  \aj, 136, 375

\bibitem[{{Navarro} {et~al.}(1996{\natexlab{a}}){Navarro}, {Eke}, \&
  {Frenk}}]{Nava96b}
{Navarro}, J.~F., {Eke}, V.~R., \& {Frenk}, C.~S. 1996{\natexlab{a}}, \mnras,
  283, L72

\bibitem[{{Navarro} {et~al.}(1996{\natexlab{b}}){Navarro}, {Frenk}, \&
  {White}}]{Nava96a}
{Navarro}, J.~F., {Frenk}, C.~S., \& {White}, S.~D.~M. 1996{\natexlab{b}},
  \apj, 462, 563

\bibitem[{{Navarro} {et~al.}(2004){Navarro}, {Hayashi}, {Power},
  {et~al.}}]{Navar04}
{Navarro}, J.~F., {Hayashi}, E., {Power}, C., {et~al.} 2004, \mnras, 349, 1039

\bibitem[{{Navarro} {et~al.}(2010){Navarro}, {Ludlow}, {Springel},
  {et~al.}}]{Navar10}
{Navarro}, J.~F., {Ludlow}, A., {Springel}, V., {et~al.} 2010, \mnras, 402, 21

\bibitem[{{Newman} {et~al.}(2013{\natexlab{a}}){Newman}, {Treu}, {Ellis}, \&
  {Sand}}]{Newman13b}
{Newman}, A.~B., {Treu}, T., {Ellis}, R.~S., \& {Sand}, D.~J.
  2013{\natexlab{a}}, \apj, 765, 25

\bibitem[{{Newman} {et~al.}(2013{\natexlab{b}}){Newman}, {Treu}, {Ellis},
  {et~al.}}]{Newman13a}
{Newman}, A.~B., {Treu}, T., {Ellis}, R.~S., {et~al.} 2013{\natexlab{b}}, \apj,
  765, 24

\bibitem[{{Noordermeer} {et~al.}(2008){Noordermeer}, {Merrifield}, \&
  {Arag{\'o}n-Salamanca}}]{Noord08}
{Noordermeer}, E., {Merrifield}, M.~R., \& {Arag{\'o}n-Salamanca}, A. 2008,
  \mnras, 388, 1381

\bibitem[{{Oh} {et~al.}(2008){Oh}, {de Blok}, {Walter}, {Brinks}, \&
  {Kennicutt}}]{Oh08}
{Oh}, S., {de Blok}, W.~J.~G., {Walter}, F., {Brinks}, E., \& {Kennicutt},
  R.~C. 2008, \aj, 136, 2761

\bibitem[{{Oh} {et~al.}(2011{\natexlab{a}}){Oh}, {Brook}, {Governato},
  {et~al.}}]{Oh11b}
{Oh}, S.-H., {Brook}, C., {Governato}, F., {et~al.} 2011{\natexlab{a}}, \aj,
  142, 24

\bibitem[{{Oh} {et~al.}(2011{\natexlab{b}}){Oh}, {de Blok}, {Brinks},
  {et~al.}}]{Oh11a}
{Oh}, S.-H., {de Blok}, W.~J.~G., {Brinks}, E., {et~al.} 2011{\natexlab{b}},
  \aj, 141, 193

\bibitem[{{Paturel} {et~al.}(2003){Paturel}, {Petit}, {Prugniel},
  {et~al.}}]{Patu03}
{Paturel}, G., {Petit}, C., {Prugniel}, P., {et~al.} 2003, \aap, 412, 45

\bibitem[{{Peter} {et~al.}(2013){Peter}, {Rocha}, {Bullock}, \&
  {Kaplinghat}}]{Peter13}
{Peter}, A.~H.~G., {Rocha}, M., {Bullock}, J.~S., \& {Kaplinghat}, M. 2013,
  \mnras, 430, 105

\bibitem[{{Pomp{\'e}ia} {et~al.}(2008){Pomp{\'e}ia}, {Hill}, {Spite},
  {et~al.}}]{Pompe08}
{Pomp{\'e}ia}, L., {Hill}, V., {Spite}, M., {et~al.} 2008, \aap, 480, 379

\bibitem[{{Pontzen} \& {Governato}(2012)}]{Pontz12}
{Pontzen}, A., \& {Governato}, F. 2012, \mnras, 421, 3464

\bibitem[{{Portinari} {et~al.}(2004){Portinari}, {Sommer-Larsen}, \&
  {Tantalo}}]{Port04}
{Portinari}, L., {Sommer-Larsen}, J., \& {Tantalo}, R. 2004, \mnras, 347, 691

\bibitem[{{Press} {et~al.}(1992){Press}, {Teukolsky}, {Vetterling}, \&
  {Flannery}}]{Press92}
{Press}, W.~H., {Teukolsky}, S.~A., {Vetterling}, W.~T., \& {Flannery}, B.~P.
  1992, {Numerical recipes in FORTRAN. The art of scientific computing}

\bibitem[{{Prugniel} \& {Soubiran}(2001)}]{Prug01}
{Prugniel}, P., \& {Soubiran}, C. 2001, \aap, 369, 1048

\bibitem[{{Rhee} {et~al.}(2004){Rhee}, {Valenzuela}, {Klypin}, {Holtzman}, \&
  {Moorthy}}]{Rhee04}
{Rhee}, G., {Valenzuela}, O., {Klypin}, A., {Holtzman}, J., \& {Moorthy}, B.
  2004, \apj, 617, 1059

\bibitem[{{Richardson} \& {Fairbairn}(2014)}]{Richa14}
{Richardson}, T., \& {Fairbairn}, M. 2014, ArXiv e-prints

\bibitem[{{Roberts} \& {Whitehurst}(1975)}]{Roberts75}
{Roberts}, M.~S., \& {Whitehurst}, R.~N. 1975, \apj, 201, 327

\bibitem[{{Rocha} {et~al.}(2013){Rocha}, {Peter}, {Bullock},
  {et~al.}}]{Rocha13}
{Rocha}, M., {Peter}, A.~H.~G., {Bullock}, J.~S., {et~al.} 2013, \mnras, 430,
  81

\bibitem[{{Rubin} \& {Ford}(1970)}]{Rubin70}
{Rubin}, V.~C., \& {Ford}, Jr., W.~K. 1970, \apj, 159, 379

\bibitem[{{Rubin} {et~al.}(1978{\natexlab{a}}){Rubin}, {Ford}, {Strom},
  {Strom}, \& {Romanishin}}]{Rubin78a}
{Rubin}, V.~C., {Ford}, Jr., W.~K., {Strom}, K.~M., {Strom}, S.~E., \&
  {Romanishin}, W. 1978{\natexlab{a}}, \apj, 224, 782

\bibitem[{{Rubin} {et~al.}(1980){Rubin}, {Ford}, \& {.~Thonnard}}]{Rubin80}
{Rubin}, V.~C., {Ford}, W.~K.~J., \& {.~Thonnard}, N. 1980, \apj, 238, 471

\bibitem[{{Rubin} {et~al.}(1978{\natexlab{b}}){Rubin}, {Thonnard}, \&
  {Ford}}]{Rubin78b}
{Rubin}, V.~C., {Thonnard}, N., \& {Ford}, Jr., W.~K. 1978{\natexlab{b}},
  \apjl, 225, L107

\bibitem[{{Salasnich} {et~al.}(2000){Salasnich}, {Girardi}, {Weiss}, \&
  {Chiosi}}]{Salas00}
{Salasnich}, B., {Girardi}, L., {Weiss}, A., \& {Chiosi}, C. 2000, \aap, 361,
  1023

\bibitem[{{Sand} {et~al.}(2002){Sand}, {Treu}, \& {Ellis}}]{Sand02}
{Sand}, D.~J., {Treu}, T., \& {Ellis}, R.~S. 2002, \apjl, 574, L129

\bibitem[{{Sand} {et~al.}(2004){Sand}, {Treu}, {Smith}, \& {Ellis}}]{Sand04}
{Sand}, D.~J., {Treu}, T., {Smith}, G.~P., \& {Ellis}, R.~S. 2004, \apj, 604,
  88

\bibitem[{{Schiavon}(2007)}]{Schi07}
{Schiavon}, R.~P. 2007, \apjs, 171, 146

\bibitem[{{Schoenmakers} {et~al.}(1997){Schoenmakers}, {Franx}, \& {de
  Zeeuw}}]{Shoen97}
{Schoenmakers}, R.~H.~M., {Franx}, M., \& {de Zeeuw}, P.~T. 1997, \mnras, 292,
  349

\bibitem[{{Sellwood}(2003)}]{Sellw03}
{Sellwood}, J.~A. 2003, \apj, 587, 638

\bibitem[{{Sellwood}(2008)}]{Sellw08}
---. 2008, \apj, 679, 379

\bibitem[{{Sellwood} \& {S{\'a}nchez}(2010)}]{Sell10}
{Sellwood}, J.~A., \& {S{\'a}nchez}, R.~Z. 2010, \mnras, 404, 1733

\bibitem[{{Shapiro} {et~al.}(2003){Shapiro}, {Gerssen}, \& {van der
  Marel}}]{Shap03}
{Shapiro}, K.~L., {Gerssen}, J., \& {van der Marel}, R.~P. 2003, \aj, 126, 2707

\bibitem[{{Simon} {et~al.}(2003){Simon}, {Bolatto}, {Leroy}, \&
  {Blitz}}]{Simo03}
{Simon}, J.~D., {Bolatto}, A.~D., {Leroy}, A., \& {Blitz}, L. 2003, \apj, 596,
  957

\bibitem[{{Simon} {et~al.}(2005){Simon}, {Bolatto}, {Leroy}, {Blitz}, \&
  {Gates}}]{Simo05}
{Simon}, J.~D., {Bolatto}, A.~D., {Leroy}, A., {Blitz}, L., \& {Gates}, E.~L.
  2005, \apj, 621, 757

\bibitem[{{Spano} {et~al.}(2008){Spano}, {Marcelin}, {Amram}, {Carignan},
  {Epinat}, \& {Hernandez}}]{Span08}
{Spano}, M., {Marcelin}, M., {Amram}, P., {Carignan}, C., {Epinat}, B., \&
  {Hernandez}, O. 2008, \mnras, 383, 297

\bibitem[{{Spekkens} {et~al.}(2005){Spekkens}, {Giovanelli}, \&
  {Haynes}}]{Spek05}
{Spekkens}, K., {Giovanelli}, R., \& {Haynes}, M.~P. 2005, \aj, 129, 2119

\bibitem[{{Spekkens} \& {Sellwood}(2007)}]{Spek07}
{Spekkens}, K., \& {Sellwood}, J.~A. 2007, \apj, 664, 204

\bibitem[{{Spergel} \& {Steinhardt}(2000)}]{Sperg00}
{Spergel}, D.~N., \& {Steinhardt}, P.~J. 2000, Physical Review Letters, 84,
  3760

\bibitem[{{Stil} \& {Israel}(2002)}]{Stil02a}
{Stil}, J.~M., \& {Israel}, F.~P. 2002, \aap, 389, 29

\bibitem[{{Strigari} {et~al.}(2007){Strigari}, {Kaplinghat}, \&
  {Bullock}}]{Strig07}
{Strigari}, L.~E., {Kaplinghat}, M., \& {Bullock}, J.~S. 2007, \prd, 75, 061303

\bibitem[{{Swaters} \& {Balcells}(2002)}]{Swat02}
{Swaters}, R.~A., \& {Balcells}, M. 2002, \aap, 390, 863

\bibitem[{{Swaters} {et~al.}(2003{\natexlab{a}}){Swaters}, {Madore}, {van den
  Bosch}, \& {Balcells}}]{Swat03}
{Swaters}, R.~A., {Madore}, B.~F., {van den Bosch}, F.~C., \& {Balcells}, M.
  2003{\natexlab{a}}, \apj, 583, 732

\bibitem[{{Swaters} {et~al.}(2003{\natexlab{b}}){Swaters}, {Verheijen},
  {Bershady}, \& {Andersen}}]{Swat03b}
{Swaters}, R.~A., {Verheijen}, M.~A.~W., {Bershady}, M.~A., \& {Andersen},
  D.~R. 2003{\natexlab{b}}, \apjl, 587, L19

\bibitem[{{Taylor} {et~al.}(2005){Taylor}, {Jansen}, {Windhorst}, {Odewahn}, \&
  {Hibbard}}]{Taylo05}
{Taylor}, V.~A., {Jansen}, R.~A., {Windhorst}, R.~A., {Odewahn}, S.~C., \&
  {Hibbard}, J.~E. 2005, \apj, 630, 784

\bibitem[{{Thomas} {et~al.}(2003){Thomas}, {Maraston}, \& {Bender}}]{Thom03}
{Thomas}, D., {Maraston}, C., \& {Bender}, R. 2003, \mnras, 343, 279

\bibitem[{{Thomas} {et~al.}(2005){Thomas}, {Maraston}, {Bender}, \& {Mendes de
  Oliveira}}]{Thom05}
{Thomas}, D., {Maraston}, C., {Bender}, R., \& {Mendes de Oliveira}, C. 2005,
  \apj, 621, 673

\bibitem[{{Thomas} {et~al.}(2011){Thomas}, {Saglia}, {Bender},
  {et~al.}}]{Thoma11}
{Thomas}, J., {Saglia}, R.~P., {Bender}, R., {et~al.} 2011, \mnras, 415, 545

\bibitem[{{Tody}(1986)}]{Tody86}
{Tody}, D. 1986, in Society of Photo-Optical Instrumentation Engineers (SPIE)
  Conference Series, Vol. 627, Society of Photo-Optical Instrumentation
  Engineers (SPIE) Conference Series, ed. D.~L. {Crawford}, 733

\bibitem[{{Tonini} {et~al.}(2006){Tonini}, {Lapi}, \& {Salucci}}]{Tonini06}
{Tonini}, C., {Lapi}, A., \& {Salucci}, P. 2006, \apj, 649, 591

\bibitem[{{Trujillo-Gomez} {et~al.}(2013){Trujillo-Gomez}, {Klypin}, {Colin},
  {et~al.}}]{Truji14}
{Trujillo-Gomez}, S., {Klypin}, A., {Colin}, P., {et~al.} 2013, ArXiv e-prints

\bibitem[{{Tully} {et~al.}(2009){Tully}, {Rizzi}, {Shaya}, {et~al.}}]{Tully09}
{Tully}, R.~B., {Rizzi}, L., {Shaya}, E.~J., {et~al.} 2009, \aj, 138, 323

\bibitem[{{Valenzuela} {et~al.}(2014){Valenzuela}, {Hernandez-Toledo}, {Cano},
  {et~al.}}]{Valen14}
{Valenzuela}, O., {Hernandez-Toledo}, H., {Cano}, M., {et~al.} 2014, \aj, 147,
  27

\bibitem[{{Valenzuela} {et~al.}(2007){Valenzuela}, {Rhee}, {Klypin},
  {et~al.}}]{Valen07}
{Valenzuela}, O., {Rhee}, G., {Klypin}, A., {et~al.} 2007, \apj, 657, 773

\bibitem[{{van Albada} \& {Sancisi}(1986)}]{vanA86}
{van Albada}, T.~S., \& {Sancisi}, R. 1986, Royal Society of London
  Philosophical Transactions Series A, 320, 447

\bibitem[{{van den Bosch} \& {Swaters}(2001)}]{vdB01}
{van den Bosch}, F.~C., \& {Swaters}, R.~A. 2001, \mnras, 325, 1017

\bibitem[{{van den Bosch} {et~al.}(2008){van den Bosch}, {van de Ven},
  {Verolme}, {Cappellari}, \& {de Zeeuw}}]{vdBo08}
{van den Bosch}, R.~C.~E., {van de Ven}, G., {Verolme}, E.~K., {Cappellari},
  M., \& {de Zeeuw}, P.~T. 2008, \mnras, 385, 647

\bibitem[{{van der Kruit} \& {de Grijs}(1999)}]{vdKr99}
{van der Kruit}, P.~C., \& {de Grijs}, R. 1999, \aap, 352, 129

\bibitem[{{van der Marel}(1994)}]{vdM94}
{van der Marel}, R.~P. 1994, \mnras, 270, 271

\bibitem[{{Vanderplaats}(1973)}]{Vand73}
{Vanderplaats}, G.~N. 1973, in NASA Technical Memo, NASA TM X-62282

\bibitem[{{Vazdekis} {et~al.}(2003){Vazdekis}, {Cenarro}, {Gorgas}, {Cardiel},
  \& {Peletier}}]{Vazd03}
{Vazdekis}, A., {Cenarro}, A.~J., {Gorgas}, J., {Cardiel}, N., \& {Peletier},
  R.~F. 2003, \mnras, 340, 1317

\bibitem[{{Walker} {et~al.}(2010){Walker}, {McGaugh}, {Mateo}, {Olszewski}, \&
  {Kuzio de Naray}}]{Walke10}
{Walker}, M.~G., {McGaugh}, S.~S., {Mateo}, M., {Olszewski}, E.~W., \& {Kuzio
  de Naray}, R. 2010, \apjl, 717, L87

\bibitem[{{Walker} \& {Pe{\~n}arrubia}(2011)}]{Walke11}
{Walker}, M.~G., \& {Pe{\~n}arrubia}, J. 2011, \apj, 742, 20

\bibitem[{{Walter} {et~al.}(2008){Walter}, {Brinks}, {de Blok}, {Bigiel},
  {Kennicutt}, {Thornley}, \& {Leroy}}]{Walt08}
{Walter}, F., {Brinks}, E., {de Blok}, W.~J.~G., {Bigiel}, F., {Kennicutt},
  R.~C., {Thornley}, M.~D., \& {Leroy}, A. 2008, \aj, 136, 2563

\bibitem[{{Watkins} {et~al.}(2013){Watkins}, {van de Ven}, {den Brok}, \& {van
  den Bosch}}]{Watki13}
{Watkins}, L.~L., {van de Ven}, G., {den Brok}, M., \& {van den Bosch},
  R.~C.~E. 2013, \mnras, 436, 2598

\bibitem[{{Weinberg} \& {Katz}(2002)}]{Weinb02}
{Weinberg}, M.~D., \& {Katz}, N. 2002, \apj, 580, 627

\bibitem[{{Weinberg} \& {Katz}(2007{\natexlab{a}})}]{Weinb07a}
---. 2007{\natexlab{a}}, \mnras, 375, 425

\bibitem[{{Weinberg} \& {Katz}(2007{\natexlab{b}})}]{Weinb07b}
---. 2007{\natexlab{b}}, \mnras, 375, 460

\bibitem[{{Westfall} {et~al.}(2011){Westfall}, {Bershady}, {Verheijen},
  {et~al.}}]{West11}
{Westfall}, K.~B., {Bershady}, M.~A., {Verheijen}, M.~A.~W., {et~al.} 2011,
  \apj, 742, 18

\bibitem[{{Williams} {et~al.}(2009){Williams}, {Bureau}, \&
  {Cappellari}}]{Will09}
{Williams}, M.~J., {Bureau}, M., \& {Cappellari}, M. 2009, \mnras, 400, 1665

\bibitem[{{Worthey} \& {Ottaviani}(1997)}]{Wort97}
{Worthey}, G., \& {Ottaviani}, D.~L. 1997, \apjs, 111, 377

\bibitem[{{Wyithe} {et~al.}(2001){Wyithe}, {Turner}, \& {Spergel}}]{Wyit01}
{Wyithe}, J.~S.~B., {Turner}, E.~L., \& {Spergel}, D.~N. 2001, \apj, 555, 504

\bibitem[{{Zacharias} {et~al.}(2005){Zacharias}, {Monet}, {Levine},
  {et~al.}}]{Zacha05}
{Zacharias}, N., {Monet}, D.~G., {Levine}, S.~E., {et~al.} 2005, VizieR Online
  Data Catalog, 1297, 0

\bibitem[{{Zhao}(1996)}]{Zhao96}
{Zhao}, H. 1996, \mnras, 278, 488

\bibitem[{{Zibetti} {et~al.}(2009){Zibetti}, {Charlot}, \& {Rix}}]{Zibe09}
{Zibetti}, S., {Charlot}, S., \& {Rix}, H.-W. 2009, \mnras, 400, 1181

\end{thebibliography}
\bibliographystyle{apj}
\appendix
\section{Composite Kinematic Templates}
\label{app:temp}
We have created composite kinematic templates from an 
empirical database of stellar spectra \citep{Prug01}. There are two reasons that recommend the 
composite method. First, the averaging of many stars will reduce the systematics from any non-representative 
abundances or unusual rotation in a single template stars. Second, the issue of template selection needs to be both 
flexible and constrained. Abundance, dust, and age gradients may exist, in which case the LOSVD ought to be allowed 
to select different weights amongst an inclusive template set for each bin. Alternatively, the LOSVD program 
ought not to have enough freedom to distribute template weights at an extreme level between neighboring bins. 
The youngest stars that contribute significantly to the integrated light in these galaxies, mostly A stars, 
have intrinsically broader spectral features than, say, K dwarfs. Allowing one bin to select only A stars and 
then allowing its neighbor to select only K stars is unphysical. A natural way to 
accommodate these requirements is to form a small number of composite templates.
\par
\begin{figure}
\centering
\includegraphics [scale=0.9,angle=0]{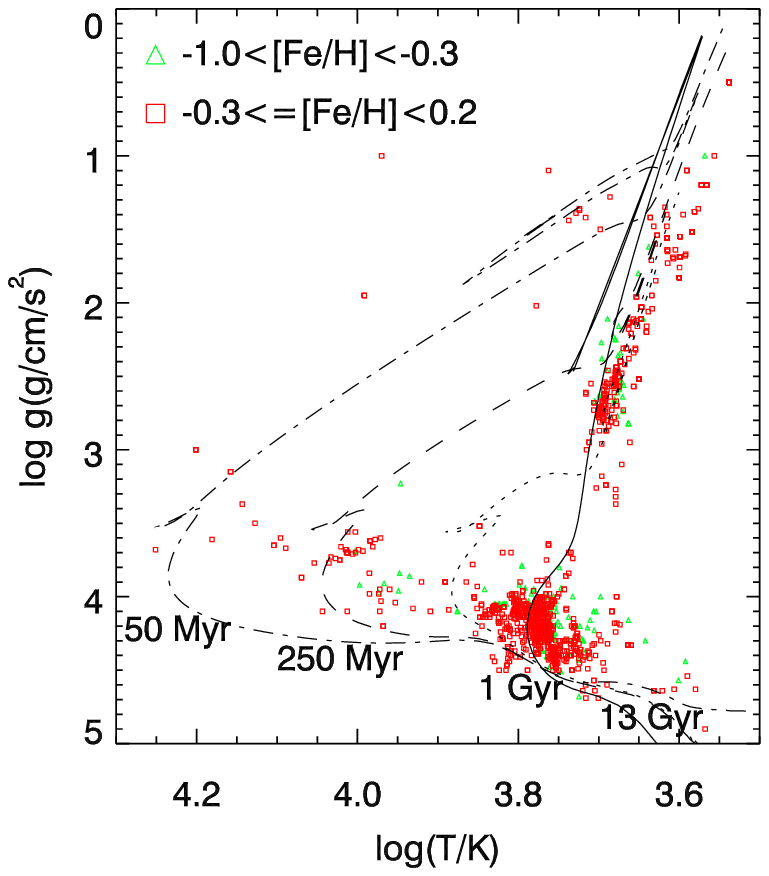}
\caption{The ELODIE stars and the \cite{Bres12} isochrones selected to 
generate composite kinematic templates. The four lines correspond to the 
listed ages for instantaneous burst. The four template spectra in Figure \ref{fig:tempspec} 
use weighted combinations of these four ages. The datapoints show spectra available 
for stacking in two metallicity bins. We have tested using the two metallicity 
bins and found no difference in the extracted kinematics. Our final stacks use [Fe/H] both 
bins. The oldest isochrones are best populated, but some stars up to the main sequence 
turnoff are available for all four ages.}
\label{fig:compiso}
\end{figure}

\begin{figure*}
\centering
\includegraphics [scale=0.9,angle=0]{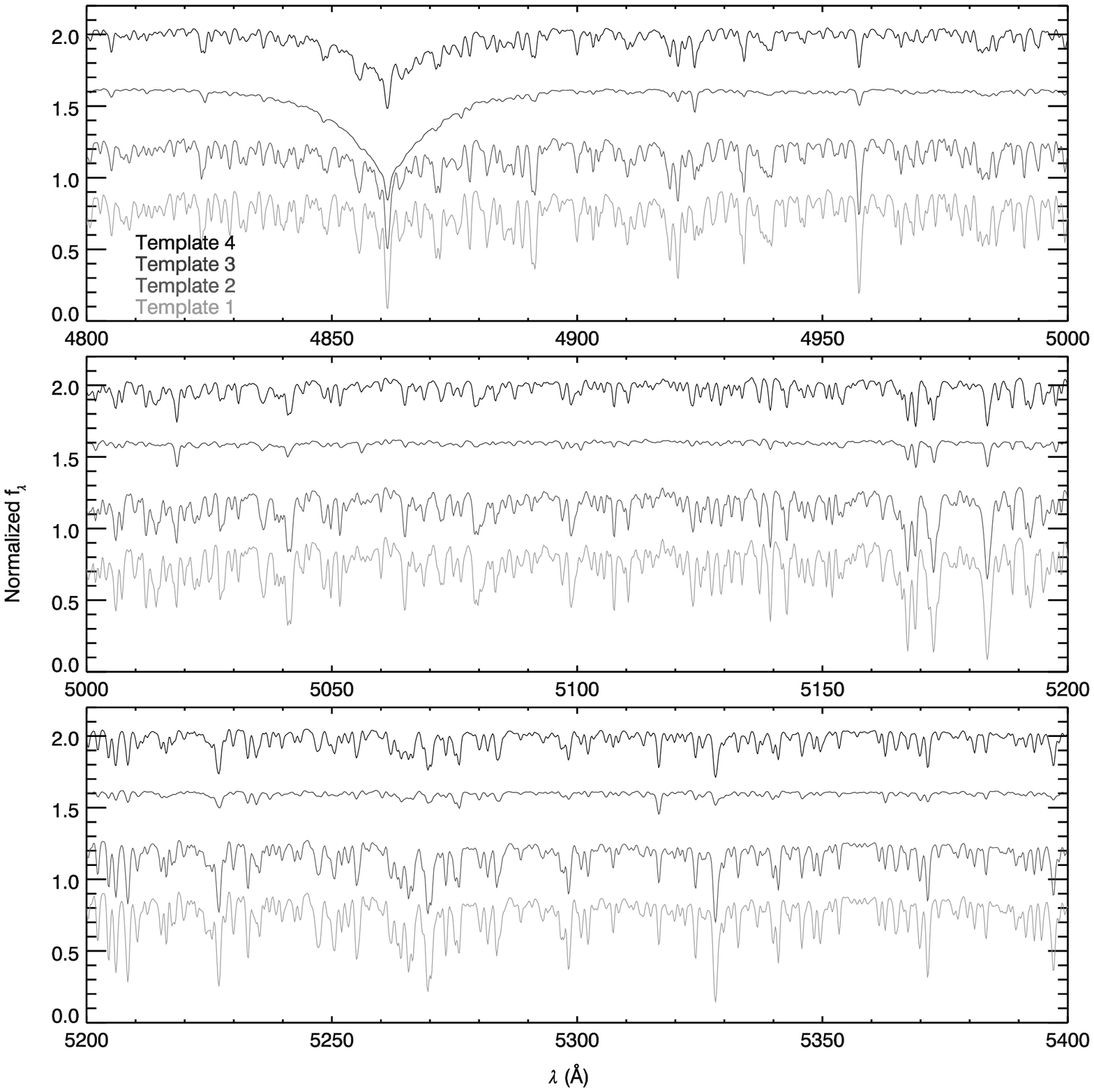}
\caption{Spectra for the four normalized, composite kinematic templates. Template one has a purely 
old stellar population, and the other three have a contribution from a progressively 
younger burst as specified in Figure \ref{fig:compiso}. From bottom to top 
are templates 1--4 with offsets of 0.4. The data are shown in their 
native R=10,000 resolution. Our stellar kinematic extraction uses these four templates with 
arbitrarily fitted weights to measure LOSVDs in each bin.}
\label{fig:tempspec}
\end{figure*}

 The composite templates were constructed by stacking stars from ELODIE \citep{Prug01} with weights. The weights 
were determined by mapping the stars' T$_{eff}$ and logg onto isochrones, populating the density along the 
isochrone according to an initial mass function, and solving for the weights that most accurately reproduced the 
model population. Similar steps are done by all stellar population synthesis models \citep[e.g.][]{Vazd03,BC03}. 
Let $k$ represent the index of the single-age populations, and f$_k$ represent the mass fraction being put into 
the isochrone. Along each isochrone $k$ there are tabulated properties at index $j$ for the 
stellar mass of a bin (m$_{j,k}$), the bin size step in mass ($\Delta$m$_{j,k}$), the logarithm of the effective 
temperate ($\log T_{eff,j,k}$), the logarithm of the surface gravity ($\log g_{j,k}$), and the absolute V-band 
magnitude (M$_{V,j,k}$). Let $i$ be the index for the subset of stars selected from ELODIE with 
m$_{V,i}$ as the apparent V-band magnitude, $\log T_{eff,i}$ as the logarithm of the effective temperature, 
$\log g_{i}$ as the logarithm of the surface gravity, $\sigma_T$ as the uncertainty on the 
$\log T_{eff,i}$, and $\sigma_g$ as the uncertainty on $\log g_{i}$. The exact values of 
the uncertainties are unimportant and serve to slightly smooth or sharpen the 
the stellar weights. We estimate $\sigma_T=0.02$ and $\sigma_g=0.3$
throughout. Let $m$ be the index for an internal sums run over the ELODIE stars, 
and let $l_{m,j,k}$ represent the relative likelihood that 
star $m$ is a good representative for mass bin $j$ according to its $T_{eff}$ and $\log g$ parameters. 
Finally, let $\xi(m_{j})\Delta m_{j,k}$ represent the number density of stars from the IMF, which 
is normalized to integrate to one solar mass. 
With this, we make a composite stack with weight $\eta_i$ for the flux-calibrated star $i$ as, 
\begin{align}
\label{eq:twei}
-2\ln l_{m,j,k}=\left(\frac{\log T_{eff,m}-\log T_{eff,j,k}}{\sigma_T}\right)^2 \nonumber \\
               +\left(\frac{\log g_{m}-\log g_{j,k}}{\sigma_g}\right)^2 \\
\eta_i=\sum_{k} \biggl( f_k \times \sum_{j} \Bigl( 10^{\frac{m_{v,i}-M_{v,j,k}}{2.5}} \times \xi(m_{j,k}) \nonumber \\
\times \frac{l_{i,j,k}}{\sum_{m} l_{m,j,k}} \times \Delta m_{j,k} \Bigl) \biggl).
\end{align}
In particular, we have used a Chabrier IMF \citep{Chab03} and the Padova isochrones \citep{Bres12}. The stellar 
parameters were taken from ELODIE's TGMET values \citep{Katz98}. We have used all stars with -1.0$<$[Fe/H]$<$0.2, and 
stars of any metallicity when T$_{eff}>$9000K. The metallicity of very hot stars has very little impact on the 
composite EWs, and the isochrones could not be well populated without the more forgiving criterion. The distribution of 
stars and the isochrones are shown in Figure \ref{fig:compiso}. Four composites 
were generated as instantaneous burst models with the following sets of mass-weighted fractions: 100\% at 13 Gyr, 
50\% each of 13 Gyr and 1 Gyr, 50\% each of 13 Gyr and 250 Myr, and 50\% each of 13 Gyr and 50 Myr. We experimented 
with additional criteria, such as drawing only -0.3$<$[Fe/H]$<$0.2 stars for the $<$1 Gyr tracks and 
-1.0$\le$[Fe/H]$<$-0.3 stars for the $\ge$1 Gyr tracks, but such changes had no measurable impact on our 
kinematic estimates. The LOSVD software can then determine optimal weights between these four templates. This 
parametrization constrains any fit to have, at minimum, a 50\% mass-weighted fraction in an old population. Templates 1 and 
3 take on EWs in Mg$_{\textrm{b}}$ that are larger and smaller, respectively, than any galaxy 
regions we observe, meaning that linear combinations of templates can match all the data. 
We show the four composite template spectra after normalization in Figure \ref{fig:tempspec}. 

\section{MGE DM Halos}
\label{app:mgedm}
\par While simple equations exist for the circular velocity of a gNFW halo \citep{Dutt05}, it is more convenient 
in the JAM framework to approximate the halo in MGE terms. This functionality is available with the IDL \texttt{MGE\_fit\_1d} 
routine provided by M.~Cappellari, and we have adapted it into FORTRAN for compatibility with our MCMC pipeline. 
The only change was with the underlying constrained minimization routine. \texttt{MGE\_fit\_1d} uses 
\texttt{MPFIT} \citep{Mark09} and its underlying MINPACK algorithm \citep{More78}, while we used a version of \texttt{conmin} 
\citep{Vand73} adapted from FORTRAN77 to FORTRAN90. We fit the MGE terms to the gNFW function over physical scales 
corresponding to one-tenth the size of a fiber and ten times the largest galacto-centric distance of any fiber. 
On average, 10 MGE terms can reproduce the gNFW function to a maximum density deviation of 5\% over this range. 

\section{Tests on vertical orbital anisotropy}
\label{app:betatest}
\par The stellar orbits for the very-late type galaxies in this study are expected to 
be significantly flattened. Early work, without kinematics, found a large scatter in the 
orbital anisotropy with a sample of massive 
spirals from $0.2<\beta_z<0.9$ \citep{vdKr99}. 
The SAURON project has found that giant, fast-rotator 
galaxies usually exhibit small, positive $\beta_z$ anisotropies \citep{Capp07}. \cite{Will09} has extended such 
studies to S0 systems and confirm that the JAM methods work well on lower mass and diskier systems. 
More directly relevant to our sample, several works have investigated SVE 
anisotropies in late-type galaxies 
by measuring both stellar and gaseous kinematics. These works operate in a cylindrical frame, and require the 
conditions behind epicycle theory to apply. When this is true and the inclination is known, 
stellar kinematic measurements along the major and minor axis completely determine the SVE 
\citep{Shap03}. While the SVE is technically constrained with these measurements, the quality of the 
constraint depends on location along the rotation curve. Along a linearly-rising portion of a 
rotation curve, the tangential and radial dispersions are nearly equal, and the dispersion-map 
will be nearly constant with azimuth angle. More information can come from measuring gas 
kinematics and using the asymmetric drift equation \citep{Binn08} as an additional constraint. This method, 
rather than the azimuthal variation, is used by the DiskMass Survey to constrain 
SVE ratios \citep{Bers10b,West11}. Most published measurements agree that late-type 
systems should have very high $\beta_z$ anisotropies, approaching the physical limit 
of $\beta_z=1$. \cite{Noord08} measure NGC~2985 (Hubble-type SAab) to have $\beta_z=0.5$. 
For UGC~463 (Hubble-type SABc), \cite{West11} find $\beta_z=0.8$. \cite{Shap03} measured 
a trend against a sample of Sa--Sbc galaxies whereby $\beta_z$ steadily rose. \cite{Gers12} 
continued this work by measuring two later types, NGC~2280 (Hubble-type Scd) and NGC~3810 
(Hubble-type Sc), and found $\beta_z=0.94$ and $\beta_z=0.92$, respectively. These two 
measurements continued the trend fit to the earlier-type galaxies (Figure 4 of \cite{Gers12}). 
While it may be reasonable to set a prior on $\beta_z$ peaked at large values, we found 
that some of our data are poorly represented by large orbital anisotropies. Therefore, we have selected a flat prior on $\beta_z$. 
Of course orbital anisotropy can also be measured by using higher order velocity moments, but this is not 
possible at our S/N and instrumental velocity resolution.  
\par
\begin{figure}
\centering
\includegraphics [scale=0.85]{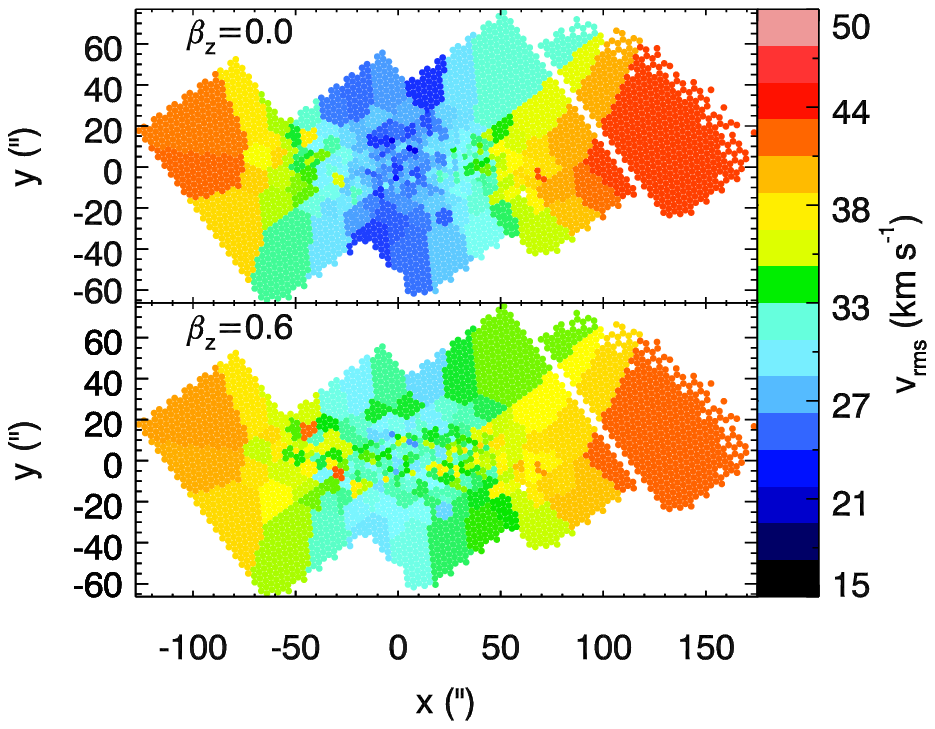}
\caption{Second-moment velocity fields simulated with JAM for a simple parameter set. Noise equal to that 
in the data has been added. Two 
values for the orbital anisotropy are shown: a low one that represents most of our data and a higher 
one as has been found by earlier studies. More flattened velocity ellipsoids generate larger dispersions at small radii. 
Steeper DM profiles will also generate larger second-moment velocities 
at small radii. The full velocity map is less degenerate between $\beta_z$ and $\gamma$ than 
a single position-velocity cut. \textit{\textbf{Top}} $\beta_z=0.0$. \textit{\textbf{Bottom}} $\beta_z=0.6$. 
The difference between the two is 
visible by eye, and our simulations can recover $\beta_z$ and $\gamma$ with little degeneracy.}\label{fig:picbeta}
\end{figure}

\begin{figure}
\centering
\includegraphics [scale=0.85]{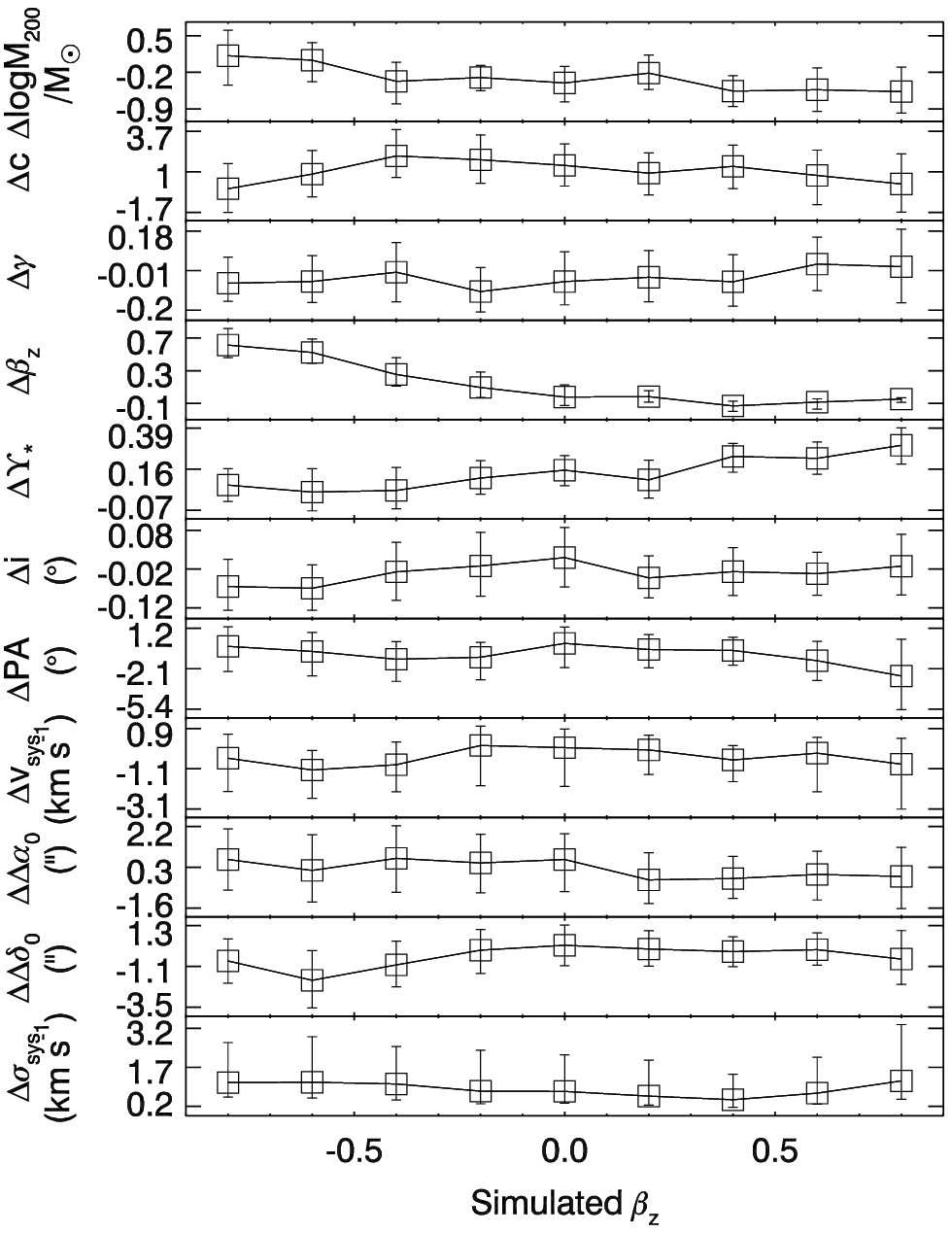}
\caption{Residuals of the extracted parameters from simulated data with a range of input orbital 
anisotropies. The data sampling and S/N ratios were the same as for the NGC~959 data. 
The error bars indicate the 68\% confidence interval from MCMC samples. The residuals for the parameters other than 
$\beta_z$ do not 
show a correlation with the true $\beta_z$. For the most negative $\beta_z$ values, the MCMC 
estimate returns a value significantly higher than the correct one, but the estimate becomes more 
accurate for the more plausible range of $\beta_z>0$ as found in real galaxies. The reason may 
be that the walkers are initialized on the positive end of $\beta_z$. We are 
mainly concerned that $\gamma$ does not show a bias in these simulations.}\label{fig:betasim}
\end{figure}

 Readers familiar with mass modeling of giant elliptical galaxies may expect a strong negative covariance 
between the estimation of orbital anisotropy and $\gamma$. For instance, \cite{vdM94} made models of M87 
for a range of orbital anisotropies, $\beta$, in his Figure 10. Models with large $\beta$ 
have higher central dispersions. However, it is important to remember that the M87 model is in a spherical 
coordinate frame with $\sigma_\phi=\sigma_\theta$, and so large values of $\beta$ only have radial 
dispersion support. Alternatively, with disky galaxies in a cylindrical frame, the ratio of 
radial and tangential 
dispersions is set by the Jeans equations and stay of order unity near the center. Therefore, 
even when $\beta_z$ is near unity in a disky system, $\beta$ will be significantly softened and 
not approach unity. We have made a numerical test on 
our ability to measure $\beta_z$ and de-couple it from $\gamma$ as follows. A model realization was generated from a 
set of parameters with $\beta_z$ values ranging from -0.8--0.8 in the bin positions of the NGC~2976 
observations. The remaining parameters were set as 
$\left\{\log M_{200}, c, \gamma, \Upsilon_{*}, i, PA, v_{sys}, \Delta\alpha_o, \Delta\delta_0, \sigma_{sys}\right\}=
\left\{11.3,15.0,0.4,0.7,62.0,-36.0,6.0,0.0,0.0,5.0\right\}$. 
The JAM models only predict the second-moment velocity, $v_{rms}^2$, while we need both $v_{los}$ 
and $\sigma$ in order to include v$_{sys}$ in our fitting process. Following the JAM formalism \citep{Capp08}, we have 
used the parameter  
\begin{equation}
\label{eq:kappa}
\kappa_{k}\equiv\frac{\left[\overline{v_\phi}\right]}{\left(\left[\overline{v^2_\phi}\right]-\left[\overline{v^2_R}\right]\right)^{1/2}} 
\end{equation} 
to divide the second-moment tangential velocity into rotation and dispersion. The line-of-sight first-moment velocity 
can then be calculated by Equation 38 of \cite{Capp08} through a double numerical integration. For this exercise, we 
have set $\kappa_k=0.7$. $\kappa_k=0.0$ gives a non-rotating system and $\kappa_k=1.0$ makes the radial and 
tangential dispersion equal. 
Since the actual fit is still to the second-moment velocity, the choice of $\kappa$ is 
unimportant. We then fit the simulated data with our MCMC pipeline. The original NGC~2976 bins and uncertainties were 
used in the likelihood fit. The second-moment velocity fields for 
two anisotropies, representing an isotropic field and the largely anisotropic 
values that previous studies have found, are shown in Figure \ref{fig:picbeta}. JAM models with 
even larger orbital anisotropies ($\beta_z=0.8$) become so extreme that the second-moment velocities peak 
in the middle. The MCMC fits do not perfectly converge on the input model for the most 
vertically biased models, but over the most likely range (positive $\beta_z$) 
the residuals are consistent with the level of observational uncertainty 
used in the simulations. We highlight several important points from this figure. First, the 
orbital anisotropy can, with limited precision, be inferred from our data in 
Figure \ref{fig:betasim}. When galaxies have truly 
large vertical orbital anisotropies ($\beta_z>0.5$), we find them in our MCMC analysis. Second, there is very little 
covariance between $\beta_z$ and $\gamma$. The more important covariances are between $\Upsilon_{*}$, $c$, $M_{200}$, and $\gamma$. 
Our results on the DM density profiles are largely independent of our results on the orbital anisotropy. 
Finally, we do not consistently find the large orbital anisotropies (in \S \ref{sec_starmodels}) that other groups have 
with similar data and systems. The reason for this potential discrepancy is not known, but we will propose some 
possibilities in \S \ref{sec:dis}. 

\section{Tests on MCMC convergence}
\label{app:MCMCtest}
\par Three parameters in \texttt{emcee} mainly determine whether the parameter space has been fully sampled and 
all minima have been found: the number of ``walkers'', the length of ``burn-in'', 
and the number of samples kept per walker. The number of walkers is suggested by the \texttt{emcee} 
documentation to be at least 
twice the number of parameter dimensions, but walkers numbering an order of magnitude more than the 
problem dimensionality are more common. Some number of the early MCMC samples are likely to represent 
the initialization choices rather than the underlying probability distribution. Usually, a number of 
these first samples are ignored in a ``burn-in'' phase. Lastly, the number of samples in the MCMC chain 
kept determine the resolution quality to the fit. We have set the number of burn-ins and the number of 
kept samples at 20 and 250 for each walker. A simple convergence test was run by varying the number of 
walkers on simulated kinematic data sets. The data were simulated for the NGC~959 footprint and S/N with 
parameters $\left\{\log M_{200}, c, \gamma, \beta_z, \Upsilon_{*}, i, PA, v_{sys}, \Delta\alpha_o, \Delta\delta_0\right\}=
\left\{11.3,15.0,0.4,0.3,0.7,55.2,67.9,595.0,0.0,0.0\right\}$. Five independent realizations were made by 
drawing normally distributed errors for both the gaseous and stellar kinematic fields. Each field was 
analyzed with 50, 100, 150, 250, and 400 walkers. The central estimates and dispersions of all parameters 
did not change for solutions with $>$100 walkers. We conservatively chose 250 walkers for our final fits.

\par The recovered parameters show several useful trends. First, some parameters, such as 
$c$, $\gamma$, $\Upsilon_{*}$, $i$, and $PA$ show slightly more scatter when using only 50 walkers. 
Along with the number of walkers, the total number of samples is decreased. The increased scatter 
means that we have not reached complete convergence in the MCMC fit. The scatter and bias stay 
flat for all simulations with more than 50 walkers. The remaining scatter at larger walker numbers 
represents the limited S/N in the observations. As a conservative choice, we have selected 250 
walkers for all the fits to real data. Second, we see that several parameters have slight biases from 
their true values. We have selected the simulation to have some parameters offset from the prior peaks and 
the posterior probabilities are being influenced by the priors for M$_{200}$ and $\Upsilon_{*}$. 
$i$ shows a small bias because of the asymmetric limits required by the MGE terms. For these three variables, 
the biases are still quite small.
\section{Age Corrections for Lick Indices}
\label{app:Lick}
\par \texttt{EZ\_AGES} operates by iteratively solving for age, 
[Fe/H], and [$\alpha$/Fe] by interpolating index measurements to the \cite{Schi07} stellar population grids. 
For our wavelengths of interest, H$\beta$ and $\langle$Fe$\rangle$ are first selected as relatively clean indicators 
of age and [Fe/H]. \texttt{EZ\_AGES} matches these two index measurements to the grids with the remaining elements 
fixed. \texttt{EZ\_AGES} can use either a solar scaled grid or an $\alpha$-enhanced grid \citep{Salas00}. 
Using the Large Magellanic Cloud (LMC) as a proxy for our galaxies, the
[$\alpha$/Fe] values of individual stars \citep{Johns06,Mucci08,Pompe08} and 
star clusters \citep{Coluc12} are at solar values until [Fe/H]$<$-1 and 
less $\alpha$-enhanced than Milky Way stars at fixed [Fe/H]. We have tried both options but found the same results with both. 
\texttt{EZ\_AGES} then selects indices that are sensitive to other elements. 
In our case, the relevant index is Mg$_b$ and the relevant abundance is [Mg/Fe]. \texttt{EZ\_AGES} re-estimates 
age and [Fe/H] in the H$\beta$-Mg$_b$ plane. If the estimates differ, primarily in estimated [Fe/H], 
the parameter [Mg/Fe] is adjusted until the two [Fe/H] estimates agree. There are no good indicators of [O/Fe] at 
low resolution, so we have adopted the default assumption of [O/Fe]=0.0. We have also let the other important 
$\alpha$-elements, Na, Si, and Ti, track Mg. \texttt{LICK\_EW} propagates observational uncertainties in the 
spectrum into Lick index uncertainties, and \texttt{EZ\_AGES} further propagates these into stellar population parameter 
uncertainties. \cite{Grav08} provide tests on globular cluster data with higher resolution spectroscopy and 
validate their accuracy to 0.15 dex in age and 0.1-0.2 dex in [Fe/H] and [Mg/Fe]. They also compare 
parameters estimated for a galaxy sample \citep{Thom05} that has been studied with an earlier inversion 
technique \citep{Thom03}. Some small zeropoint offsets are found, but the two methods generally agree. 
This algorithm has recently been employed by a group studying population gradients in the far outskirts of nearby 
elliptical galaxies \citep{Gree12,Gree13} using IFU data similar to ours. 
\par We first must make a correction for the H$\beta$ index and our limited spectral coverage. To optimize the S/N 
at Mg$b$, we set up the grism so that the full range of the H$\beta$ was not recorded. The lowest wavelength 
that all our observing runs covered, including a 5\AA\ buffer, is 4855\AA, which misses the entire blue sideband and 
some of the main index window. The core of H$\beta$ is retained in our window. However, 
the kinematic templates we use cover this whole range and do well to approximate the observed spectra. We 
therefore measure the EW of a narrower index in both our data and the matching template-based 
kinematic model, and then the full H$\beta$ index in the template-based kinematic model. Let nH$\beta$ be this 
narrower H$\beta$ index, with a bandpass of 4857.000--4876.625\AA\ and sidebands from 4855.300--4857.000\AA\ and 
4876.623--4891.600\AA. The index is measured inside the \texttt{LICK\_EW} package. We form the corrected 
measurement as, 
\begin{equation}
\label{eq:chb}
H\beta=\frac{H\beta(model)\times nH\beta(data)}{nH\beta(model)}.
\end{equation} 
\par An important complication facing us is that our sample galaxies are not well represented as old, simple stellar 
populations. \texttt{EZ\_AGES} performs best, with minimal zeropoint uncertainty, for very old ($>$10 Gyr) populations. 
Furthermore, the \cite{Schi07} grids are limited to 1--15 Gyr ages and -1.3$<$[Fe/H]$<$0.3. However, we do have 
some information on the separation of old and young stellar populations by the nature of our LOSVD template selection. We 
will leverage the kinematic template weights and measurements of Lick indices on the composite templates to approximate 
the Lick indices of the old stellar populations. We give the Lick indices and the narrow H$\beta$ as measured off the 
four kinematic templates in Table \ref{tab:licktemp}. 
\begin{deluxetable*}{lrrrrrrrrr}
\tabletypesize{\scriptsize}
\tablecaption{Index measurements for the kinematic templates.\label{tab:licktemp}}
\tablewidth{0pt}
\tablehead{
\colhead{Template} & \colhead{H$\beta$} & \colhead{nH$\beta$} & \colhead{Fe5015} & \colhead{Mg$_1$} & \colhead{Mg$_2$} & \colhead{Mg$_\textrm{b}$} & \colhead{Fe5270} & \colhead{Fe5335} & \colhead{Fe5406} \\
\colhead{} & \colhead{(\AA)} & \colhead{(\AA)} & \colhead{(\AA)} & \colhead{(mag)} & \colhead{(mag)} & \colhead{(\AA)} & \colhead{(\AA)} & \colhead{(\AA)} & \colhead{(\AA)}
}
\startdata
1  & 1.71 & 0.72 & 5.03 & 0.0068 & 0.0847 & 2.82 & 3.09 & 2.59 & 1.73 \\
2  & 3.50 & 1.08 & 3.74 & 0.0022 & 0.0527 & 1.65 & 2.10 & 1.75 & 1.05 \\
3  & 7.65 & 2.26 & 1.22 &-0.0007 & 0.0200 & 0.64 & 0.69 & 0.59 & 0.45 \\
4  & 3.80 & 1.00 & 2.33 &-0.0026 & 0.0118 & 0.25 & 0.84 & 0.83 & 0.40
\enddata
\end{deluxetable*}

\begin{deluxetable*}{lrrrrrrrrr}
\tabletypesize{\scriptsize}
\tablecaption{Luminosity weight of 13 Gyr population in each composite template.\label{tab:licklum}}
\tablewidth{0pt}
\tablehead{
\colhead{Template} & \colhead{lw(H$\beta$)} & \colhead{lw(nH$\beta$)} & \colhead{lw(Fe5015)} &\colhead{lw(Mg$_1$)} & \colhead{lw(Mg$_2$)} & \colhead{lw(Mg$_b$)} & \colhead{lw(Fe5270)} & \colhead{lw(Fe5335)} & \colhead{lw(Fe5406)}
}
\startdata
2  & 0.133 & 0.160 & 0.083 & 0.129 & 0.106 & 0.114 & 0.127 & 0.148 & 0.134 \\
3  & 0.028 & 0.036 & 0.016 & 0.087 & 0.034 & 0.027 & 0.036 & 0.031 & 0.037 \\
4  & 0.000 & 0.021 & 0.010 & 0.036 & 0.000 & 0.007 & 0.014 & 0.019 & 0.009
\enddata
\end{deluxetable*}

\par It is impossible to find valid solutions in \texttt{EZ\_AGES} for the directly measured Lick indices. Stellar populations 
younger than those in the model libraries suppress the Fe and Mg line strengths to values that fall off all available grids. 
We make a crude recovery of the old population indices as follows. In its most general form, the equivalent width 
of a composite population cannot be expressed as a linear combination of its component equivalent widths. However, when the 
continua of the populations only vary slowly over the index bandpass, such an equation is approximately correct. Let 
EW$_t$ be the total equivalent width, EW$_i$ be the equivalent width of the component populations in a particular index, 
$f_{\lambda,i}$ be the flux density, $f_{\lambda,c,i}$ be the continuum flux density, $\overline{f_{\lambda,c,i}}$ be the 
average flux density over the index bandpass, and f$_i$ be the normalized luminosity weight 
of component $i$ as found during the LOSVD fit. Then, 
\begin{align}
\label{eq:eq_corr}
EW_t&\equiv\int(1-\frac{\sum_i f_i f_{\lambda,i}}{\sum_i f_i f_{\lambda,c,i}})d\lambda \nonumber \\
&\approx \frac{\sum_i f_i EW_i \overline{f_{\lambda,c,i}}}{\sum_i f_i \overline{f_{\lambda,c,i}}}.
\end{align} 
The ELODIE templates may not accurately represent the stellar population parameters contained in our target galaxies. As such, their 
absolute equivalent widths should not be used. We instead form a correction to the measured 
equivalent widths by the ratio of the old population template and the total template equivalent widths. We count both the 
13 and 3 Gyr populations as old, as \texttt{EZ\_AGES} has models that bracket both, and form the following corrected ratio for 
each atomic-type equivalent width. Since the kinematic templates are 
continuum normalized, the average continua fall out of the equation. 
Let lw$_{i,j}$ be the luminosity weight of the 13 Gyr population in index j of 
template i (with lw$\equiv$1 for i$=1$) from Table \ref{tab:licklum}, $\mu_i$ be the 
LOSVD-measured template weight, and tEW$_{i,j}$ be the template's equivalent width from 
Table \ref{tab:licktemp}. We form corrected equivalent widths, EW$_j$, from the directly measured values, mEW$_j$, as
\begin{align}
\label{eq:lick_old}
EW_{j}=mEW_j \times \nonumber \\
\frac{\sum_{i=1,2} \mu_i tEW_{i,j} + \sum_{i=3,4} \mu_i lw_{i,j} tEW_{i,j}}{(\sum_{i=1,2} \mu_i + \sum_{i=3,4} \mu_i lw_{i,j})\times(\sum_{i} \mu_i tEW_{i,j})}.
\end{align} 
Mg$_1$ and Mg$_2$ are molecular-type equivalent widths with units of magnitudes, and we measure their values by first 
transforming the magnitudes into linear units, applying the linear correction, and 
then transforming back into magnitudes. The S/N of the index measurements in individual fibers or even the kinematic bins are 
too noisy to be useful. So, we have made one bin per galaxy. 
The two most common ways that the data fall off \texttt{EZ\_AGES} grids is that 
H$\beta$ is too low or $\langle$Fe$\rangle$ is too low. If we average all the data and push H$\beta$ up onto the grid to the location of 
a 10 Gyr age, we find [Fe/H]$\sim$-1 and [Mg/Fe]$\sim$0.5. This value of [Mg/Fe] is surprisingly high considering 
the values measured for the LMC. The uncorrected values also fall off the grids, but with 
Mg$_{\textrm{b}}$ and often $\langle$Fe$\rangle$ that are too small for the 
lowest abundance models. 

\end{document}